\documentclass[12pt]{article}
\usepackage[left=1in,top=1in,right=1in,bottom=1in,nohead]{geometry}
\usepackage{amsmath,amssymb,comment}
\usepackage{mathtools}
\usepackage{rotating}
\usepackage{graphicx}

\usepackage{amsmath,amssymb}
\usepackage{amssymb}
\usepackage{bm}                % Bold font in formula
\usepackage{mathrsfs}      %HUA TI
\usepackage{amsmath, amsthm}
\usepackage{graphicx,epsf,subfigure,pstricks,pst-node,psfrag,amsthm,amssymb,amsmath}
\usepackage{float}
\usepackage{array}
\usepackage{authblk}
\usepackage{natbib}
\usepackage{url}
\usepackage{hyperref}
\newtheorem{theorem}{Theorem}

\begin{document}
\title{Graph signal denoising using $t$-shrinkage priors}
\date{} 
\author[1]{Sayantan Banerjee}
\author[2,*]{Weining Shen}

\affil[1]{Operations Management and Quantitative Techniques Area, Indian Institute of Management Indore, M.P., India}
\affil[2]{Department of Statistics, University of California, Irvine, CA, USA}  
\affil[*]{Corresponding author; \href{mailto:weinings@uci.edu}{weinings@uci.edu}}
\maketitle

\begin{abstract}
We study the graph signal denoising problem by estimating a piecewise constant signal over an undirected graph. We propose a new Bayesian approach that first converts a general graph to a chain graph via the depth first search algorithm, and then imposes a heavy-tailed $t$-shrinkage prior on the differences between consecutive signals over the induced chain graph.  We show that the posterior computation can be conveniently conducted by fully exploring the conjugacy structure in the model. We also  derive the posterior contraction rate for the proposed estimator, and show that this rate is optimal up to a logarithmic factor, besides automatically adapting to the unknown edge sparsity level of the graph. We demonstrate the excellent empirical performance of the proposed method via extensive simulation studies and applications to stock market data. 
\end{abstract}

\begin{quotation}

\noindent {\it Keywords:} Adaptive estimation; Bayesian shrinkage; depth first search; graph denoising; heavy tailed prior; posterior contraction rate.
\end{quotation}\par

\section{Introduction}
Let $G = (V,E)$ be an undirected graph with node (or, vertex) set $V = \{1,\ldots,n\} $ and edge set $E$. The main focus of this paper is the estimation of a piecewise constant signal $\bm{\theta} = (\theta_1,\ldots,\theta_n)^T \in \mathbb{R}^n$ defined over $G$. In particular, we consider the following model,
\begin{align}\label{eq:model}
y_i = \theta_{i} + \epsilon_i,\,i = 1,\ldots,n; ~~ \epsilon_i \stackrel{\text{i.i.d}}{\sim} N(0,\sigma^2),
\end{align}
where  $\theta_{i}$ is the signal associated with the node $i$ for every $i=1,\ldots,n$, $\epsilon_1,\ldots,\epsilon_n$ are independent and identically distributed (i.i.d) random Gaussian noises with an unknown variance parameter $\sigma^2$, and $y_1,\ldots,y_n$ are the observed data points. By assuming that $\bm{\theta}$ has a {\it piecewise constant} structure, we mean that for any pair of indexes $(i,j) \in E$, there is a chance that the associated signals satisfy $\theta_{i} = \theta_{j}$. 

Model \eqref{eq:model} is often referred to as a graph denoising model, and it is closely connected to the vast literature on change point detection  \citep{tartakovsky2014sequential,brodsky2013nonparametric} and shape constrained regression \citep{guntuboyina2018nonparametric}, especially when $G$ is a chain graph. In general, the graph denoising model has been widely studied in various areas such as image segmentation and denoising \citep{besag1986statistical}, signal processing \citep{gavish2010multiscale}, and network analysis \citep{crovella2003graph,shuman2013emerging}. In the literature, regularization has become a popular strategy to solve the graph denoising problem \citep{barron1999risk,birge2007minimal}. In particular, \citet{Fan2018} proposed an $\ell_0$-edge denoising method by considering the following minimization problem,  
\begin{align}\label{l0}
 \arg \min_{\bm{\theta} \in \mathbb{R}^n} \dfrac{1}{2}\|\bm{y} - \bm{\theta}\|_2^2 + \lambda \sum_{(i,j) \in E} I(\theta_i \neq \theta_j),
\end{align}
where $\bm{y} = (y_1,\ldots,y_n)^T$ is the observed data, and $\lambda > 0$ is the tuning parameter. Also, throughout the paper, we use the notation $\|\bm{x}\|_2$ to denote the $L_2$-norm of a vector $\bm{x}$, and $I(\cdot)$ is the indicator function. Due to the combinatorial nature of the objective function in \eqref{l0}, it is usually very difficult to find the exact solution to the minimization for general graphs. \citet{Fan2018} hence developed an efficient approximation algorithm for solving \eqref{l0} within a polynomial time and showed that the resulting solution achieved the same risk bound (up to a constant factor) with that of the exact minimizer of \eqref{l0}. Alternatively, \citet{Padilla2017} proposed a fused Lasso approach by considering an $L_1$ convex relaxation of \eqref{l0}, i.e., 
\begin{align}\label{l1}
 \arg \min_{\bm{\theta} \in \mathbb{R}^n} \dfrac{1}{2}\|\bm{y} - \bm{\theta}\|_2^2 + \lambda \sum_{(i,j) \in E} |\theta_i - \theta_j|. 
\end{align}
The key idea of that paper is to first perform a depth-first search (DFS) \citep{tarjan1972depth} that converts an arbitrary graph to a linear chain, and then apply the fused Lasso \citep{rudin1992nonlinear,tibshirani2005sparsity} for denoising purpose (also known as total variation denoising). The authors also provided a theoretical justification for using DFS, that is, they showed that the total variation in the signal defined over the chain graph obtained from DFS is at most twice as its total variation over the original graph. Based on this result, the authors obtained an upper bound for the mean squared error (MSE) of their proposed signal estimators and established minimax optimality for the obtained bound with respect to tree-structured graphs. 
%Recent work in the frequentist literature has proposed an algorithm called the DFS fused lasso, which tries to use the computational advantage of fused lasso over chain graphs to solve the fused lasso problem for arbitrary graphs. 

Despite their success in computation and theory, it still remains unclear how the aforementioned regularization methods can be used to conduct statistical inference for graph signal denoising problems; and this is the main focus of our paper. In particular, we propose a simple-yet-useful Bayesian solution to graph denoising problem. The main idea is to first perform DFS that converts a general graph to a chain graph, and then conduct Bayesian fusion estimation over the DFS-induced chain graph by shrinking the difference between consecutive pairs of signals towards $0$. One advantage of adopting a full Bayesian framework is that it automatically accounts for the uncertainty from both DFS and fusion estimation. Moreover, the inference can be conveniently conducted based on posterior samples. For Bayesian fusion estimation, 
we choose to use the recently studied $t$-shrinkage prior \citep{song2017nearly, Song2019}, which has been shown to have a better theoretical (e.g., in terms of posterior contraction rate) and empirical performance \citep{bhattacharya2015dirichlet,castillo2015bayesian} than the conventional Laplace prior in Bayesian fused Lasso \citep{kyung2010penalized}. We show that the posterior computation can be conveniently implemented using a Gibbs sampler thanks to the conjugacy structure in our model. We also derive a posterior contraction rate for the estimated signals with respect to the class of ``edge sparse'' graphs. To be precise, we show that when $G$ satisfies $\sum_{(i,j) \in E} I(\theta_i \neq \theta_j) \leq s$, then with posterior probability tending to one, the mean squared error for $\bm{\theta}$ is of order $ s \log n/ n $, which only differs with the oracle rate $s /n$ (assuming that the true graph sparsity structure is known) by a logarithmic factor in $n$. This result provides a useful theoretical justification for our proposed method, and also reveals a desirable adaptive estimation property since we do not require the knowledge of the graph edge sparsity parameter $s$ when constructing the prior while the posterior can automatically achieve the optimal rate of convergence with respect to different values of $s$. 

Despite the fast growing literature on change point detection and shape-constrained regression for linear chains, the exact minimax rate for estimating a piecewise constant or monotonic signal over the linear chain is only obtained  recently by \citet{gao2020estimation}. In the Bayesian domain, several approaches have been proposed to fit such models, such as the recursion method \citep{hutter2007exact}, Bayesian fused Lasso \citep{kyung2010penalized}, and Bayesian trend filtering  \citep{roualdes2015bayesian}. The posterior contraction property for those methods, however, remains largely unknown until recently. For example, \citet{Song2019} proposed to fit a
Bayesian piecewise constant model using a heavy-tailed shrinkage prior and obtained the posterior contraction rate for the estimated signals. \citet{liu2017empirical}  considered a Bayesian piecewise polynomial model using an empirical Bayesian approach and obtained some posterior contraction rate and  structure recovery results. Bayesian piecewise constant models on general graphs, to the best of our knowledge, still remain unsolved to date; and our theoretical contribution in this paper is to fill this gap.

The rest of the paper is organized as follows. We describe the proposed methodology in Section \ref{sec:method} and discuss an efficient posterior computational algorithm in Section \ref{sec:compute}. We then present the theoretical results in Section \ref{sec:theory}, and demonstrate the excellent empirical performance of our method via simulations and real data examples in Sections \ref{sec:simulation} and \ref{sec:real}, respectively. We discuss several future work directions in Section \ref{sec:dis}. Additional numerical results and proof of theorems are given in the Appendix.

\section{Bayesian Model Specification}\label{sec:method}
We propose a two-step Bayesian approach for solving the piecewise constant model in \eqref{eq:model}. We first perform a depth-first search (DFS) over the graph $G$ to convert it into a chain graph, denoted by $G_C=(V,E_C)$, where $V = \{1,\ldots,n\}$ is the same set of nodes in $G$, and $E_C$ represents a new set of $(n-1)$ edges in $G_C$. Denote the root node for the DFS search by $r \in V$ and let $\theta_r$ be its associated signal. Since the choice of the root node $r$ is random, the DFS-induced graph $G_C$ is not unique. However, in our implementation, we can choose a set of different root nodes and then pool the corresponding posterior samples of $\bm{\theta}$ together for inference purpose. Doing so will also account for the randomness associated with DFS in the final results. As discussed in \citet{Padilla2017}, there are two benefits by converting the original graph to a chain graph using DFS. The first reason is the obvious computational benefit - by performing DFS, whose computational complexity is of order $O(|E|)$, we only need to search over a chain graph instead of a potentially complex graph $G$. The second reason is that the estimated $\bm{\theta}$ obtained from $G_C$ is essentially as accurate as that from the original graph $G$ because the total variation in $\bm{\theta}$ over the DFS-induced chain graph is bounded by the twice of that on the original graph. This property provides an intuitive justification for using DFS and will later help us derive the posterior contraction property for the proposed Bayesian estimator. 

Once we obtain the DFS-induced graph $G_C$, the next step is to conduct Bayesian analysis of $\bm{\theta}$ over $G_C$. Note that the fused Lasso penalty term in \eqref{l1} has a natural Bayesian analogue as the negative logarithm of the prior density corresponding to a Laplace distribution. However, it has been shown in the literature  \citep{bhattacharya2015dirichlet,castillo2015bayesian} that the induced posterior corresponding to a Laplace prior may have undesired posterior contraction property (e.g., a sub-optimal contraction rate). Therefore, motivated by the recent work in \citet{song2017nearly} and \citet{Song2019}, we consider a heavy-tailed prior for the difference between signals over edges to facilitate better convergence. In particular, we use a $t$-shrinkage prior on the successive differences of signals, i.e., $\theta_i - \theta_j$ for every $(i,j) \in E_C$ defined on the induced chain graph. Note that we also need to specify a prior on the signal corresponding to the root node $\theta_r$; and we use a normal prior for convenience. The prior specification is thus given as:
\begin{eqnarray}
	\theta_{r}\mid \sigma^2 &\sim& N(0,\lambda_0\sigma^2),\nonumber\\
	(\theta_i - \theta_j) \mid \sigma^2 &\stackrel{\text{ind}}{\sim}& t_{\nu}(m \sigma^2), ~~\text{for every~}~(i,j) \in E_C,\nonumber \\
	\sigma^2 &\sim& \text{IG}(a_\sigma,b_\sigma),
	\label{eqn:t-prior}
\end{eqnarray}
where IG denotes an inverse Gamma distribution, $\lambda_0$, $a_{\sigma}$ and $b_{\sigma}$ are positive hyperparameters, and $t_{\nu}(\xi)$ denotes a $t$-distribution with $\nu$ degrees of freedom and scale parameter $\xi$. Similarly with the Laplace distribution, the $t$-distribution can be expressed as a Normal-scale mixture with inverse Gamma mixing weights. Therefore we can further write the prior in \eqref{eqn:t-prior} as
\begin{eqnarray}\label{eq4}
\theta_{r}\mid \sigma^2 &\sim& N(0,\lambda_0\sigma^2),\nonumber\\
(\theta_i - \theta_j) \mid \sigma^2,\lambda_{k_{(i,j)}} &\stackrel{\text{ind}}{\sim}& N(0,\lambda_{k_{(i,j)}}\sigma^2), ~~\text{for every~}~(i,j)  \in E_C,\nonumber \\
\lambda_{k_{(i,j)}} &\stackrel{\text{i.i.d}}{\sim}& \text{IG}(a_t,b_t),~~\text{for every~}~(i,j)  \in E_C, \nonumber \\
\sigma^2 &\sim& \text{IG}(a_\sigma,b_\sigma),
\end{eqnarray}
where the hyperparameters $a_t$ and $b_t$ satisfy $\nu = 2a_t$ and $m = \sqrt{b_t/a_t}$, and $k_{(i,j)}$ is the index for the edge that connects nodes $i$ and $j$. 

The choice of hyperparameters in a hiearchical Bayesian model setup plays a crucial role. In our case, the hyperparamters corresponding to the Inverse Gamma prior for the variance parameter $\sigma^2$ are chosen in such a way that the prior is almost non-informative, e.g., $a_\sigma = b_\sigma = 0.5.$ For the choice of the hyperparameters $a_t$ and $b_t$, we follow the approach in \citet{Song2019}, i.e., we choose $a_t = 2$, and the scale parameter $m$ in the $t$-prior is chosen so as to satisfy $P(|t_{\nu}(m)| \geq \sqrt{\log(n)/n}) \approx n^{-1}$.
In the numerical studies, we also compare our method with Laplace-prior based method, in which we choose the hyperparameter $\lambda$ as $\lambda = \sqrt{2\log(n)}$ and the value of $\lambda_0$ as $5$ based on the recommendation in \citet{Song2019}. 

\section{Posterior computation}\label{sec:compute}
One of the nice features of the proposed method is that the posterior calculation can be implemented in a relatively easy way by exploring the conjugacy structure in the model. Given the Gaussian likelihood and prior specification in the previous section, 
the conditional posterior distribution of the parameters of interest can be conveniently obtained by the Gibbs sampling as follows, 
\begin{eqnarray}
\lambda_{k_{(i,j)}} \mid \cdot &\sim& \text{IG}\left(a_t + \dfrac{1}{2}, b_t + \dfrac{(\theta_i - \theta_j)^2}{2\sigma^2}\right),\, k_{(i,j)} \in \{1,\ldots,n-1\},~\text{for every}~ (i,j) \in E_C, \nonumber \\
\sigma^2 \mid \cdot & \sim & \text{IG}\left(a_\sigma + n,
b_\sigma + \dfrac{\theta_r^2}{2\lambda_0} + \dfrac{1}{2}\|\bm{y}-\bm{\theta}\|_2^2 + 
\sum_{(i,j) \in E_C} \dfrac{(\theta_i - \theta_j)^2}{2\lambda_{k_{(i,j)}}}\right),\nonumber \\
\theta_i\mid \cdot &\sim& N(\mu_i,\nu_i),\, i = \{1,\ldots,n\}\setminus\{r\},
\end{eqnarray}
where $\mid \cdot$ means conditional on the rest of other parameters and the data, $r$ is the root node for DFS, and 
\begin{eqnarray}
\nu_i^{-1} &=& \dfrac{1}{\sigma^2} + \dfrac{1}{\lambda_{k_{(i,j)}}\sigma^2}I((i,j) \in E_C) + \dfrac{1}{\lambda_{k_{(j,i)}}\sigma^2}I((j,i) \in E_C), \nonumber \\
\mu_i &=& \nu_i\left(\dfrac{y_i}{\sigma^2} + \dfrac{\theta_j}{\lambda_{k_{(i,j)}}\sigma^2}I((i,j) \in E_C) + \dfrac{\theta_j}{\lambda_{k_{(j,i)}}\sigma^2}I((j,i) \in E_C) \right) \nonumber
\end{eqnarray}
for every $i \in \{1,\ldots,n\}\setminus \{r\}$, and $k_{(i,j)}$ is the edge number ranging from $1$ to $(n-1)$ in $G_C$ such that $(i,j) \in E_C$.

%To specify the priors on the parameters related to the DFS-chain graph $G_C$ induced from $G$, we first need to label the edges and nodes of $G_C=(V,E_C)$ as outlined above. The corresponding edge-incidence matrix is $\nabla_{G_C}$. Since $G_C$ is a chain-graph, $\nabla_{G_C}$ is a $(n-1) \times n$ matrix corresponding to $(n-1)$ edges and $n$ nodes. We denote $k_{(i,j)} \in \{1,\ldots,n-1\}$ to be the row number corresponding to the edge $(i,j) \in E_C$, that is, $(\nabla_{G_C})_{k_{(i,j)},i}, (\nabla_{G_C})_{k_{(i,j)},j} \in \{-1,1\},$ rest of the columns in that row being zero.

\section{Theoretical guarantees}\label{sec:theory}
In this section, we study the posterior concentration property for the proposed method. Let $\bm{\theta}^* = (\theta_1^*,\ldots,\theta_n^*)^T$ and $\sigma^*$ be the respective true values for the signal $\bm{\theta}$ and the noise standard deviation. We make the following assumptions, 
\begin{enumerate}
    \item[(C1)] Define the number of blocks in $\bm{\theta}^*$ by $s =  \sum_{(i,j) \in E} I(\theta_i^* \neq \theta_j^*)$. We assume $s = o(n/\log n)$. 
    \item[(C2)] For $a_t$ and $b_t$ defined in \eqref{eq4}, we assume that $a_t/b_t = n^c$ for some positive constant $c$. 
    \item[(C3)] For every $i,j$ such that $(i,j) \in E$, we assume  $\log (|\theta_i^* - \theta_j^*| / \sigma^*) = O(\log n)$. 
    \item[(C4)] For every $i=1,\ldots,n$, we assume $\theta_i^{*2} / (\lambda_0 \sigma^{*2}) + \log(\lambda_0) = O(\log n)$, where $\lambda_0$ is defined in \eqref{eqn:t-prior}.
\end{enumerate}
 Then we have the following theorem that gives the posterior contraction rate of the proposed estimator. The proof is given in the Appendix. 
\begin{theorem}\label{thm1}
Suppose that Conditions (C1)--(C4) hold, then for any constant $M > 0$, the posterior distribution of $\bm{\theta}$, denoted by $\Pi( \cdot \mid \bm{y})$, satisfies
$$\Pi(\|\bm{\theta} - \bm{\theta^*}\|_2/\sqrt{n} \leq M \sigma^* \sqrt{s \log n /n } \mid \bm{y}) \rightarrow 1, ~~\text{as}~~ n \rightarrow \infty,$$
where the convergence holds in probability and also with respect to $L_1$-distance. 
\end{theorem}
Theorem \ref{thm1} shows that the posterior convergence rate for the average $L_2$-estimation error $\|\bm{\theta} - \bm{\theta^*}\|_2/\sqrt{n}$ is of order $\sigma^* \sqrt{s \log n/n}$. Note that when the graph sparse structure is known, then the oracle rate of contraction is $O(\sigma^* \sqrt{s/n})$. Therefore the proposed Bayesian shrinkage method achieves the optimal rate of convergence except the logarithm term in $n$, and this rate is adaptive to the unknown graph edge sparsity level $s$. This result agrees with the findings in \citet{Song2019} for linear chains, and also connects to the rate obtained by DFS fused Lasso for general graphs in \citet{Padilla2017} in the sense that those two rates are equivalent except the $\log n$ factor. It is worthy discussing and comparing our assumptions (C1)--(C4) with those in \citet{Padilla2017}. In particular, we do not require the minimal signal strength assumption, but we do need the additional assumptions on the prior hyperparameters as in (C2) and the maximum signal size as in (C3) and (C4). Assumptions (C1)-(C4) are in general mild and commonly used in the  literature. For example, (C1) refers to the usual ``edge sparse'' graphs in \citet{Fan2018} and the upper bound on $s$ seems reasonable given that the signal has a length of $n$ in total. Assumption (C2) ensures the $t$-shrinkage prior to assign sufficiently amount of prior mass to the tails for optimal convergence as discussed in  \citet{Song2019}. Assumptions (C3) and (C4) essentially require the true signals and their paired difference to be bounded by a diverging sequence of numbers as $n$ increases, which is easily satisfied for most examples in practice.

\section{Simulation studies}\label{sec:simulation}
We carry out extensive simulation studies to evaluate the empirical performance of our proposed DFS $t$-fusion approach, and compare with two alternative approaches: (i) Laplace fusion approach, where the Laplace prior is used instead of the $t$-shrinkage prior in model \eqref{eqn:t-prior} and the rest steps are the same with our approach, and (ii) $L_1$ fusion, which is the DFS fused Lasso approach as proposed in \citet{Padilla2017}. We consider three types of the graphs, linear chain graph, 2-D lattice graph, and tree-like graph. When implementing our method, except for the linear chain graph, we consider three random root notes for the DFS step and then pool the posterior samples of $\bm{\theta}$ together for inference purpose. All the computations were performed in R \citep{R-Team} on a computing server with two 512 GB RAM, AMD Opteron 6378 processors. For evaluating the performance of our proposed method and comparing with competing approaches, we consider the mean squared error (MSE) and adjusted MSE, respectively defined as
\begin{eqnarray}
\mathrm{MSE} &=& \|\hat{\bm{\theta}} - \bm{\theta_0}\|_2^2/n,\nonumber \\
\mathrm{adj. MSE} &=& \|\hat{\bm{\theta}} - \bm{\theta}_0\|_2^2/\|\bm{\theta}_0\|_2^2,\nonumber 
\end{eqnarray}
where $\hat{\bm{\theta}}$ is the estimated signal, and $\bm{\theta}_0$ is the true signal.

\subsection{Linear Chain graphs}
We first consider the estimation of a piecewise constant function on linear chain graphs. This problem corresponds to the classical change-point detection problem, in which case we can directly use the true graph structure without running the DFS algorithm. For a chain with $n = 100$ nodes, three different signals with varyingly spaced pieces are considered -- (i) 10 evenly spaced pieces, each having a length of 10, (ii) 10 unevenly spaced pieces, with the smaller components having a length of 5, and (iii) 10 very unevenly spaced pieces, with the smaller components having a length of 2. The response variable $y$ is simulated from $N(\bm{\theta_0}, \sigma^2 I_n)$, where $\bm{\theta_0}$ is the true signal vector and $\sigma \in \{0.1, 0.3, 0.5\}.$  

\begin{figure}
	\begin{tabular}{cc}
		\includegraphics[width=75mm]{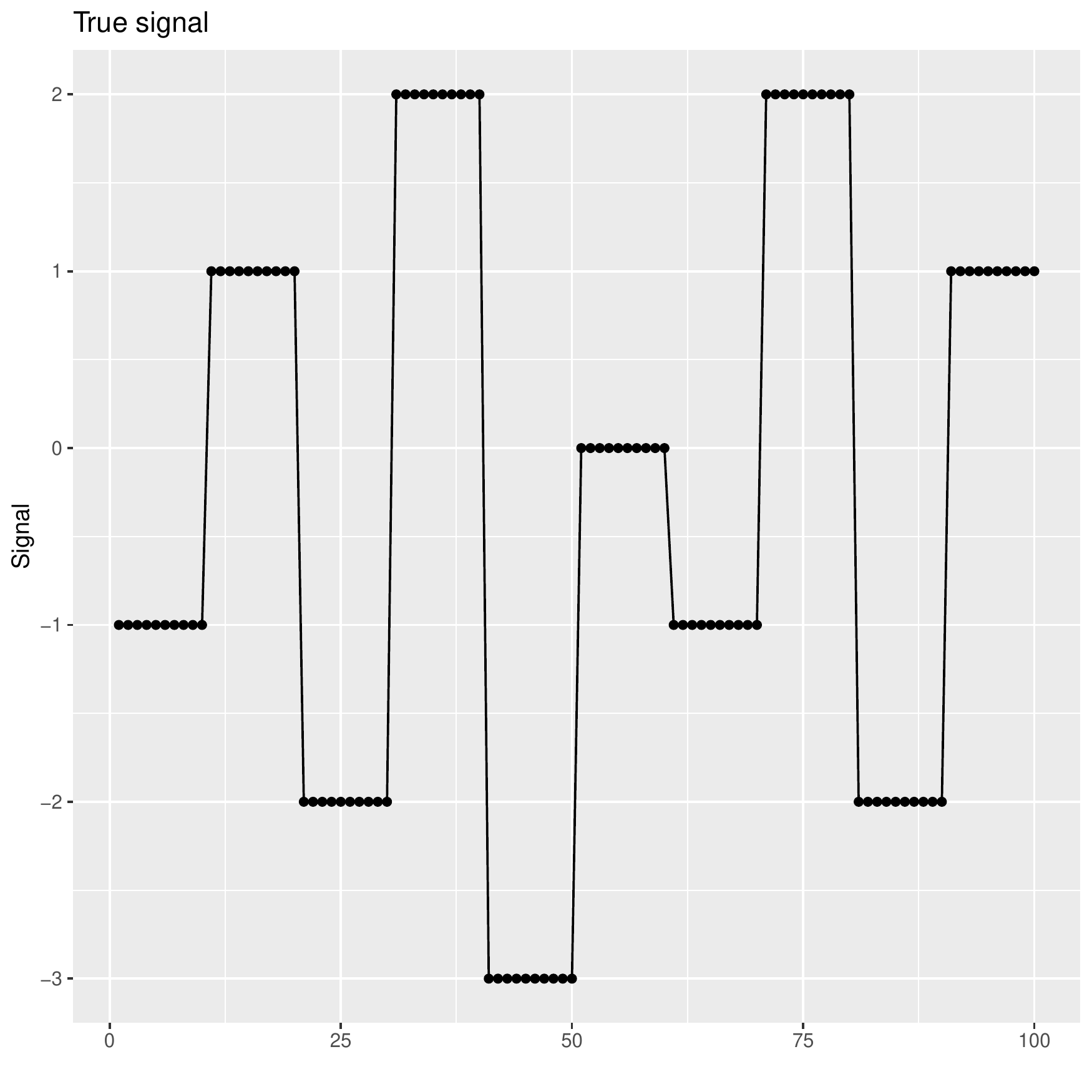} &   \includegraphics[width=75mm]{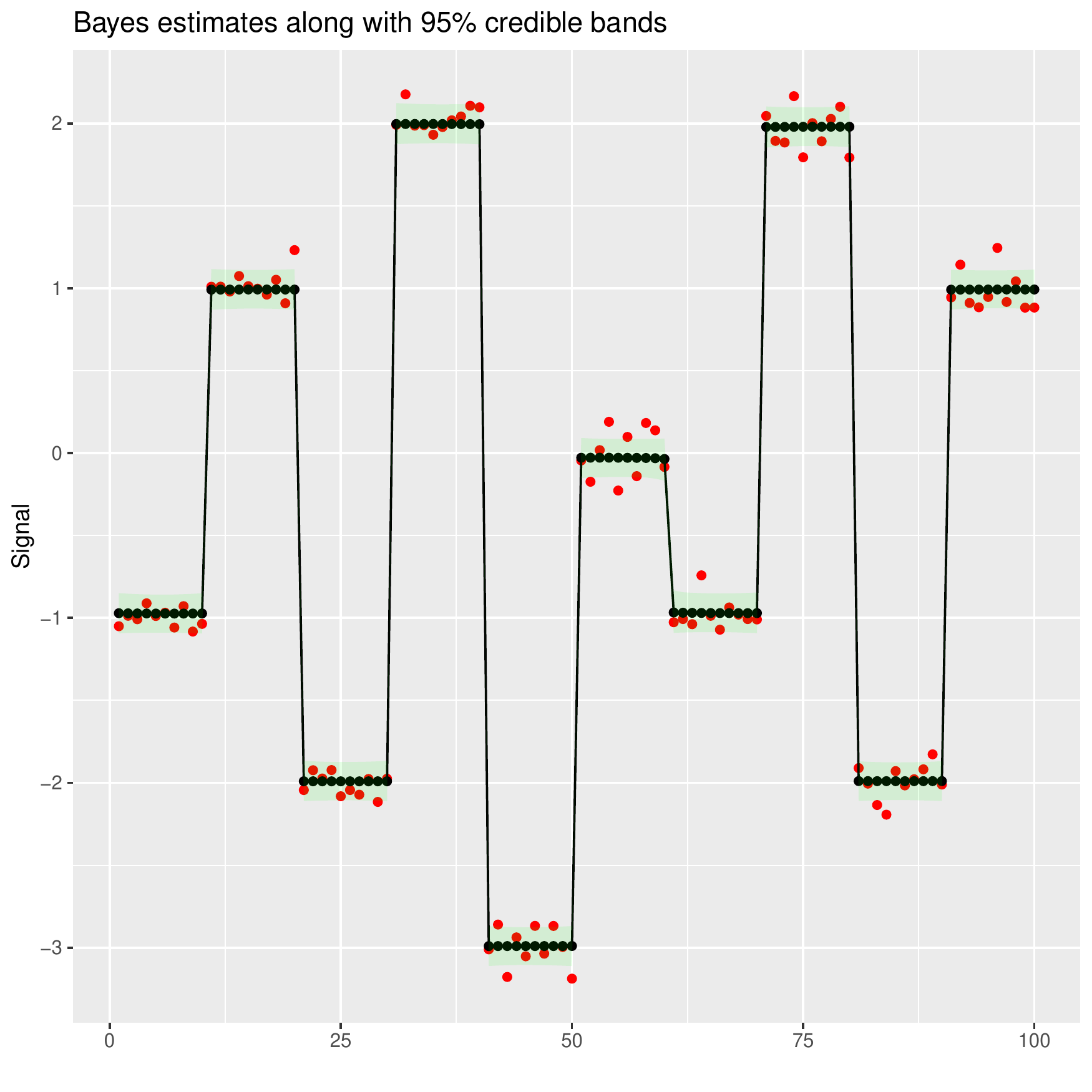} \\
		(a) True signal & (b) $t$-fusion estimates \\[6pt]
		\includegraphics[width=75mm]{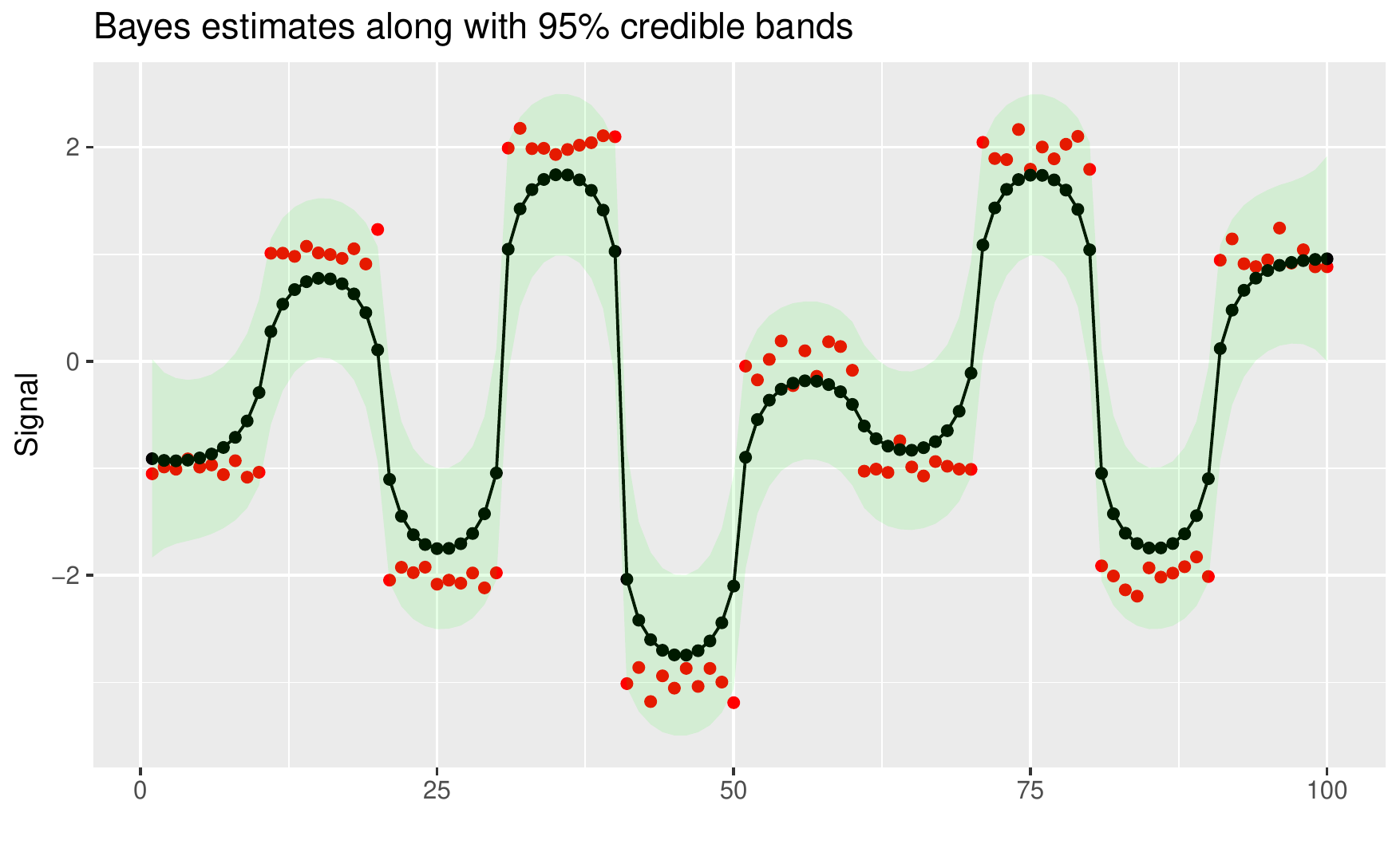} &   \includegraphics[width=75mm]{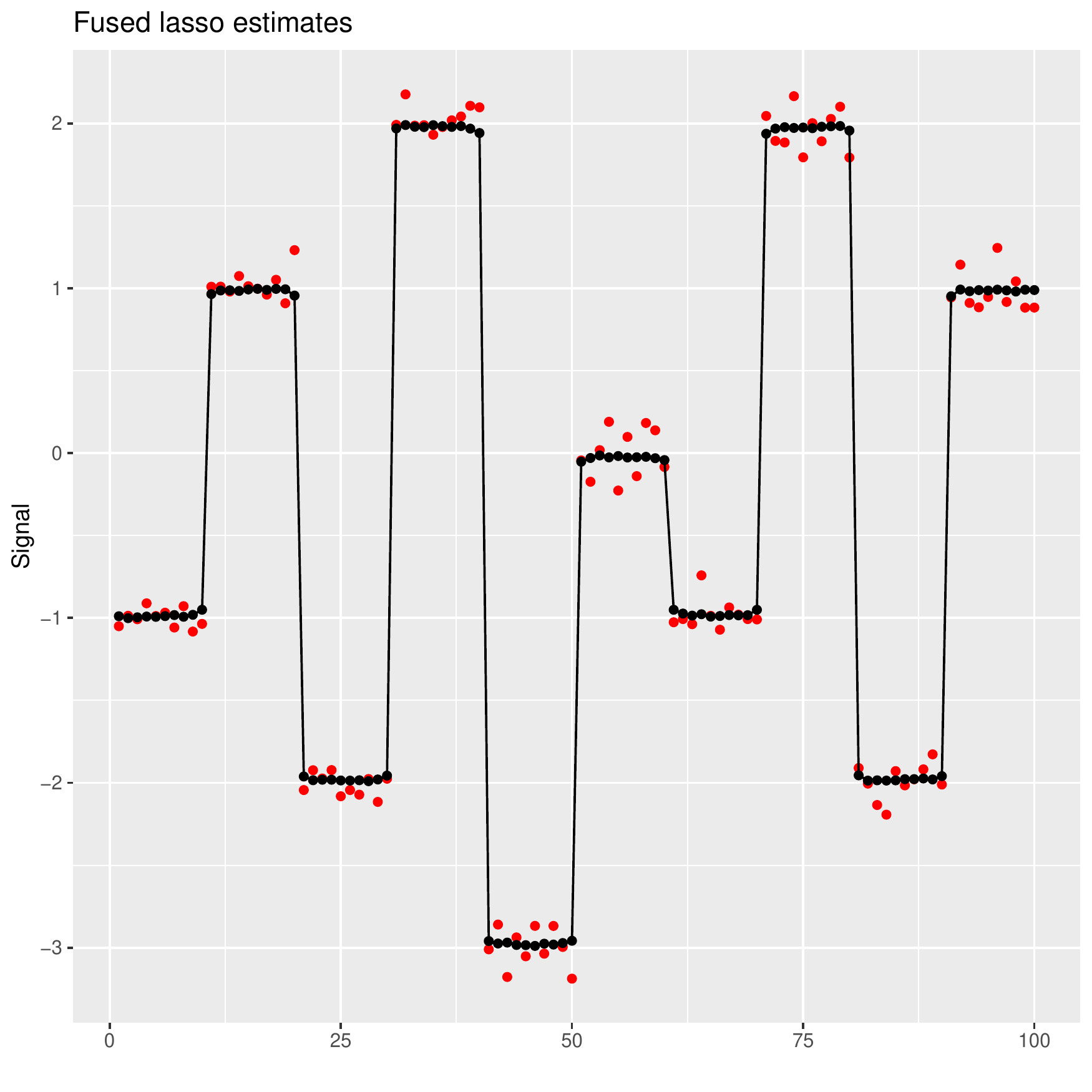} \\
		(c) Laplace fusion estimates & (d) $L_1$ fusion estimates\\[6pt]
	\end{tabular}
	\caption{Signal denoising performance for linear chain graphs with evenly spaced signals and $\sigma = 0.1.$: data in red dots, point estimates in black dots, and 95\% credible bands of the Bayesian procedures in green. }\label{fig1}
\end{figure}

We first present the estimated signals obtained from three methods in Figure \ref{fig1} when the true signal is evenly spaced and $\sigma = .1$. It is clear that both the proposed $t$-fusion prior and fused Lasso are capable of recovering the underlying piecewise constant block structure very well while Laplace prior does not have a satisfactory performance in identifying those ``blocks''. Also, the proposed approach can conveniently obtain posterior credible bands that are much narrower than those from Laplace prior while $L_1$ fused Lasso is only able to provide point estimations. These observations confirm the excellent performance of the use of $t$ prior in terms of shrinkage effect and signal recovery. We also present pictures for other cases (unevenly spaced true signal with higher $\sigma$ values) in Figures \ref{sim:linear-uneven}--\ref{sim:linear-veryuneven2} in the Appendix. Similar findings are observed.  

We then summarize the mean squared error (MSE) and adjusted MSE, and their associated standard errors (in brackets)  over 100 Monte-Carlo replications in Table~\ref{table:lin-chain-MSE}. We find that our proposed method has achieved the highest accuracy for most cases when the noise level is reasonably low, i.e., $\sigma = .1$ and $.3$, but does not perform as well when $\sigma = .5$. One possible explanation is that as the noise level increases, the observed data deviates more from a piecewise constant signal and hence it becomes more challenging to recover the true block structure. As our method uses a heavy-tailed shrinkage prior, it naturally prefers a more sparse structure with less blocks and hence the estimation accuracy deteriorates.

\begin{table}[h]
	\begin{tabular}{ll|ll|ll|ll}
		\hline 
		&       & \multicolumn{2}{c}{$t$-fusion} & \multicolumn{2}{c}{Laplace fusion} & \multicolumn{2}{c}{$L_1$ fusion}  \\
		Signal      & $\sigma$ & MSE            &  adj MSE      & MSE              & adj MSE         & MSE          & adj MSE         \\
		\hline 
		& 0.1   & 0.002(0.001)    & 0.001(0.000) & 0.232(0.008)   & 0.081(0.016) & 
		0.005(0.003)   &  0.002(0.001)   \\
		even        & 0.3   & 0.037(0.017) & 0.013(0.006)    & 
		0.265(0.020)    & 0.093(0.019) & 0.037(0.020)   &    0.013(0.007)   \\
		& 0.5   & 0.262(0.308)  & 0.092(0.108) & 0.323(0.036) &  0.113(0.024)  & 0.098(0.051) & 0.034(0.018)      \\
		&&&&&&& \\
		& 0.1   & 0.002(0.001)    & 0.001(0.001) & 0.274(0.008) &   0.113(0.017)   & 0.006(0.003) & 0.002(0.001)    \\
		uneven      & 0.3   & 0.043(0.014) &  0.018(0.007)   & 0.309(0.018) &  0.128(0.020)   & 0.039(0.019) & 0.016(0.008)        \\
		& 0.5   & 0.524(0.457)    &  0.216(0.186) & 0.368(0.032) & 0.152(0.025) & 0.099(0.048) & 0.041(0.019)   \\
		&&&&&&& \\
		& 0.1   & 0.007(0.006) & 0.003(0.003)    & 0.337(0.006) &  0.156(0.019)    & 0.009(0.002)   & 0.004(0.001)    \\
		v uneven & 0.3   & 0.192(0.240)  &  0.088(0.110) & 0.365(0.015)   & 0.169(0.022)  & 0.063(0.023)   & 0.029(0.010)      \\
		& 0.5   & 0.989(0.111) & 0.459(0.075)    & 0.411(0.025) &  0.191(0.028)  & 0.139(0.059)   &  0.064(0.027)   \\
		\hline 
	\end{tabular}
\caption{Linear Chain Graphs: MSE and adjusted MSE (their associated standard errors) for $t$-fusion, Laplace fusion, and $L_1$ fusion method, when the true signal is evenly spaced (``even''), unevenly spaced (``uneven''), and very unevenly spaced (``v uneven'').}
\label{table:lin-chain-MSE}
\end{table}

\subsection{2-D lattice graphs}
For the second simulation scenario, we explore 2-D lattice graphs. We consider a grid graph of dimension 16, with piecewise constant signals given by $\theta_{0,ij} = \kappa$ for points $(i,j)$ within a distance of 4 units from the center of the grid, and $\theta_{0,ij} = 0$ otherwise.
The observations $y_{ij}$ are simulated independently from $N(\theta_{0,ij}, \sigma^2)$ with $\sigma = 0.3$, and the parameter $\kappa$ that controls the signal strength is chosen as $\kappa \in \{1,5,10\}$.

\begin{figure}
	\centering
	\begin{tabular}{cc}
		\includegraphics[width=65mm]{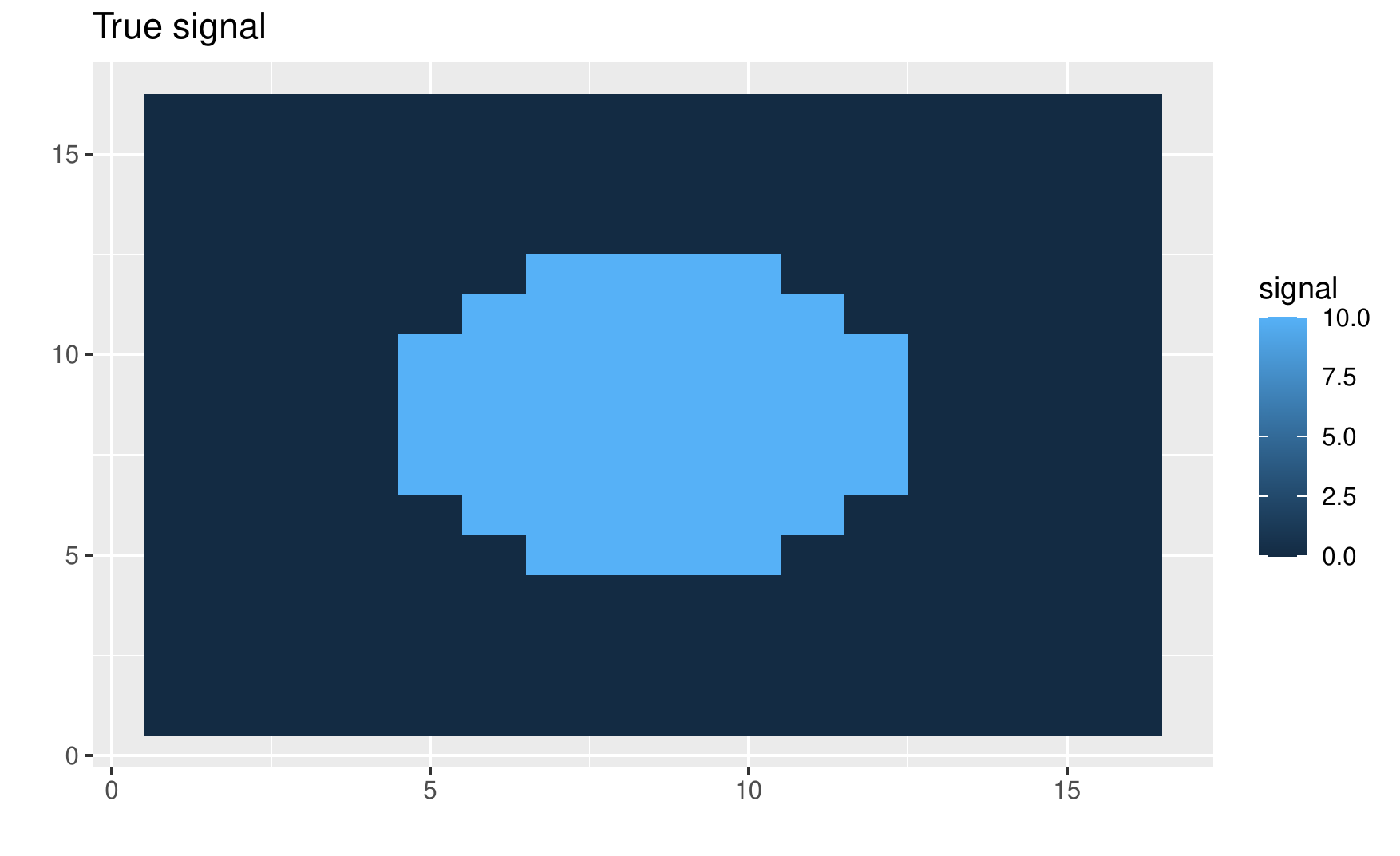} &   \includegraphics[width=65mm]{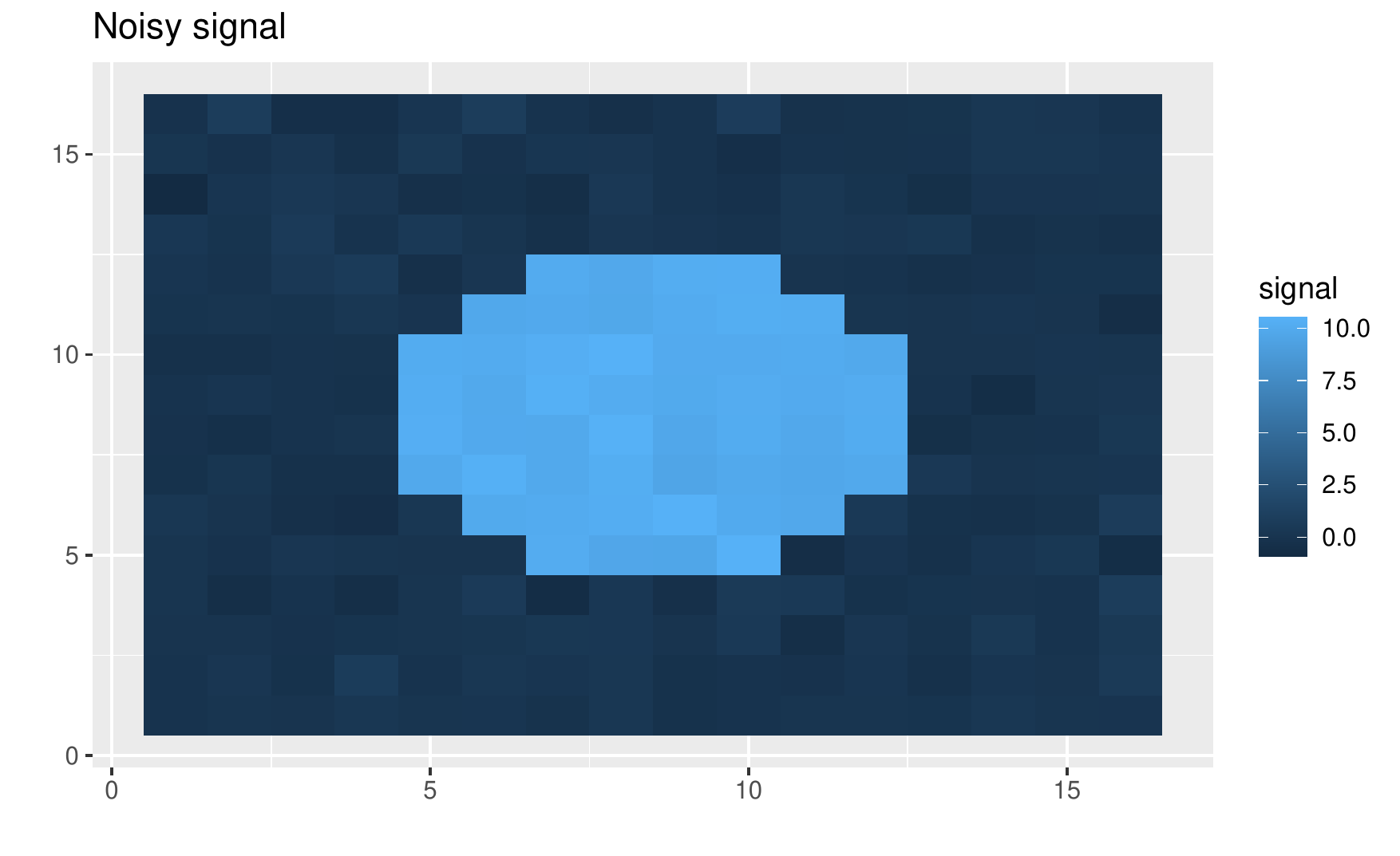} \\
		(a) True signal & (b) Noisy signal \\[1pt]
		\includegraphics[width=65mm]{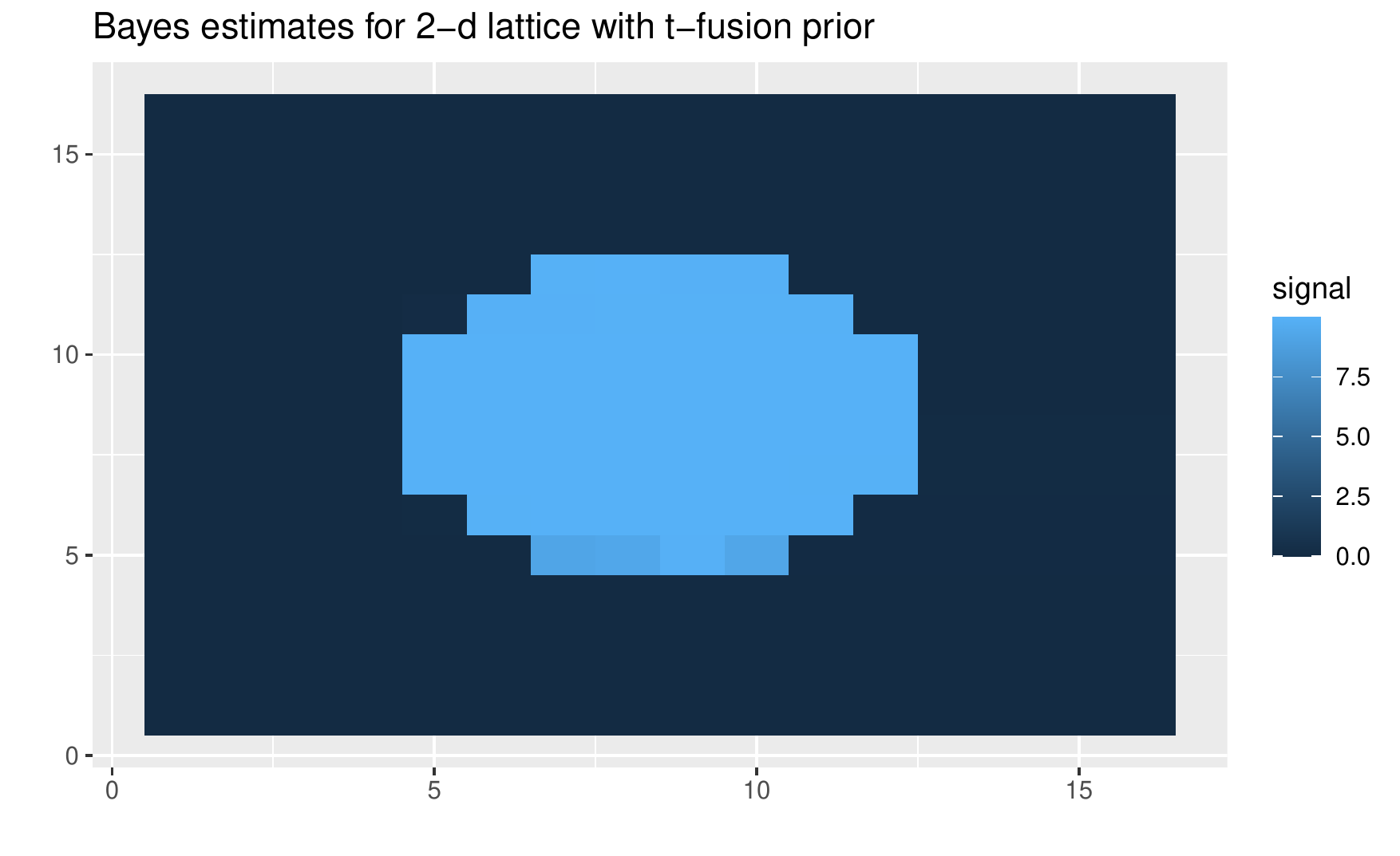} &   \includegraphics[width=65mm]{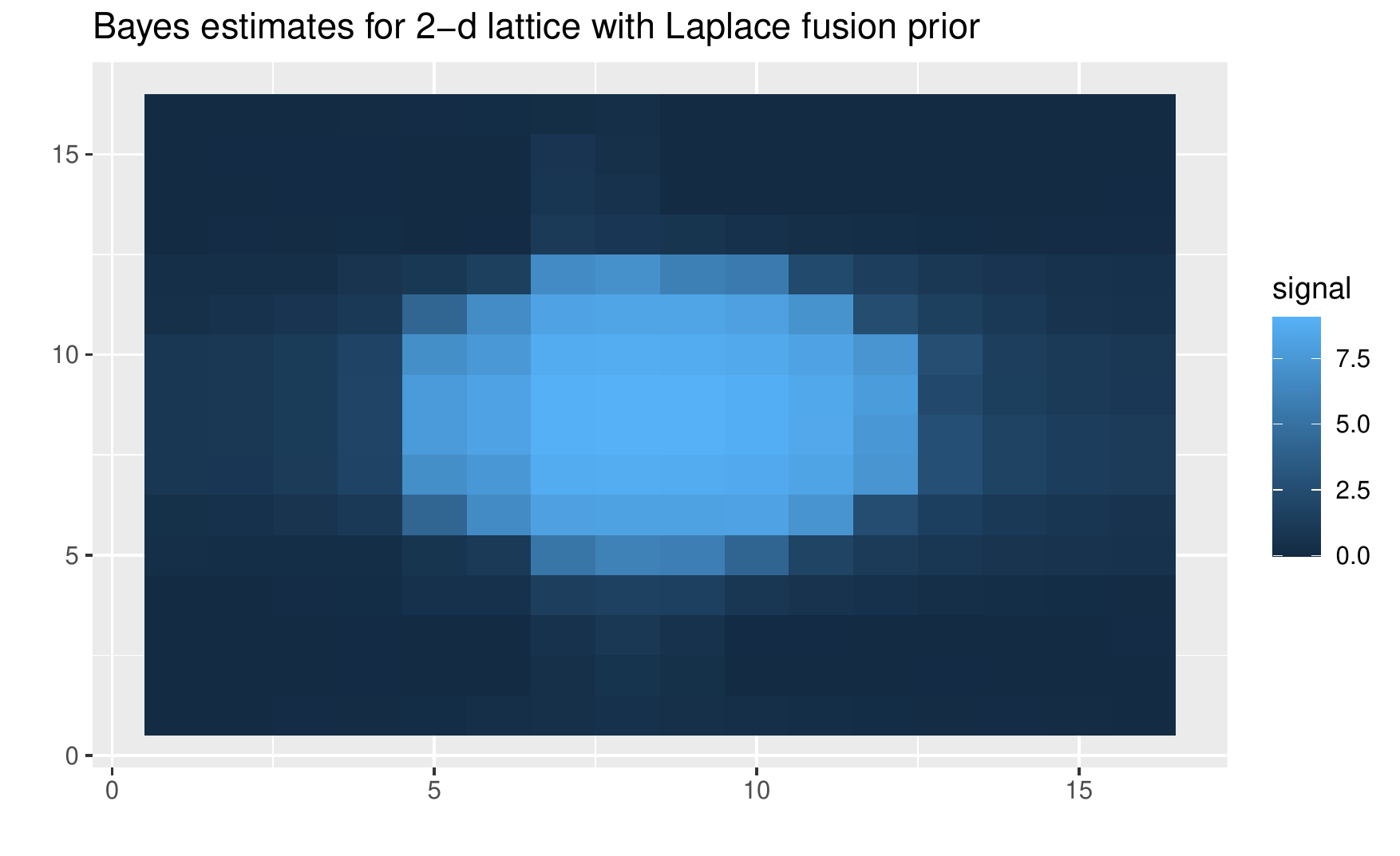} \\
		(c) $t$-fusion estimates & (d) Laplace fusion estimates\\[1pt]
		\multicolumn{2}{c}{\includegraphics[width=65mm]{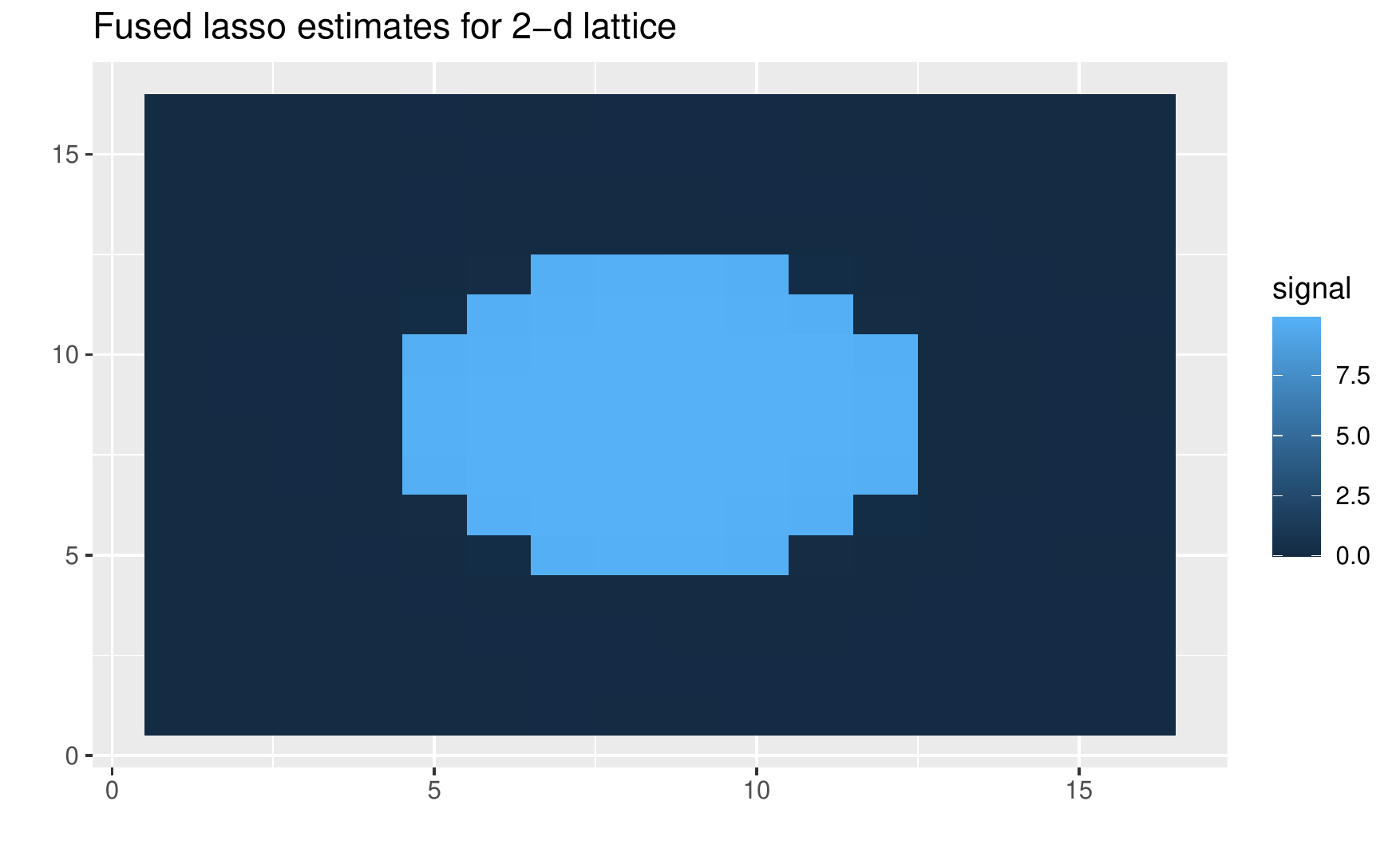} }\\
		\multicolumn{2}{c}{(e) $L_1$ fusion estimates}
	\end{tabular}
	\caption{Signal denoising performance for 2-D lattice graphs with strong signal strength ($\kappa = 10$).}\label{fig2}
\end{figure}

We first present the plots of the true signal and the recovered signals obtained from three methods in Figure \ref{fig2} when $\kappa = 10$. Similarly with the linear chain graphs, we find that both our method and $L_1$ fused Lasso manage to correctly identify the latent structure in the signal while Laplace prior method has an unsatisfactory performance in identifying the boundary of the signals. Plots for $\kappa = \{1,5\}$ are provided as Figure \ref{sim:2d-weak} and \ref{sim:2d-med} in the Appendix; and we observe a similar pattern there.

Next we summarize both the MSE and adjusted MSE for all three methods in   Table~\ref{table:lattice-MSE} based on 100 Monte-Carlo replications. As the signal becomes stronger, all three methods have a better accuracy in terms of smaller adjusted MSE values. Overall, $L_1$ fusion has a slightly lower estimation error than our method although this advantage becomes less noticeable as the signal becomes strong. The use of $t$ prior has a much better signal estimation accuracy than that of the Laplace prior, which again confirms the benefit of using a heavy-tailed shrinkage prior when the true signal itself is indeed piecewise constant.

\begin{table}[h]
	\centering
	\begin{tabular}{lll|ll|ll}
		\hline 
		 & \multicolumn{2}{c}{$t$-fusion} & \multicolumn{2}{c}{Laplace} & \multicolumn{2}{c}{$L_1$ fusion} \\
		Signal  & MSE           & adj MSE       & MSE          & adj MSE       & MSE           & adj MSE        \\
		\hline 
		weak   & 0.031(0.010)  & 0.039(0.012)    & 0.107(0.007) &  0.131(0.009)   & 0.012(0.005) &   0.014(0.006)    \\
		medium & 0.047(0.015)  & 0.009(0.003)   & 0.509(0.015) &   0.100(0.003)  & 0.012(0.005) &  0.002(0.001)    \\
		strong & 0.028(0.033) & 0.001(0.002)    & 1.920(0.031) &  0.094(0.002)   & 0.012(0.005) &  0.001(0.000) \\
		\hline   
	\end{tabular}
	\caption{2-D lattice Graphs: MSE and adjusted MSE (their associated standard errors) for $t$-fusion, Laplace fusion, and $L_1$ fusion method, for weak ($\kappa = 1$), medium ($\kappa = 5$), and strong ($\kappa = 10$) signal strength.}
	\label{table:lattice-MSE}
\end{table}

\begin{figure}
	\centering
	\begin{tabular}{cc}
		\includegraphics[width=65mm]{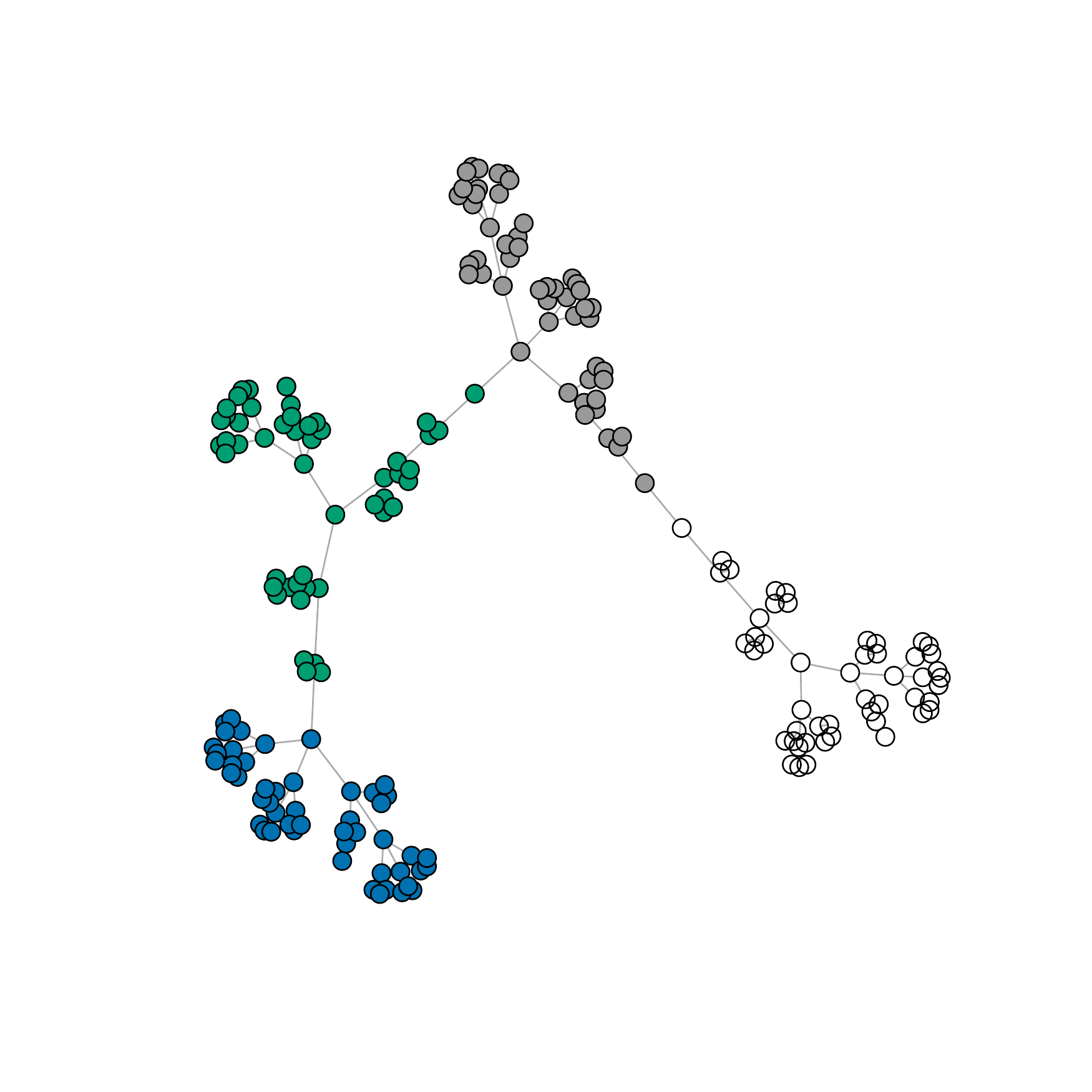} &   \includegraphics[width=65mm]{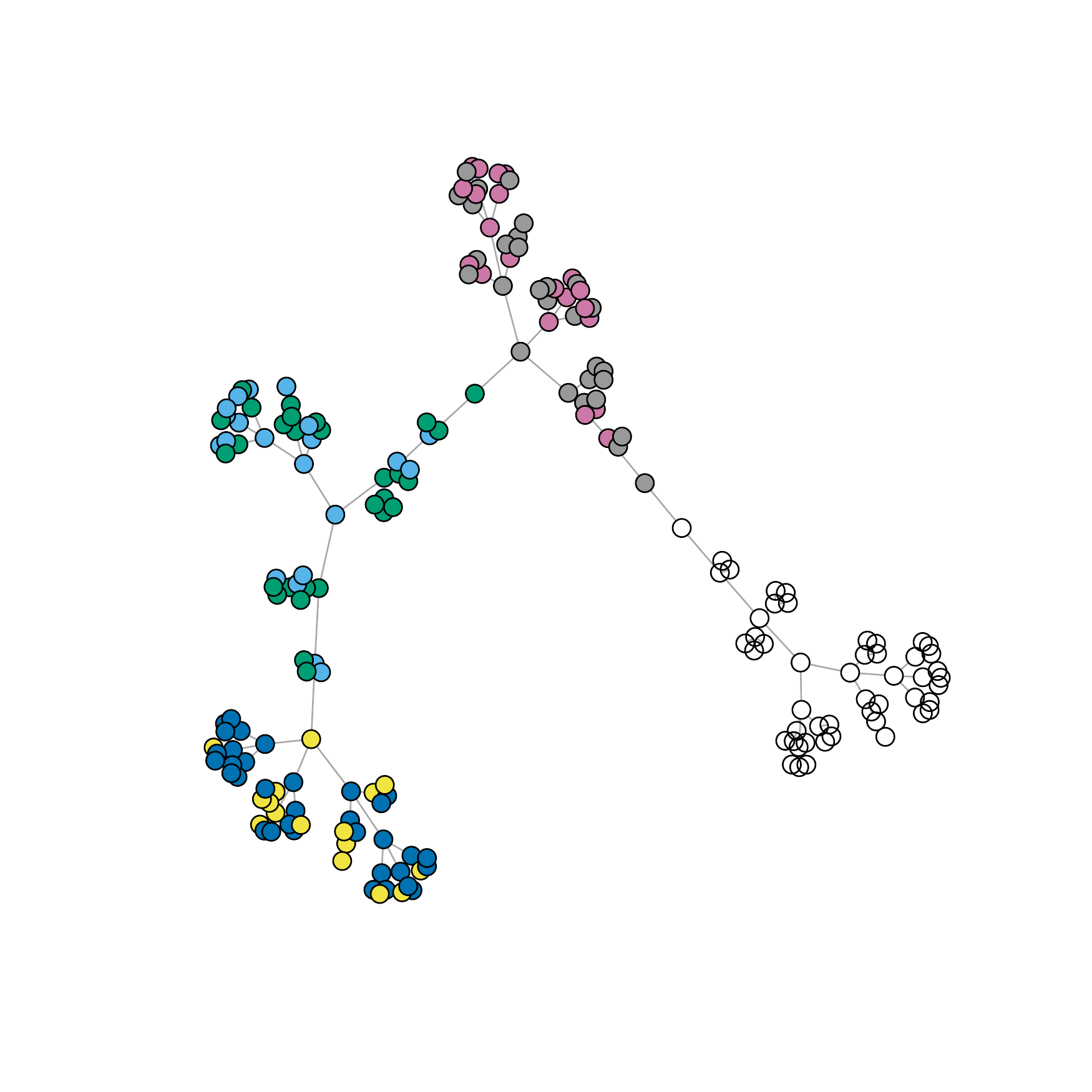} \\
		(a) True signal & (b) Noisy signal \\[1pt]
		\includegraphics[width=65mm]{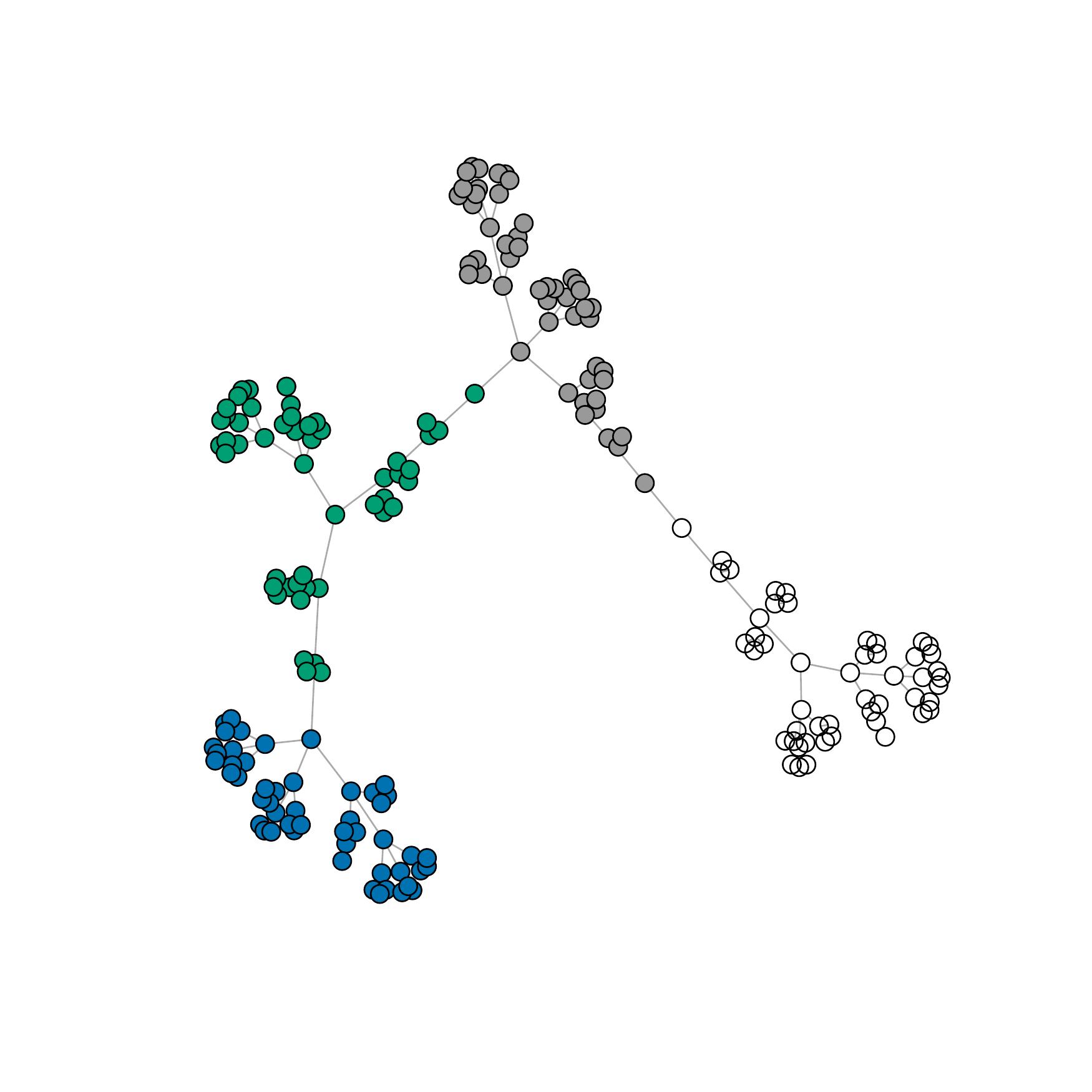} &   \includegraphics[width=65mm]{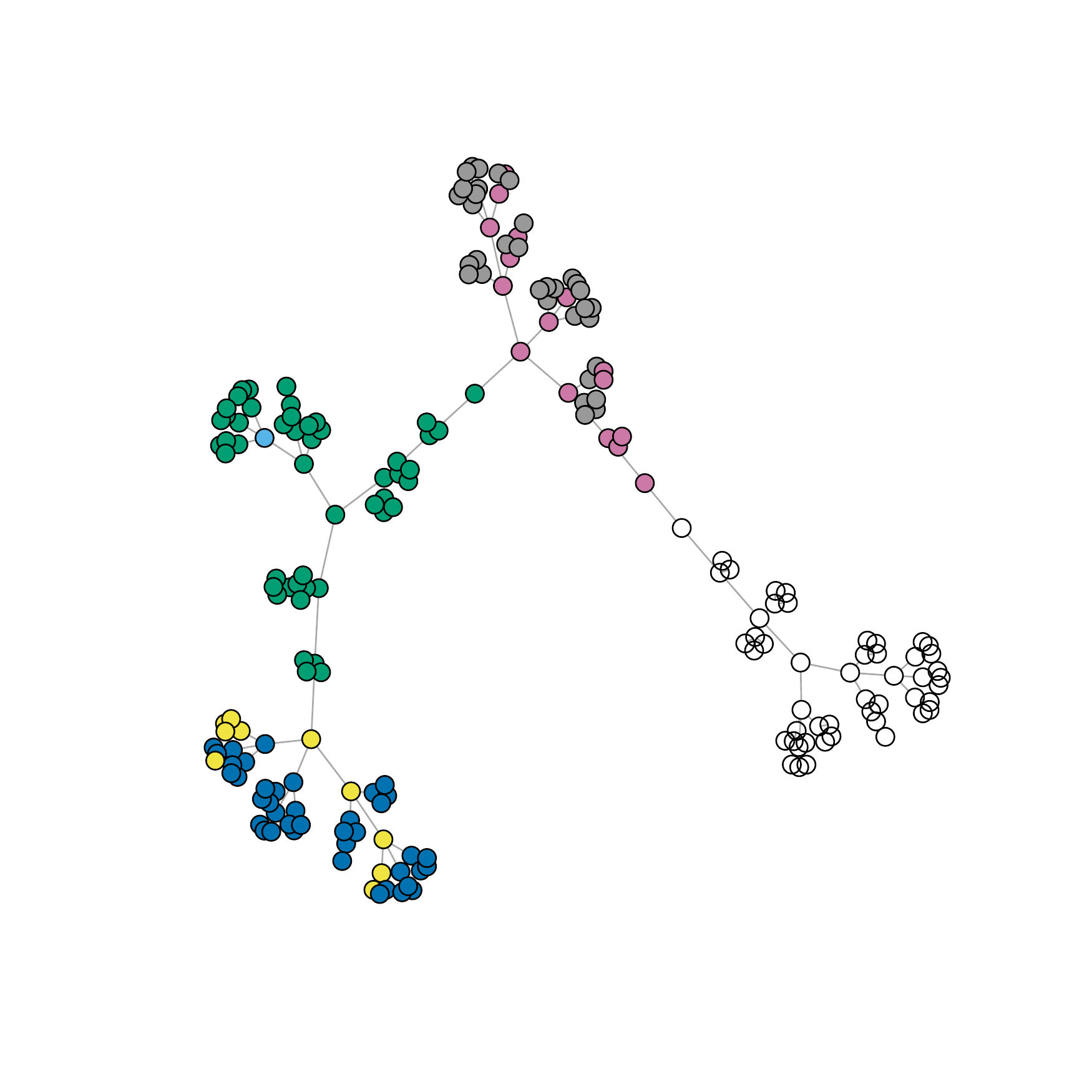} \\
		(c) $t$-fusion estimates & (d) Laplace fusion estimates\\[1pt]
		\multicolumn{2}{c}{\includegraphics[width=65mm]{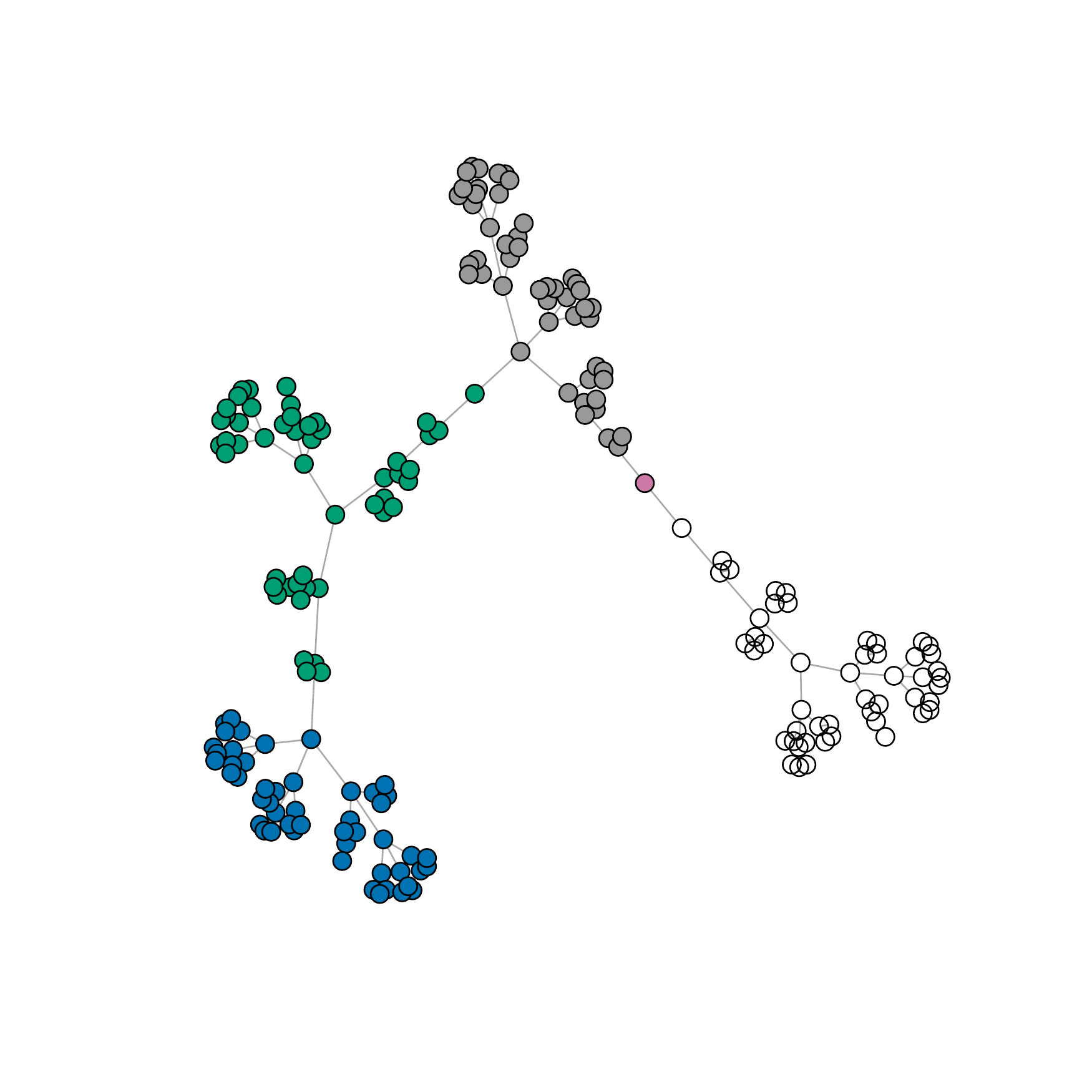} }\\
		\multicolumn{2}{c}{(e) $L_1$ fusion estimates}
	\end{tabular}
	\caption{Signal denoising performance for tree graphs.}\label{fig3}
\end{figure}

\subsection{Tree graphs}
Finally we consider signals defined on tree-structured graphs. We consider four trees, namely, $T_1,\ldots,T_4$, each having 50 nodes and 3 children per node. A connected tree-like graph $G_T$ having 200 nodes is then formed by randomly selecting a node from $T_i$ and $T_{i+1},\, i = 1,2,3$ and placing an edge between the respective nodes. The true signal values $\theta_0$ defined on the trees $T_1,\ldots,T_4$ are given by $1,-1,4,-4$ respectively. We generate the data on the graph $G_T$ from $N(\theta_0, \sigma^2)$ with $\sigma = 0.3$. The signal denoising performance is presented in Figure \ref{fig3} for three methods. Similarly with the previous two scenarios, both our method and $L_1$ fusion have an excellent signal recovery performance while Laplace fusion tends to produce over-clustered results. This finding is also confirmed in Table \ref{tab:treegraph}, where the estimation accuracy of $t$-fusion and $L_1$-fusion is better than that of the Laplace prior. 
\begin{table}[h]
	\centering
	\begin{tabular}{lcc}
		\hline 
		Method    & MSE   & adj.MSE \\
		\hline 
		$t$-fusion  & 0.004 (0.003) & 0.0004 (0.0003)   \\
		Laplace   & 0.049 (0.006) & 0.0060 (0.0007)   \\
		$L_1$-fusion & 0.005 (0.002) & 0.0006 (0.0003) \\
		\hline 
	\end{tabular}
\caption{MSE and adjusted MSE (standard error) for tree graphs.}\label{tab:treegraph}
\end{table}

\section{Real data application}\label{sec:real}

Change-point detection in financial data is a widely studied problem. We consider the problem of change-point detection in the cumulative stock index returns of two different indices in two different periods, one for the Dow Jones Industrial Average (DJI) and the other for the Shanghai Stock Exchange Index (SSE)\footnote{The datasets are publicly available at \url{https://finance.yahoo.com/}}. We carefully choose the two different time periods so as to include known stock-market crash phenomena. To be precise, we study the DJI for the period of January 2007 to January 2010, that encompasses the 2008 Global Recession period, and the SSE for the period January 2013 to April 2016, that encompasses the 2015 China Stock Market crash. 
%The cumulative stock returns for DJI and SSE corresponding to the respective time periods are displayed in Figures~\ref{fig:dji-returns} and \ref{fig:sse-returns}.

\begin{comment}
\begin{figure}
	\centering
	\includegraphics[height=3in,width=5in]{DJI-returns}
	\caption{Cumulative Stock returns for the Dow Jones Index (DJI) in the period of January 2007 to January 2010.}
	\label{fig:dji-returns}
\end{figure}

\begin{figure}
	\centering
	\includegraphics[height=3in, width=5in]{SSE-returns}
	\caption{Cumulative Stock returns for the Shanghai Stock Exchange (SSE) in the period of January 2013 to April 2016.}
	\label{fig:sse-returns}
\end{figure}
\end{comment}

\begin{figure}
	\centering
	\begin{tabular}{c}
		\includegraphics[height=2.3in, width=4.5in]{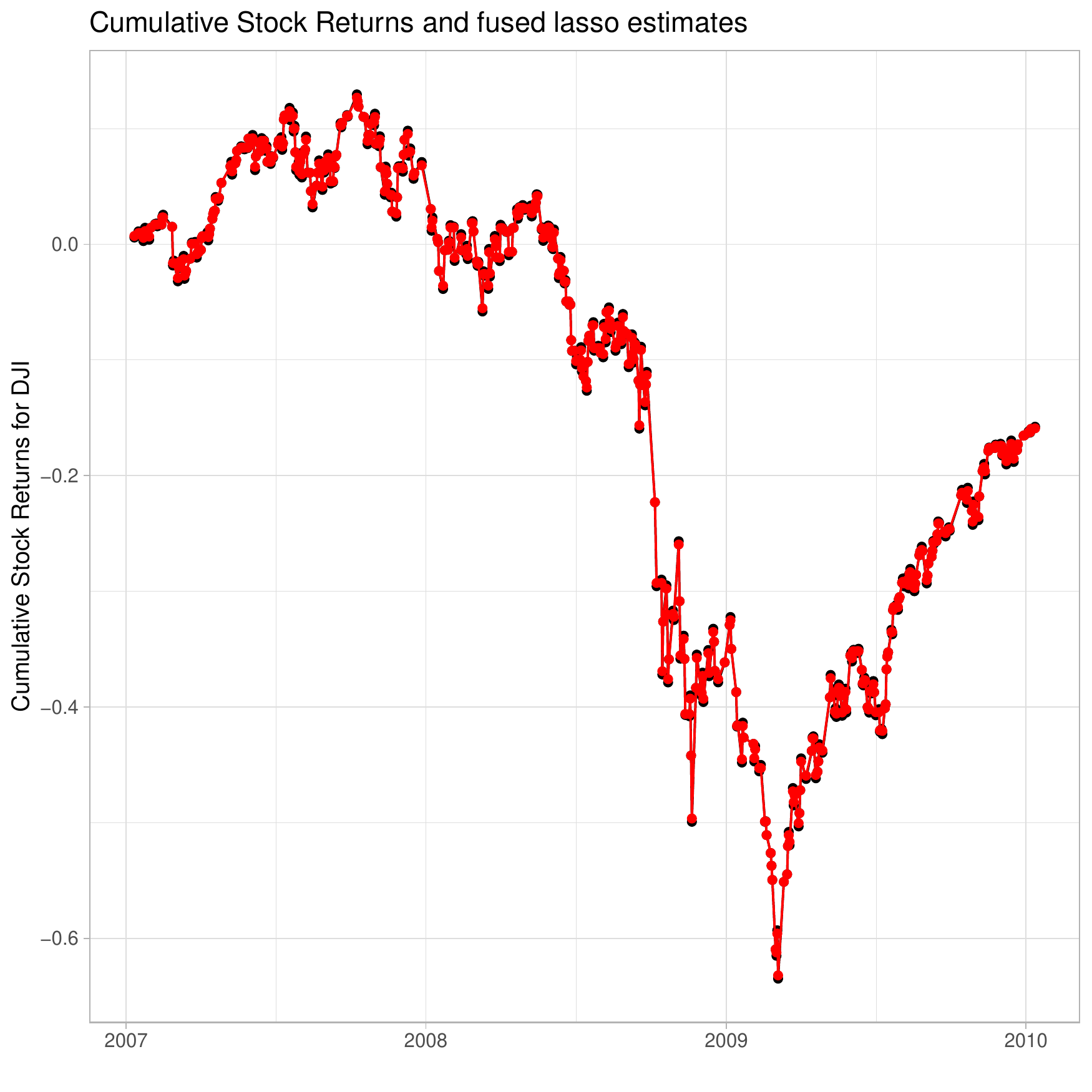}  \\
		(a) Fused Lasso solution  \\[1pt]
 	\includegraphics[height=2.3in, width=4.5in]{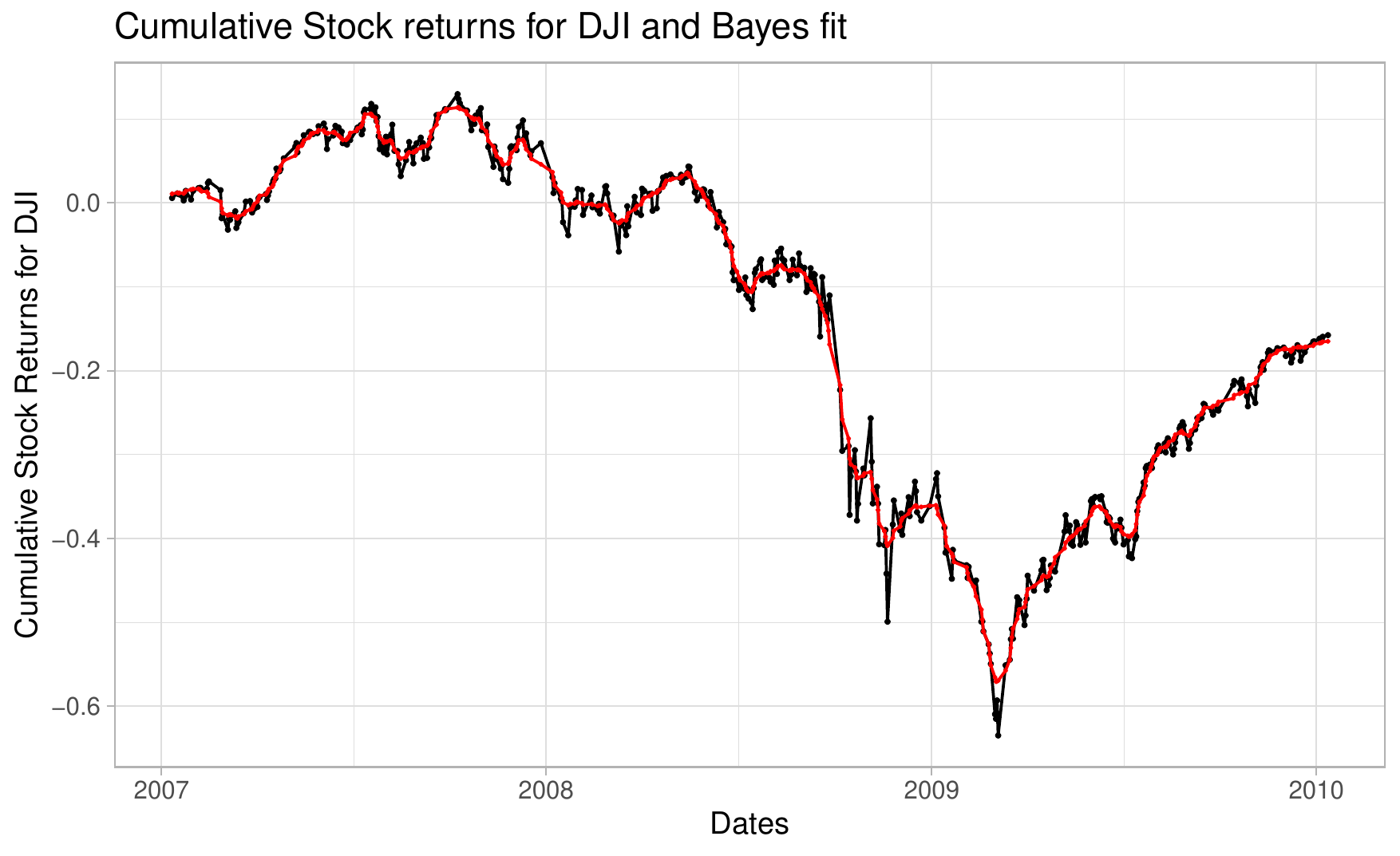} \\
		(b) Bayesian Laplace-fusion model \\[1pt]
	 	\includegraphics[height=2.3in, width=4.5in]{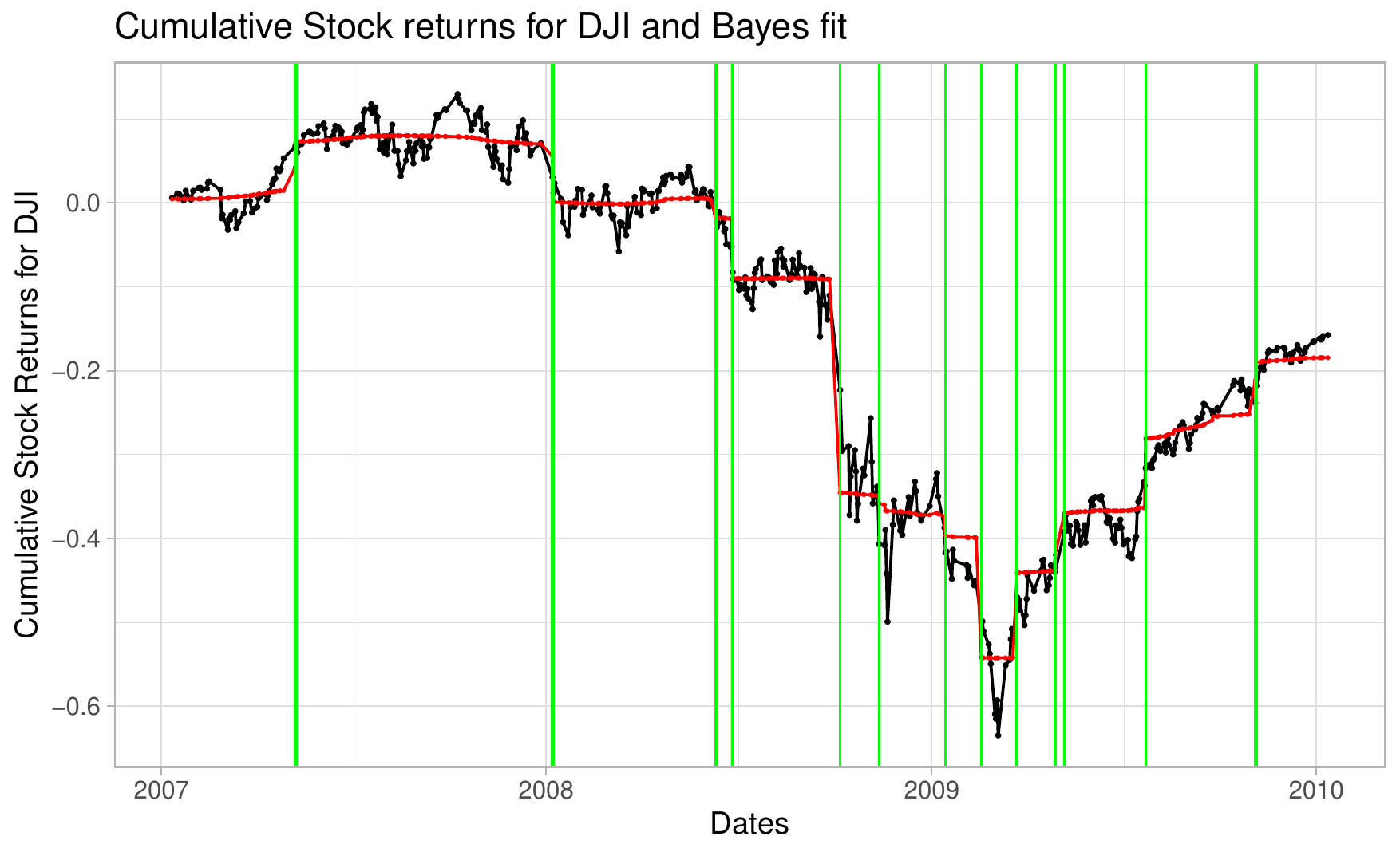} \\
	 	(c) Proposed DFS $t$-fusion method
	\end{tabular}
	\caption{Estimation results for cumulative Stock returns for the Dow Jones Industrial Average (DJI) in the period of January 2007 to January 2010 by three methods. The black lines refer to the observed data. The green lines in (c) indicate the dates where a significant change in the means occur. }\label{fig: data2_1}
\end{figure}

\begin{figure}
	\centering
	\begin{tabular}{c}
		\includegraphics[height=2.3in, width=4.5in]{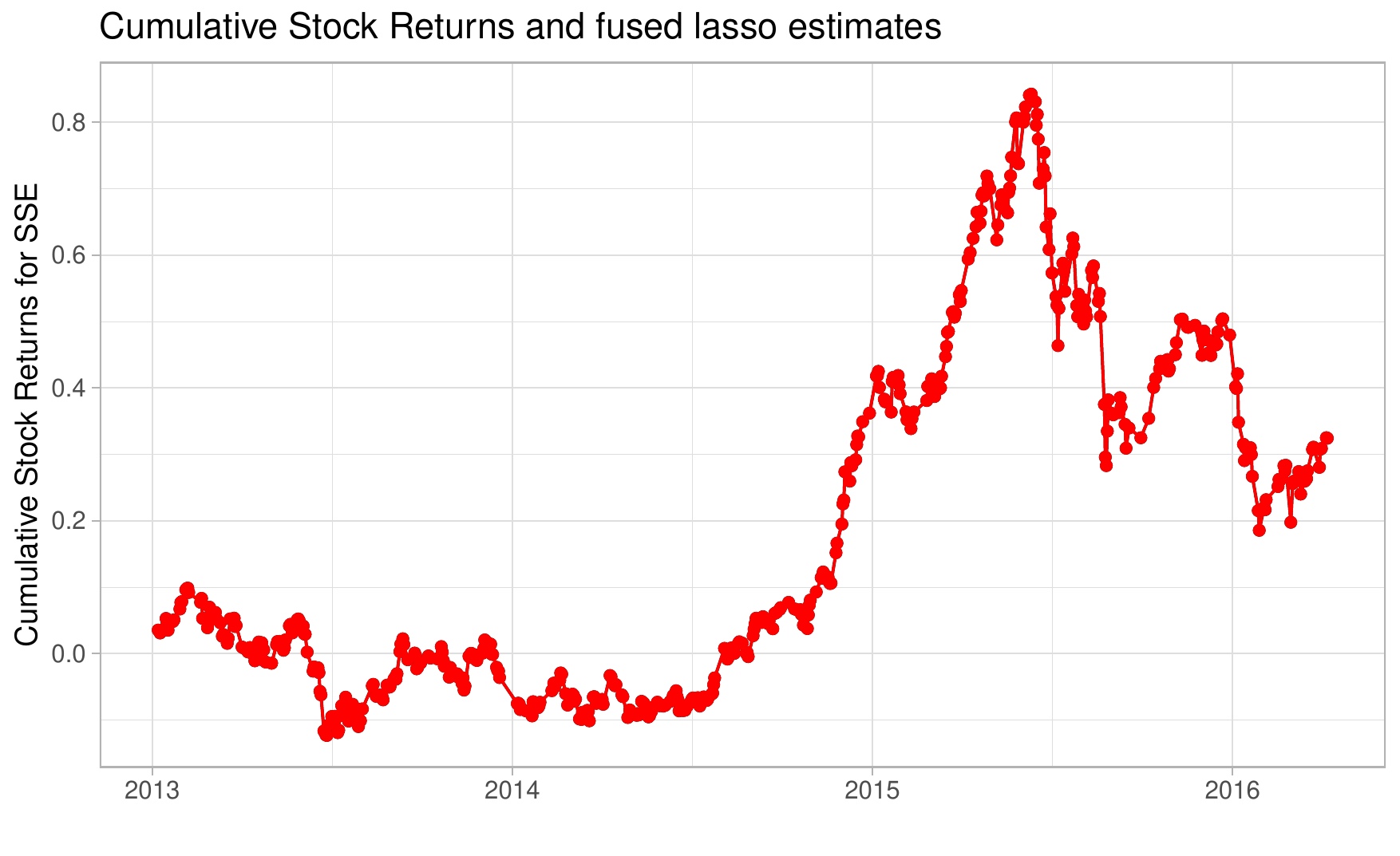}  \\
		(a) Fused Lasso solution  \\[1pt]
 	\includegraphics[height=2.3in, width=4.5in]{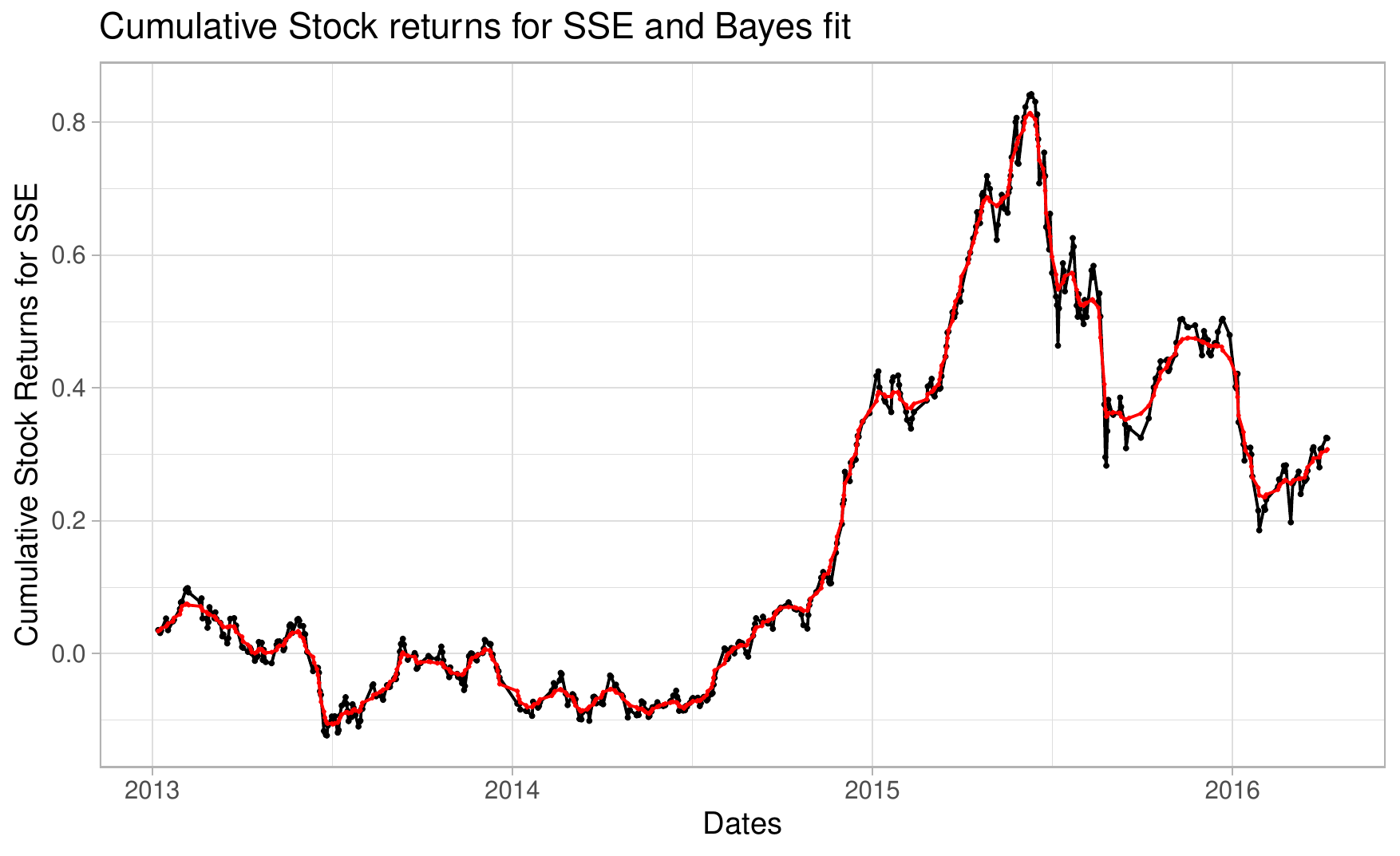} \\
		(b) Bayesian Laplace-fusion model \\[1pt]
	 	\includegraphics[height=2.3in, width=4.5in]{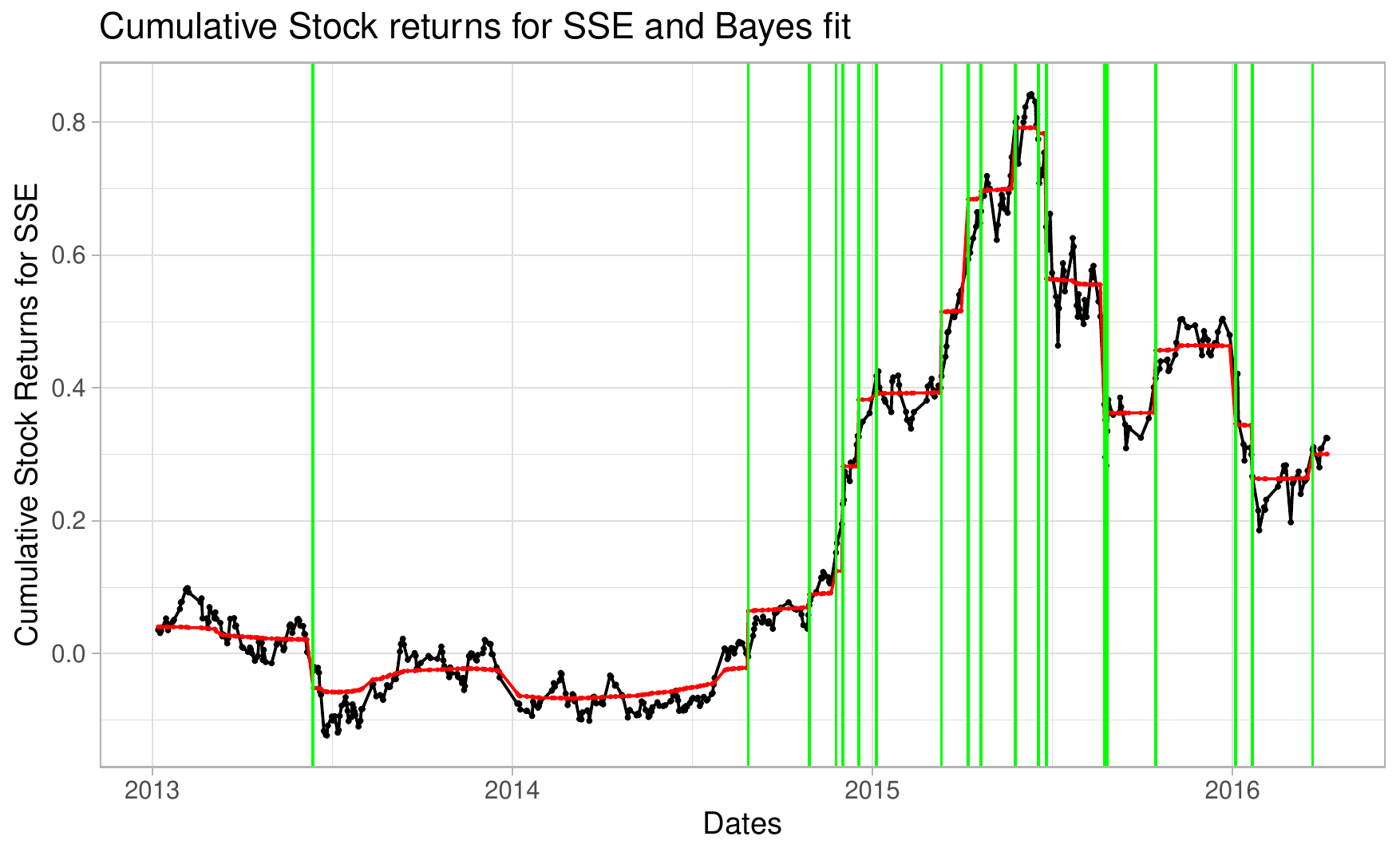} \\
	 	(c) Proposed DFS $t$-fusion method
	\end{tabular}
	\caption{Estimation results for cumulative Stock returns for the Shanghai Stock Exchange (SSE) in the period of January 2013 to April 2016 by three methods. The black lines refer to the observed data. The green lines in (c) indicate the dates where a significant change in the means occur.}\label{fig: data2_2}
\end{figure}

We apply the graph signal denoising method using both the $t$-fusion prior and the Laplace prior, along with the frequentist fused Lasso method on both data sets. The results are shown in Figure \ref{fig: data2_1} and \ref{fig: data2_2}. We find that the fused Lasso solution suffers from severe over-fitting problems, and a similar pattern is observed in the Bayesian Laplace fusion prior as well (see (a) and (b) in Figure~\ref{fig: data2_1} and \ref{fig: data2_2}). In contrast, the proposed Bayesian $t$-fusion method produces almost piecewise constant estimates. Following the suggestions of \citet{Song2019}, we ``sparsify" the continuous posterior accounting for the prior concentration. Specifically, we estimate the quantity $I(\theta_i = \theta_j)$ for $i \neq j$ by $I(|\hat{\theta}_i - \hat{\theta}_j)|/\hat{\sigma} \leq mt_{1/2n})$, where $t_{1/2n}$ is the $(1 - 1/2n)$-quantile of a $t$ distribution with $2a_t$ degrees of freedom and $\hat{\theta}, \hat{\sigma}$ are Bayes estimates given by the respective posterior means. It is clear that this sparsification step is very useful for locating the blocks (or approximate pieces) precisely, which in turn, provide the change points in the times series. 

The change points obtained from the sparsification of the estimates capture major financial events during the time periods. During the global financial crisis of 2007-2008, large financial institutions across the United States failed around mid-September 2008, followed by failure of major banks in Europe. Price of stocks and other commodities plummeted sharply across the globe. The-then U.S. President George W. Bush created a Troubled Asset Relief Program (TARP) for purchasing failing bank assets, via signing the Emergency Economic Stabilization Act into law. The effects of this 2008 Global Recession were disastrous, with the world economy taking a serious blow. The resulting period is referred to as the `Great Recession', that officially spanned the period of December 2007 to June 2009 in the United States. The DFS $t$-fusion method captures these major changes in the movement of stocks, with change-points detected at the beginning of the recession (late 2007), followed by multiple change-points during late 2008, and at periods when the recession officially ended.

The 2015-2016 Chinese Stock Market crash started with a stock market bubble during mid-June 2015, after which the prices crashed sharply throughout July and August 2015. Following a steep sell-off in the Chinese stock market on January 2016, global markets felt severe tremors. The proposed method of ours captures these major change-points in the stock market movements, with several change-points detected en-route June 2015, indicating the rapid speed at which the stock market bubble was building up, followed by the series of crashes post June 2015. The January 2016 global meltdown period has been captured as well.

In a recent work on dynamic change-point detection in financial networks, \cite{banerjee2020change} analyzed fully connected time-varying networks based on major stock returns in the Asia-Pacific region along with stock market indices from the U.S. and the U.K. Their method was able to detect major financial events in different time periods, including the two regimes considered above. Our proposed approach works with a single time-series at a time, but alongside detecting known stock-market crash phenomena, it additionally captures ripples in the market both apriori and aposteriori the critical events. Thus, our method may be utilized for serving as potential warning prior to major disruptions in global financial markets.

\begin{comment}
\begin{figure}
	\centering
	\includegraphics[height=3in, width=5in]{fused-DJI}
	\caption{Fused lasso solution for fitting cumulative Stock returns for the Dow Jones Index (DJI) in the period of January 2007 to January 2010.}
	\label{fig:dji-returns-fused}
\end{figure}

\begin{figure}
	\centering
	\includegraphics[height=3in, width=5in]{DJI-Laplace-fusion}
	\caption{Bayesian Laplace-fusion model fit for cumulative Stock returns for the Dow Jones Index (DJI) in the period of January 2007 to January 2010.}
	\label{fig:dji-returns-Lap-fusion}
\end{figure}

\begin{figure}
	\centering
	\includegraphics[height=3in, width=5in]{DJI-t-fusion}
	\caption{Bayesian $t$-fusion model fit for cumulative Stock returns for the Dow Jones Index (DJI) in the period of January 2007 to January 2010. The green lines indicate the dates where a significant change in the means occur.}
	\label{fig:dji-returns-$t$-fusion}
\end{figure}
\end{comment}

\begin{comment}

\begin{figure}
	\centering
	\includegraphics[height=3in, width=5in]{fused-SSE}
	\caption{Fused lasso solution for fitting cumulative Stock returns for the Shanghai Stock Exchange (DJI) in the period of January 2013 to April 2013.}
	\label{fig:sse-returns-fused}
\end{figure}

\begin{figure}
	\centering
	\includegraphics[height=3in, width=5in]{SSE-Laplace-fusion}
	\caption{Bayesian Laplace-fusion model fit for cumulative Stock returns for the Shanghai Stock Exchange (DJI) in the period of January 2013 to April 2013.}
	\label{fig:sse-returns-Lap-fusion}
\end{figure}

\begin{figure}
	\centering
	\includegraphics[height=3in, width=5in]{SSE-t-fusion}
	\caption{Bayesian $t$-fusion model fit for cumulative Stock returns for the Shanghai Stock Exchange (DJI) in the period of January 2013 to April 2013. The green lines indicate the dates where a significant change in the means occur.}
	\label{fig:sse-returns-$t$-fusion}
\end{figure}

\end{comment}

\section{Discussion}\label{sec:dis} 
In this paper, we propose a new Bayesian approach for analyzing the piecewise constant model over general graphs. The proposed method makes efficient use of the depth first search (DFS) algorithm and heavy-tailed shrinkage prior. Several future working directions remain open. First, our proposed method in general works for any graph-searching algorithms (instead of restricting to DFS); and it will be of interest to compare and investigate the effect of different graph-searching algorithms on the signal estimation performance. Second, the theory studied in this paper mainly focuses on signal estimation. It will be of interest to investigate the structure recovery property in the graph. We expect the results in a recent work \citep{kim2020bayesian} to be helpful. Lastly, it will be of interest to consider more general residual distributions other than the normal distribution in model \eqref{eq:model}.

\section*{Acknowledgements}

S.B. is partially supported by DST INSPIRE Faculty Award, Govt. of India, Grant No. 04/2015/002165, and IIM Indore Young Faculty Research Chair grant.

  \bibliographystyle{asa}
\bibliography{ref_test}

\begin{thebibliography}{26}
\newcommand{\enquote}[1]{``#1''}
\expandafter\ifx\csname natexlab\endcsname\relax\def\natexlab#1{#1}\fi

\bibitem[{Banerjee and Guhathakurta(2020)}]{banerjee2020change}
Banerjee, S. and Guhathakurta, K. (2020), \enquote{Change-point analysis in
  financial networks,} \textit{Stat}, 9, e269.

\bibitem[{Barron et~al.(1999)Barron, Birg{\'e}, and Massart}]{barron1999risk}
Barron, A., Birg{\'e}, L., and Massart, P. (1999), \enquote{Risk bounds for
  model selection via penalization,} \textit{Probability theory and related
  fields}, 113, 301--413.

\bibitem[{Besag(1986)}]{besag1986statistical}
Besag, J. (1986), \enquote{On the statistical analysis of dirty pictures,}
  \textit{Journal of the Royal Statistical Society: Series B (Methodological)},
  48, 259--279.

\bibitem[{Bhattacharya et~al.(2015)Bhattacharya, Pati, Pillai, and
  Dunson}]{bhattacharya2015dirichlet}
Bhattacharya, A., Pati, D., Pillai, N.~S., and Dunson, D.~B. (2015),
  \enquote{Dirichlet--Laplace priors for optimal shrinkage,} \textit{Journal of
  the American Statistical Association}, 110, 1479--1490.

\bibitem[{Birg{\'e} and Massart(2007)}]{birge2007minimal}
Birg{\'e}, L. and Massart, P. (2007), \enquote{Minimal penalties for Gaussian
  model selection,} \textit{Probability theory and related fields}, 138,
  33--73.

\bibitem[{Brodsky and Darkhovsky(2013)}]{brodsky2013nonparametric}
Brodsky, E. and Darkhovsky, B.~S. (2013), \textit{Nonparametric methods in
  change point problems}, vol. 243, Springer Science \& Business Media.

\bibitem[{Castillo et~al.(2015)Castillo, Schmidt-Hieber, Van~der Vaart,
  et~al.}]{castillo2015bayesian}
Castillo, I., Schmidt-Hieber, J., Van~der Vaart, A., et~al. (2015),
  \enquote{Bayesian linear regression with sparse priors,} \textit{The Annals
  of Statistics}, 43, 1986--2018.

\bibitem[{Crovella and Kolaczyk(2003)}]{crovella2003graph}
Crovella, M. and Kolaczyk, E. (2003), \enquote{Graph wavelets for spatial
  traffic analysis,} in \textit{IEEE INFOCOM 2003. Twenty-second Annual Joint
  Conference of the IEEE Computer and Communications Societies (IEEE Cat. No.
  03CH37428)}, IEEE, vol.~3, pp. 1848--1857.

\bibitem[{Fan and Guan(2018)}]{Fan2018}
Fan, Z. and Guan, L. (2018), \enquote{Approximate $\ell_ {0} $-penalized
  estimation of piecewise-constant signals on graphs,} \textit{The Annals of
  Statistics}, 46, 3217--3245.

\bibitem[{Gao et~al.(2020)Gao, Han, Zhang, et~al.}]{gao2020estimation}
Gao, C., Han, F., Zhang, C.-H., et~al. (2020), \enquote{On estimation of
  isotonic piecewise constant signals,} \textit{Annals of Statistics}, 48,
  629--654.

\bibitem[{Gavish et~al.(2010)Gavish, Nadler, and
  Coifman}]{gavish2010multiscale}
Gavish, M., Nadler, B., and Coifman, R.~R. (2010), \enquote{Multiscale Wavelets
  on Trees, Graphs and High Dimensional Data: Theory and Applications to Semi
  Supervised Learning.} in \textit{ICML}, pp. 367--374.

\bibitem[{Guntuboyina et~al.(2018)Guntuboyina, Sen,
  et~al.}]{guntuboyina2018nonparametric}
Guntuboyina, A., Sen, B., et~al. (2018), \enquote{Nonparametric
  shape-restricted regression,} \textit{Statistical Science}, 33, 568--594.

\bibitem[{Hutter et~al.(2007)}]{hutter2007exact}
Hutter, M. et~al. (2007), \enquote{Exact Bayesian regression of piecewise
  constant functions,} \textit{Bayesian Analysis}, 2, 635--664.

\bibitem[{Kim and Gao(2020)}]{kim2020bayesian}
Kim, Y. and Gao, C. (2020), \enquote{Bayesian model selection with graph
  structured sparsity,} \textit{Journal of Machine Learning Research}, 21,
  1--61.

\bibitem[{Kyung et~al.(2010)Kyung, Gill, Ghosh, Casella,
  et~al.}]{kyung2010penalized}
Kyung, M., Gill, J., Ghosh, M., Casella, G., et~al. (2010), \enquote{Penalized
  regression, standard errors, and Bayesian lassos,} \textit{Bayesian
  Analysis}, 5, 369--411.

\bibitem[{Liu et~al.(2020)Liu, Martin, and Shen}]{liu2017empirical}
Liu, C., Martin, R., and Shen, W. (2020), \enquote{Empirical priors and
  posterior concentration in a piecewise polynomial sequence model,}
  \textit{arXiv e-prints}, arXiv--1712.

\bibitem[{Padilla et~al.(2017)Padilla, Sharpnack, and Scott}]{Padilla2017}
Padilla, O., Sharpnack, J., and Scott, J. (2017), \enquote{The DFS fused lasso:
  Linear-time denoising over general graphs,} \textit{The Journal of Machine
  Learning Research}, 18, 6410--6445.

\bibitem[{{R Core Team}(2019)}]{R-Team}
{R Core Team} (2019), \textit{R: A Language and Environment for Statistical
  Computing}, R Foundation for Statistical Computing, Vienna, Austria.

\bibitem[{Roualdes(2015)}]{roualdes2015bayesian}
Roualdes, E.~A. (2015), \enquote{Bayesian trend filtering,} \textit{arXiv
  preprint arXiv:1505.07710}.

\bibitem[{Rudin et~al.(1992)Rudin, Osher, and Fatemi}]{rudin1992nonlinear}
Rudin, L.~I., Osher, S., and Fatemi, E. (1992), \enquote{Nonlinear total
  variation based noise removal algorithms,} \textit{Physica D: nonlinear
  phenomena}, 60, 259--268.

\bibitem[{Shuman et~al.(2013)Shuman, Narang, Frossard, Ortega, and
  Vandergheynst}]{shuman2013emerging}
Shuman, D.~I., Narang, S.~K., Frossard, P., Ortega, A., and Vandergheynst, P.
  (2013), \enquote{The emerging field of signal processing on graphs: Extending
  high-dimensional data analysis to networks and other irregular domains,}
  \textit{IEEE signal processing magazine}, 30, 83--98.

\bibitem[{Song and Cheng(2019)}]{Song2019}
Song, Q. and Cheng, G. (2019), \enquote{Bayesian Fusion Estimation via t
  Shrinkage,} \textit{Sankhya A}, 1--33.

\bibitem[{Song and Liang(2017)}]{song2017nearly}
Song, Q. and Liang, F. (2017), \enquote{Nearly optimal Bayesian shrinkage for
  high dimensional regression,} \textit{arXiv preprint arXiv:1712.08964}.

\bibitem[{Tarjan(1972)}]{tarjan1972depth}
Tarjan, R. (1972), \enquote{Depth-first search and linear graph algorithms,}
  \textit{SIAM journal on computing}, 1, 146--160.

\bibitem[{Tartakovsky et~al.(2014)Tartakovsky, Nikiforov, and
  Basseville}]{tartakovsky2014sequential}
Tartakovsky, A., Nikiforov, I., and Basseville, M. (2014), \textit{Sequential
  analysis: Hypothesis testing and changepoint detection}, CRC Press.

\bibitem[{Tibshirani et~al.(2005)Tibshirani, Saunders, Rosset, Zhu, and
  Knight}]{tibshirani2005sparsity}
Tibshirani, R., Saunders, M., Rosset, S., Zhu, J., and Knight, K. (2005),
  \enquote{Sparsity and smoothness via the fused lasso,} \textit{Journal of the
  Royal Statistical Society: Series B (Statistical Methodology)}, 67, 91--108.

\end{thebibliography}

\newpage
\section*{Appendix}
\setcounter{section}{0}
\section{Proof of Theorem \ref{thm1}}
The proof of the results is based upon the proof of Theorem 2.1 and Corollary 2.1 in \citet{Song2019} since both papers are using the $t$-shrinkage prior.
After we perform the DFS on the general graph, our proposed model becomes the piecewise constant model defined on a chain using $t$ shrinkage prior as presented in \citet{Song2019}. Under the assumption that $\sum_{(i,j) \in E} I(\theta_i^* \neq \theta_j^*) \leq s$, we have $\sum_{(i,j) \in E_C} I(\theta_i^* \neq \theta_j^*) \leq 2s$ based on Lemma 1 in \citet{Padilla2017}. Therefore we can directly apply Theorem 2.1 and Corollary 2.1 in \citet{Song2019} and obtain the rates. Note that when performing DFS, the root node can be any of the values in $\theta_1,\ldots,\theta_n$. Therefore the assumption in \citet{Song2019} are generalized to (C3)  for arbitrary pairs of connected nodes $(\theta_i,\theta_j)$ and to (C4) for every $\theta_i^*$ with $i=1,\ldots,n$. 

\section{Additional numerical results} 
Here we present additional numerical results for the Simulation Section in the main paper. 
% Uneven signals, low variance
\begin{figure}
	\begin{tabular}{cc}
		\includegraphics[width=75mm]{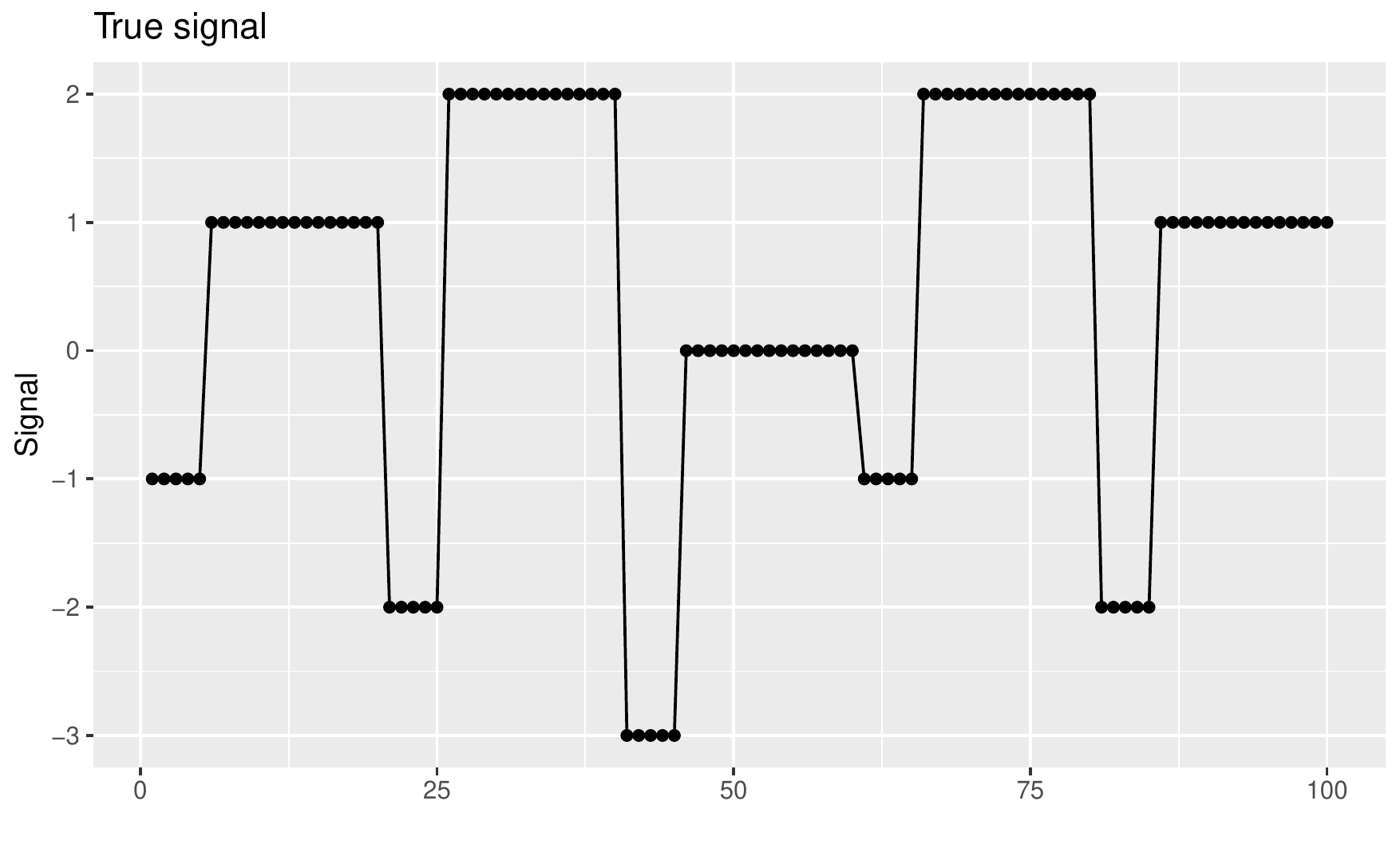} &   \includegraphics[width=75mm]{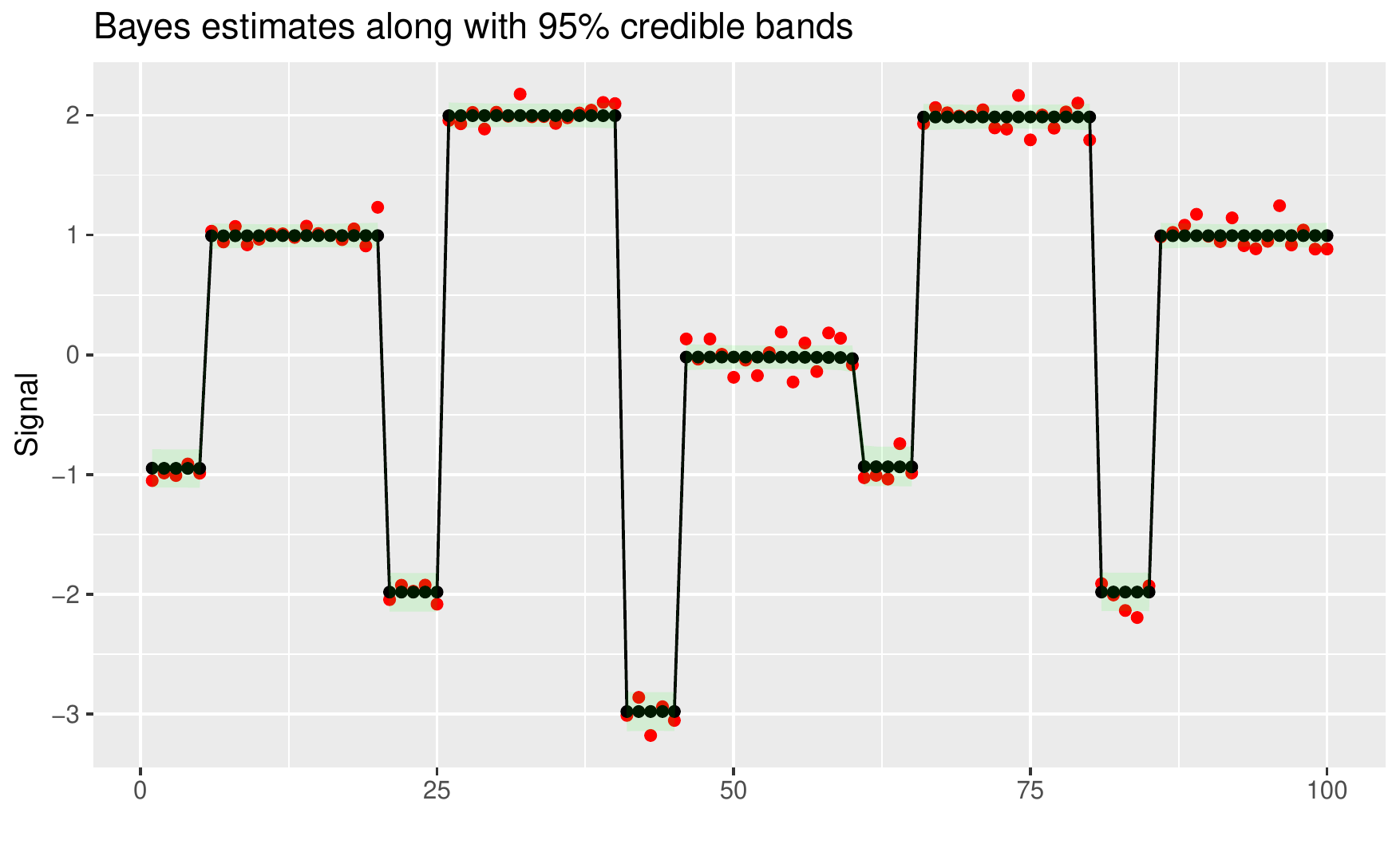} \\
		(a) True signal & (b) $t$-fusion estimates \\[6pt]
		\includegraphics[width=75mm]{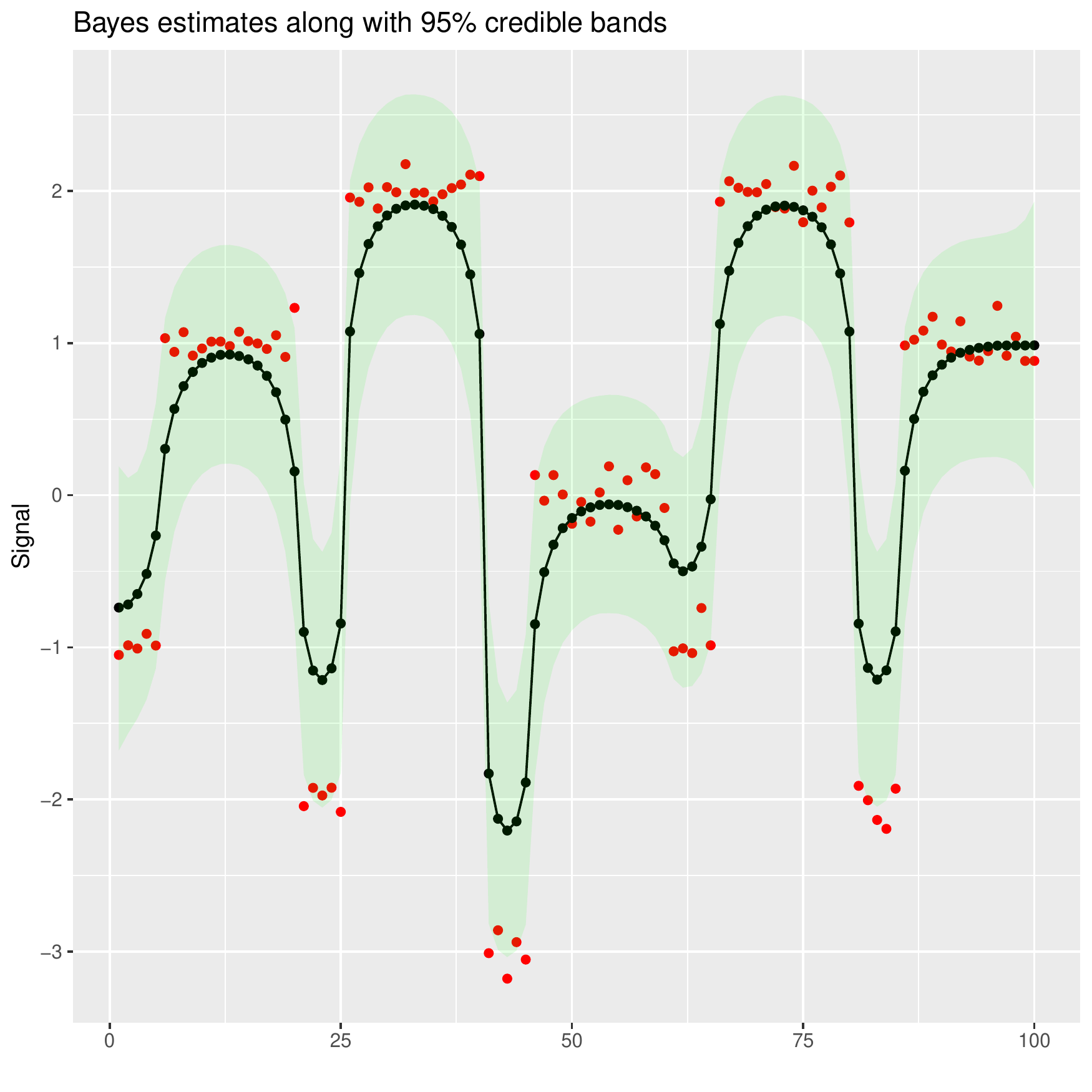} &   \includegraphics[width=75mm]{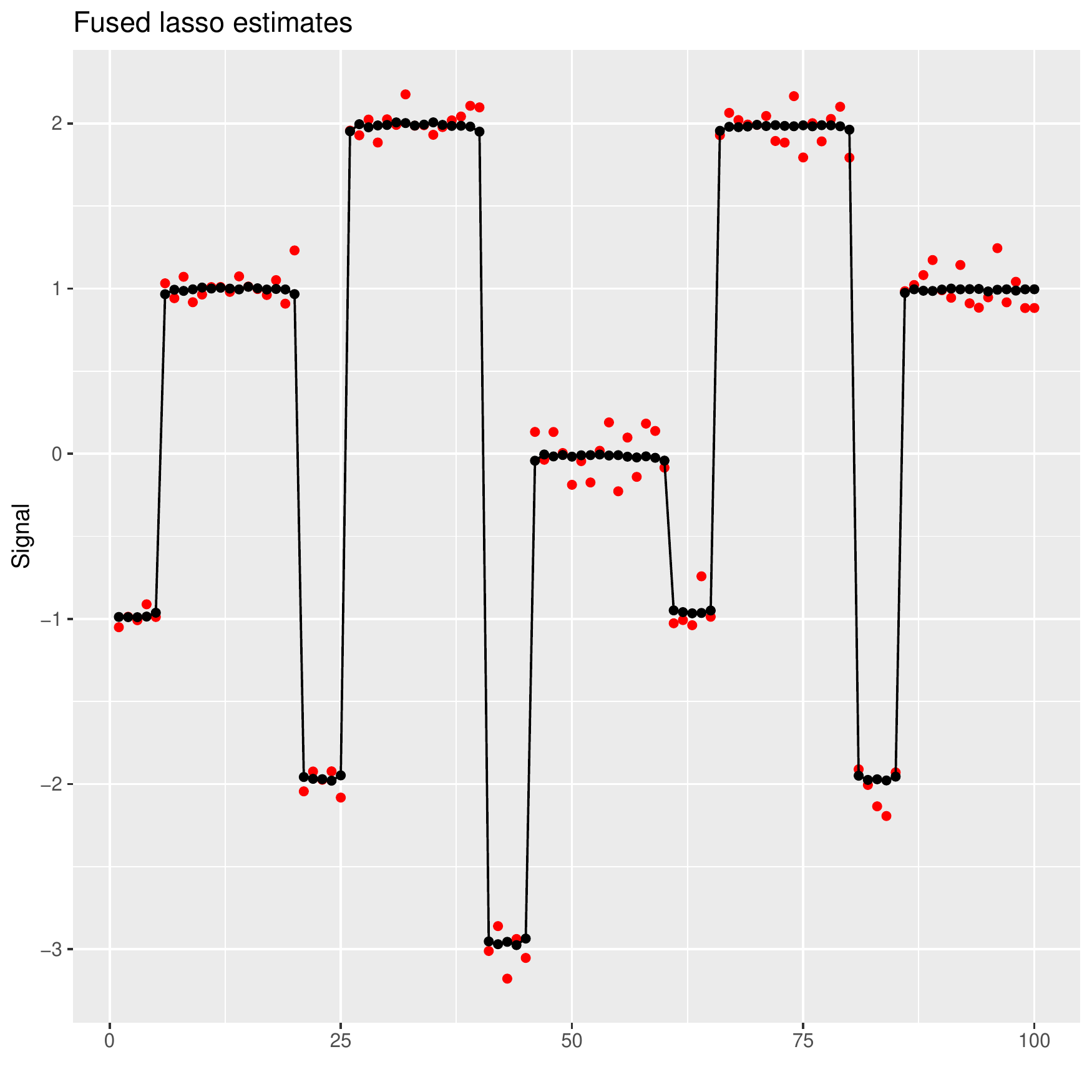} \\
		(c) Laplace fusion estimates & (d) $L_1$ fusion estimates\\[6pt]
	\end{tabular}
	\caption{Figure showing signal denoising performance for linear chain graphs in case of unevenly spaced signals and error sd $\sigma = 0.1.$ 95\% credible bands (in green) are also provided for the Bayesian procedures. The red dots correspond to observations.}\label{sim:linear-uneven}
\end{figure}

% Very uneven signals, low variance
\begin{figure}
	\begin{tabular}{cc}
		\includegraphics[width=75mm]{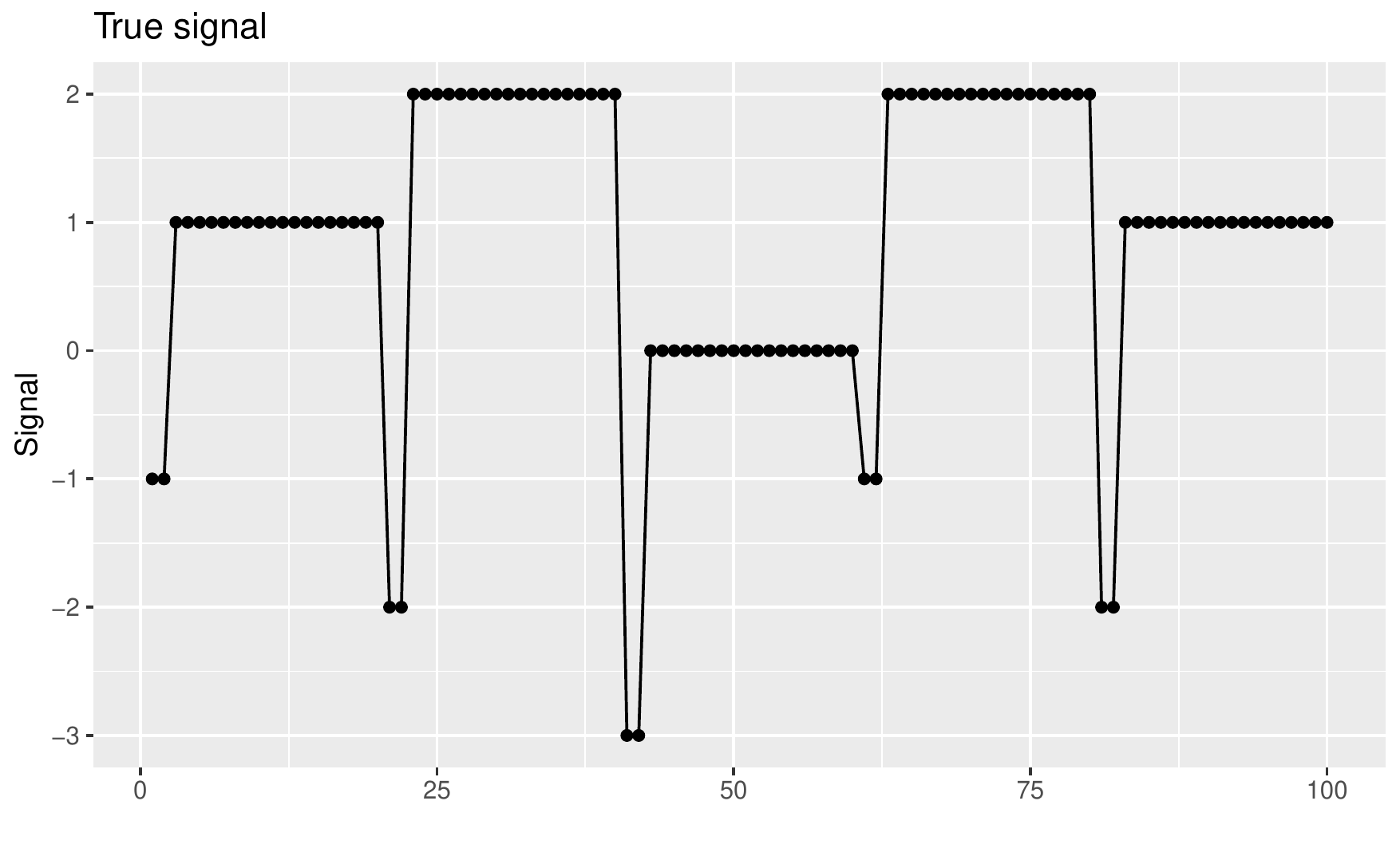} &   \includegraphics[width=75mm]{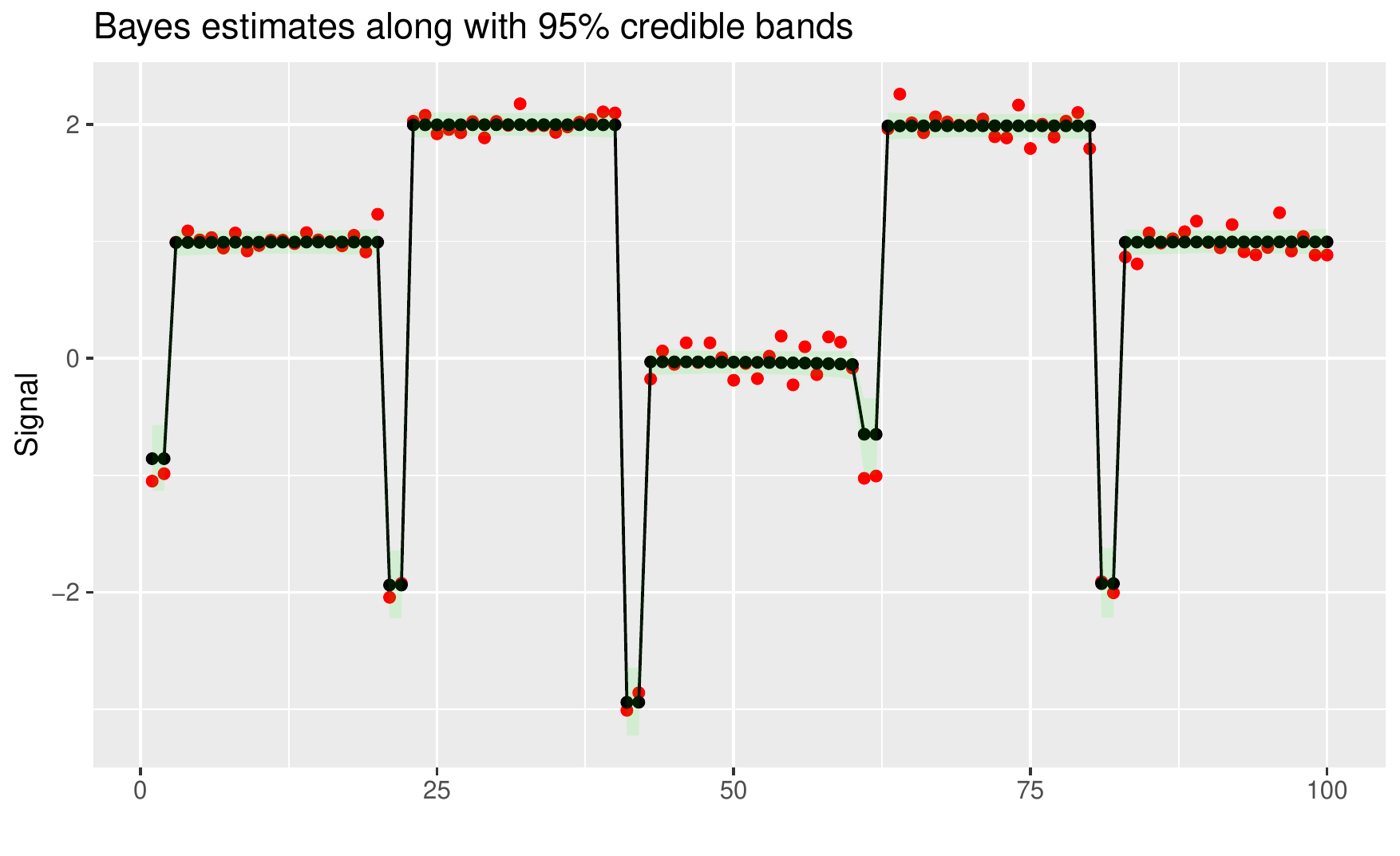} \\
		(a) True signal & (b) $t$-fusion estimates \\[6pt]
		\includegraphics[width=75mm]{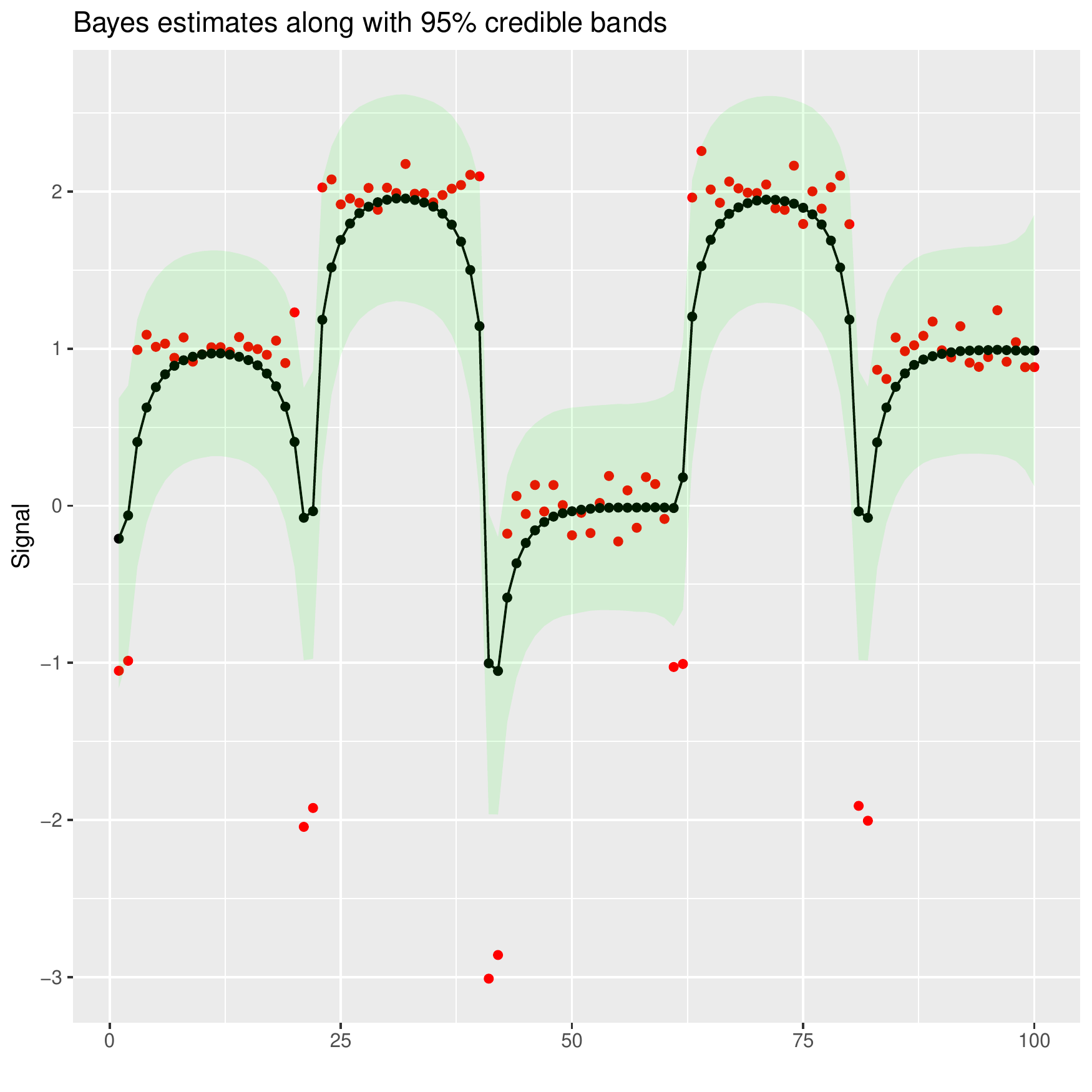} &   \includegraphics[width=75mm]{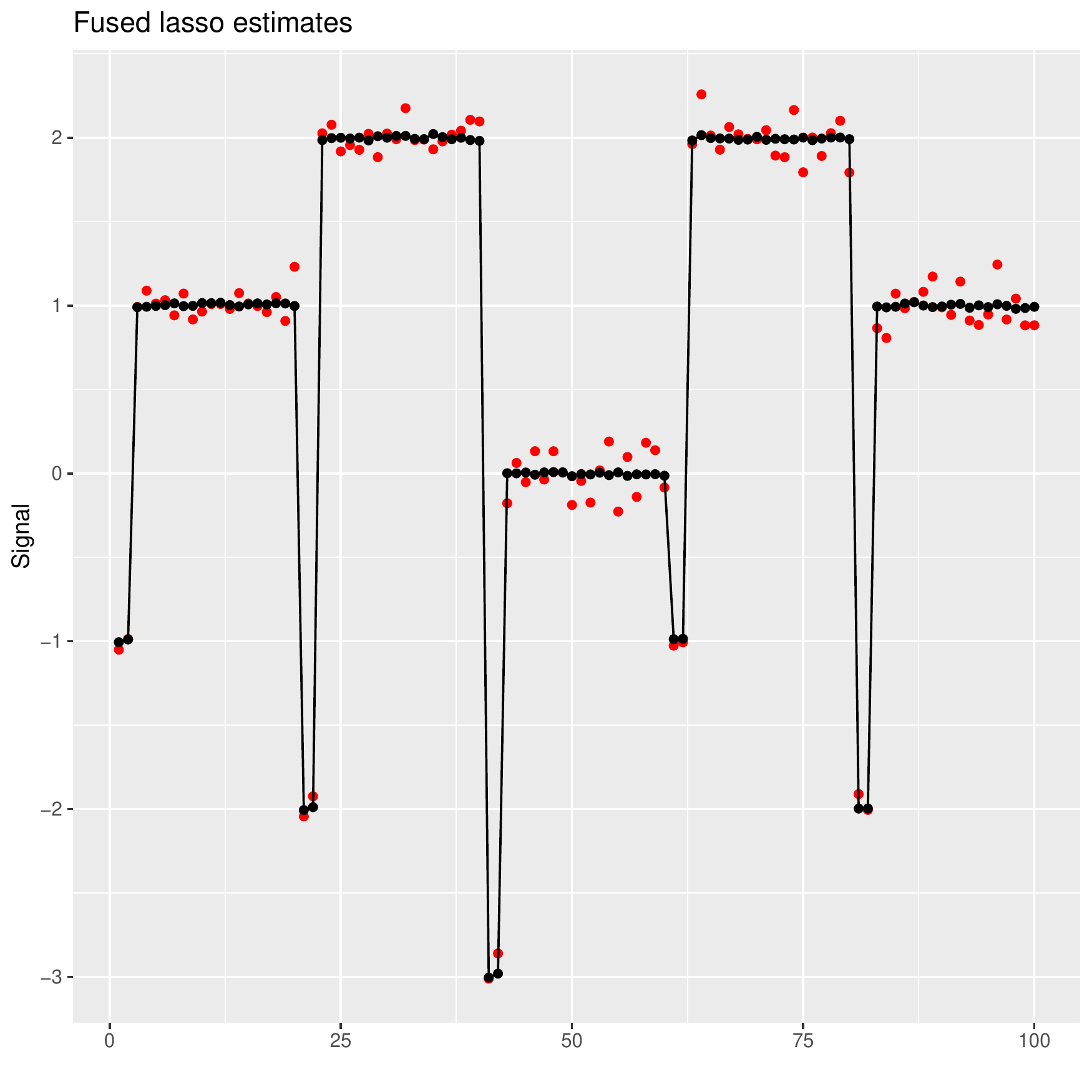} \\
		(c) Laplace fusion estimates & (d) $L_1$ fusion estimates\\[6pt]
	\end{tabular}
	\caption{Figure showing signal denoising performance for linear chain graphs in case of very unevenly spaced signals and error sd $\sigma = 0.1.$ 95\% credible bands (in green) are also provided for the Bayesian procedures. The red dots correspond to observations.}\label{sim:linear-veryuneven}
\end{figure}

% Even signals, medium variance
\begin{figure}
	\begin{tabular}{cc}
		\includegraphics[width=75mm]{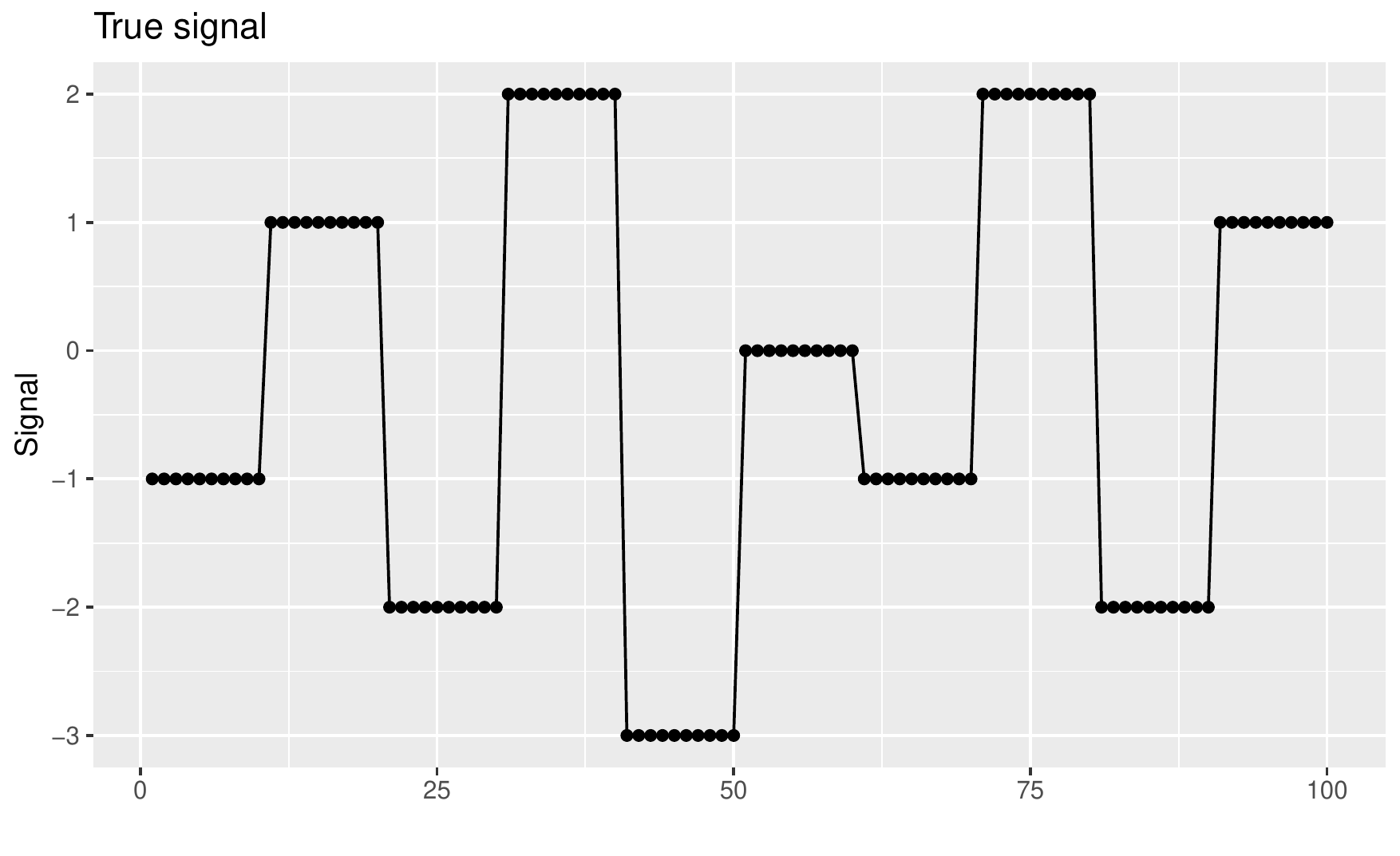} &   \includegraphics[width=75mm]{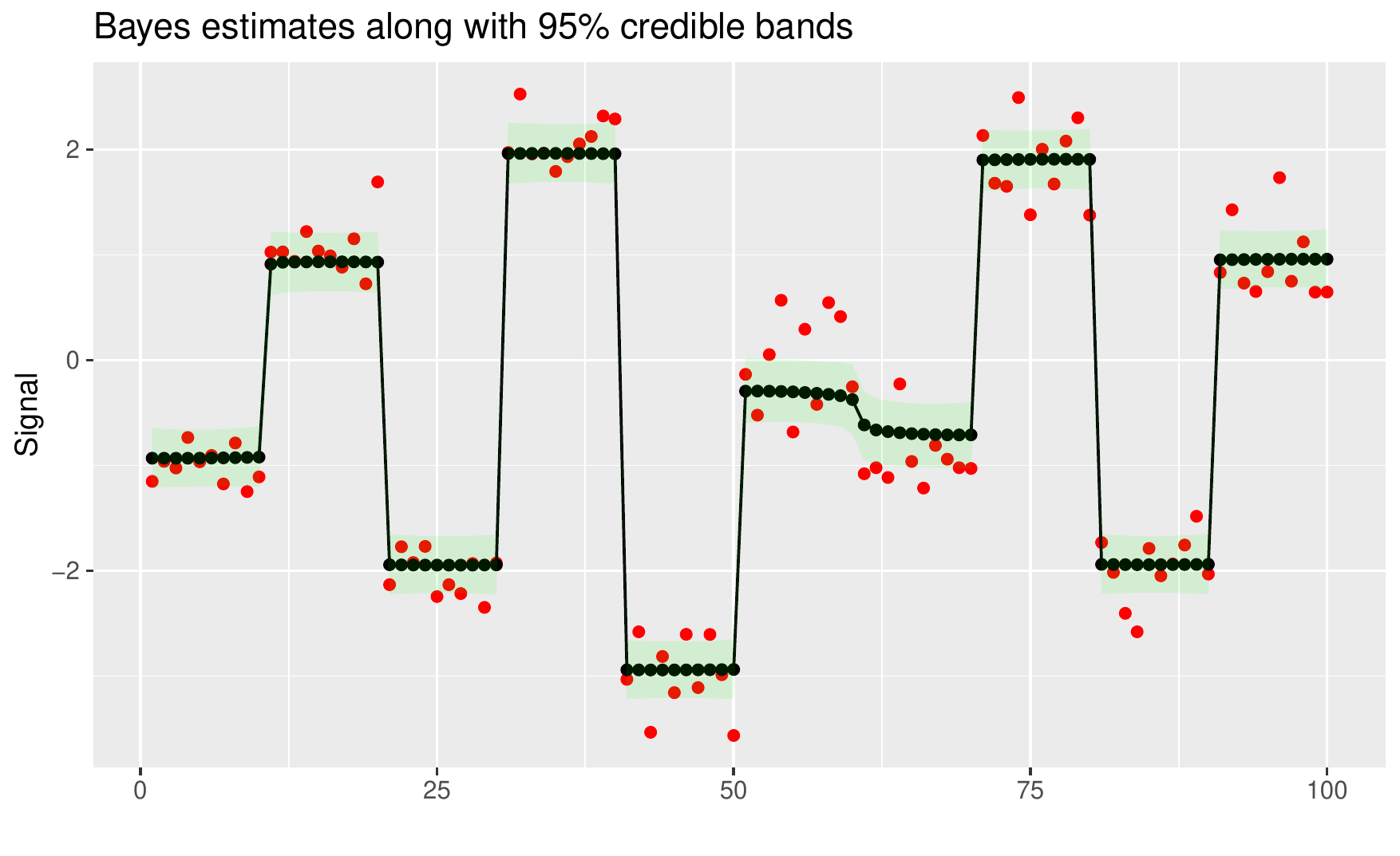} \\
		(a) True signal & (b) $t$-fusion estimates \\[6pt]
		\includegraphics[width=75mm]{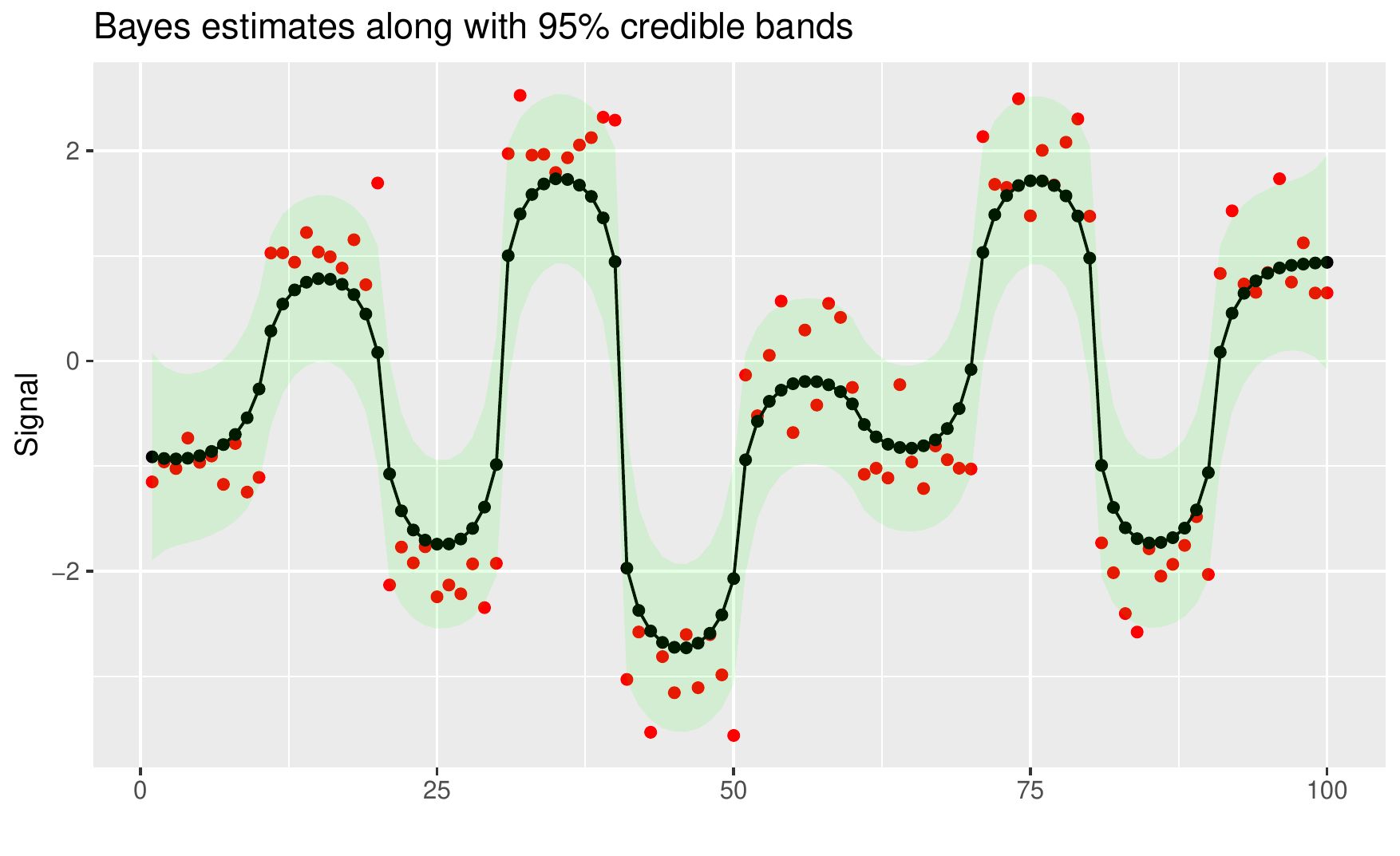} &   \includegraphics[width=75mm]{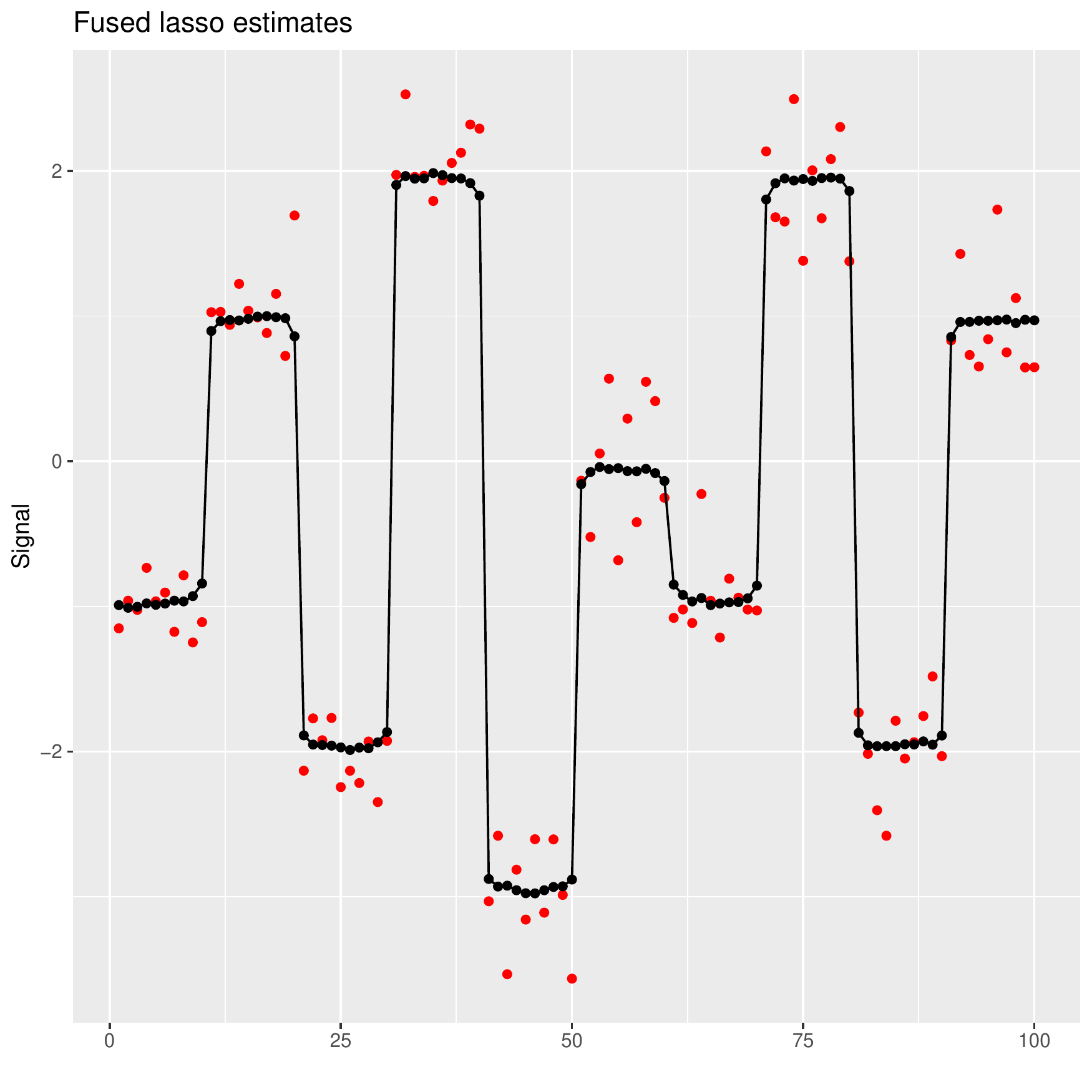} \\
		(c) Laplace fusion estimates & (d) $L_1$ fusion estimates\\[6pt]
	\end{tabular}
	\caption{Figure showing signal denoising performance for linear chain graphs in case of evenly spaced signals and error sd $\sigma = 0.3.$ 95\% credible bands (in green) are also provided for the Bayesian procedures. The red dots correspond to observations.}\label{sim:linear-even2}
\end{figure}

% Unven signals, medium variance
\begin{figure}
	\begin{tabular}{cc}
		\includegraphics[width=75mm]{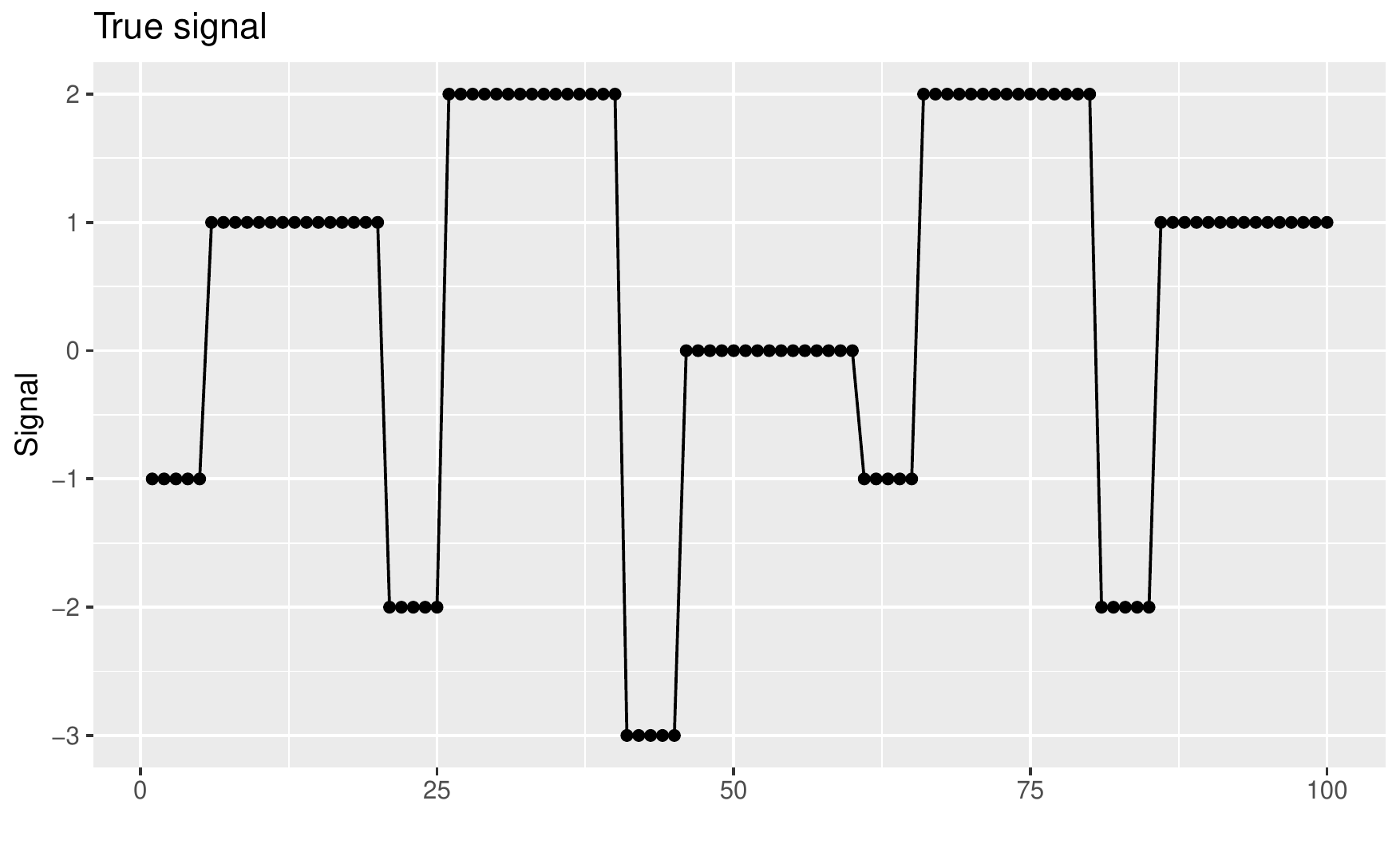} &   \includegraphics[width=75mm]{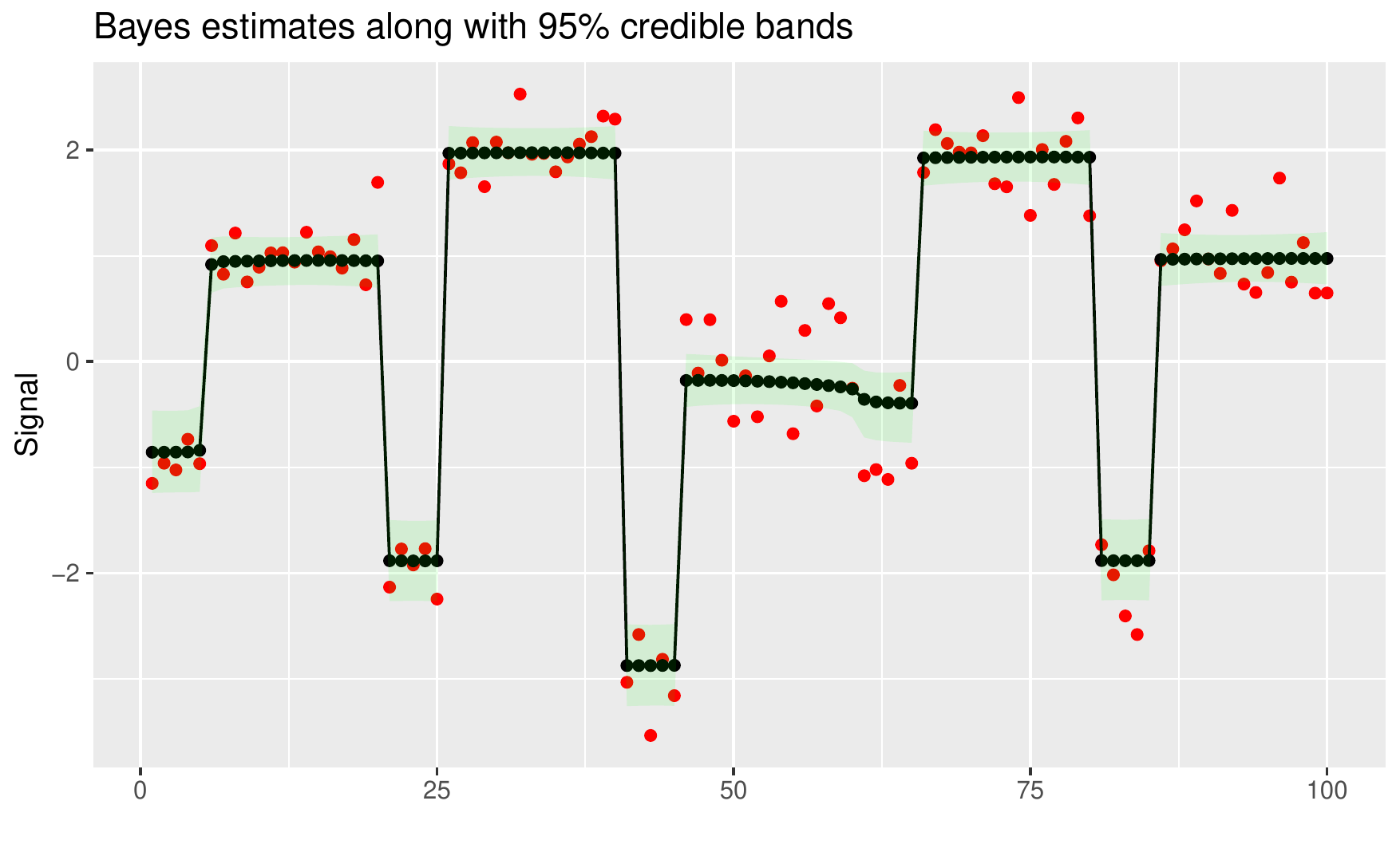} \\
		(a) True signal & (b) $t$-fusion estimates \\[6pt]
		\includegraphics[width=75mm]{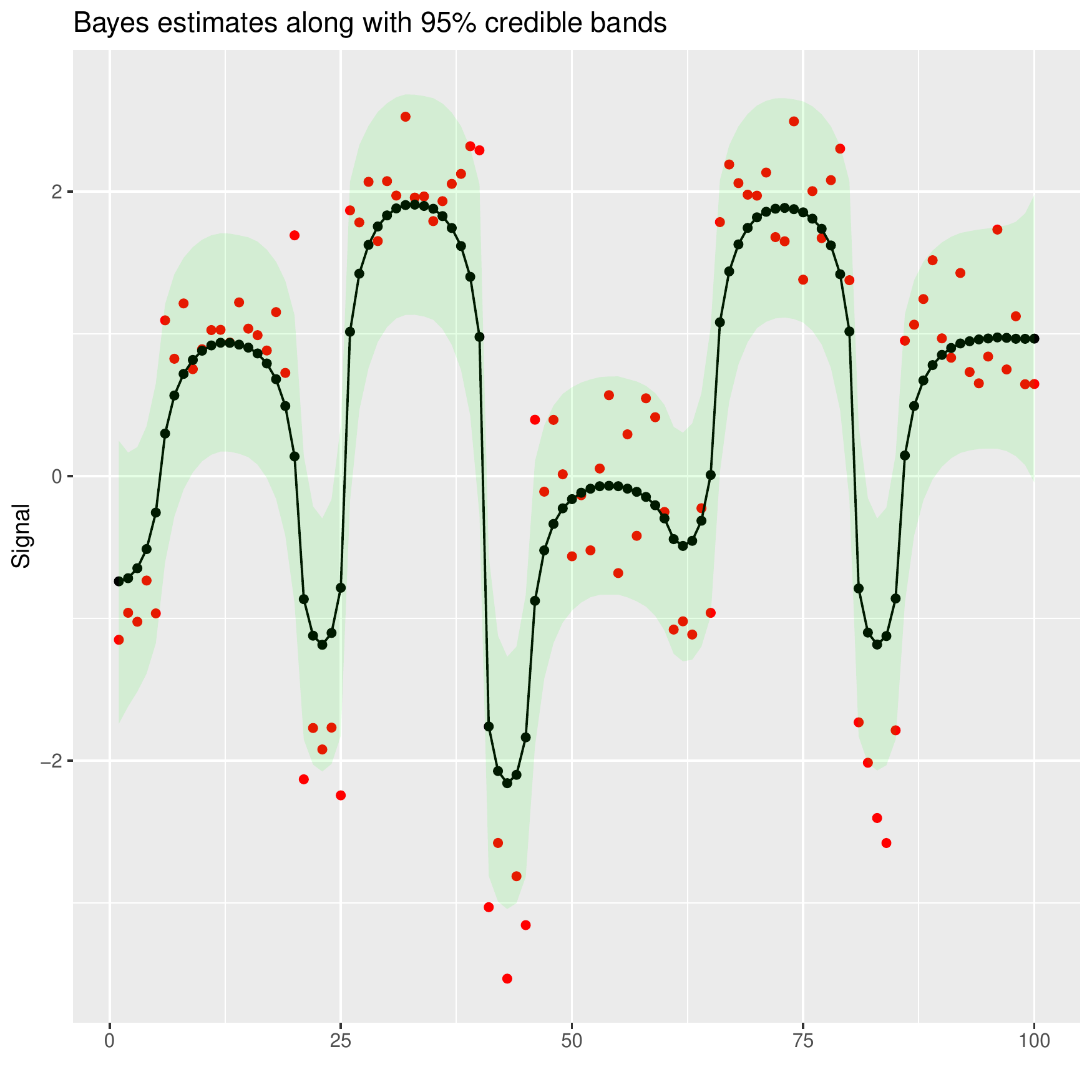} &   \includegraphics[width=75mm]{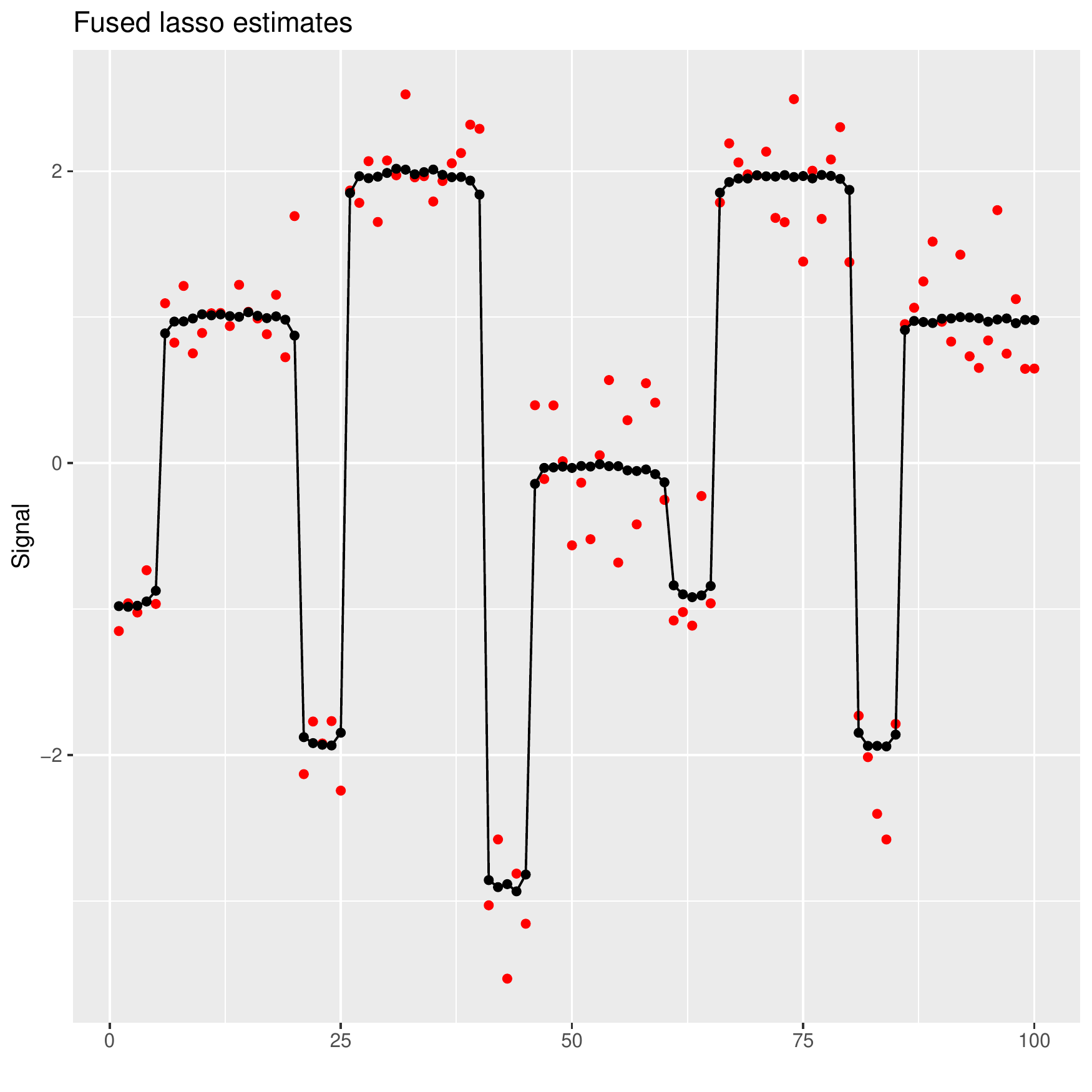} \\
		(c) Laplace fusion estimates & (d) $L_1$ fusion estimates\\[6pt]
	\end{tabular}
	\caption{Figure showing signal denoising performance for linear chain graphs in case of unevenly spaced signals and error sd $\sigma = 0.3.$ 95\% credible bands (in green) are also provided for the Bayesian procedures. The red dots correspond to observations.}\label{sim:linear-uneven2}
\end{figure}

% Very unven signals, medium variance
\begin{figure}
	\begin{tabular}{cc}
		\includegraphics[width=75mm]{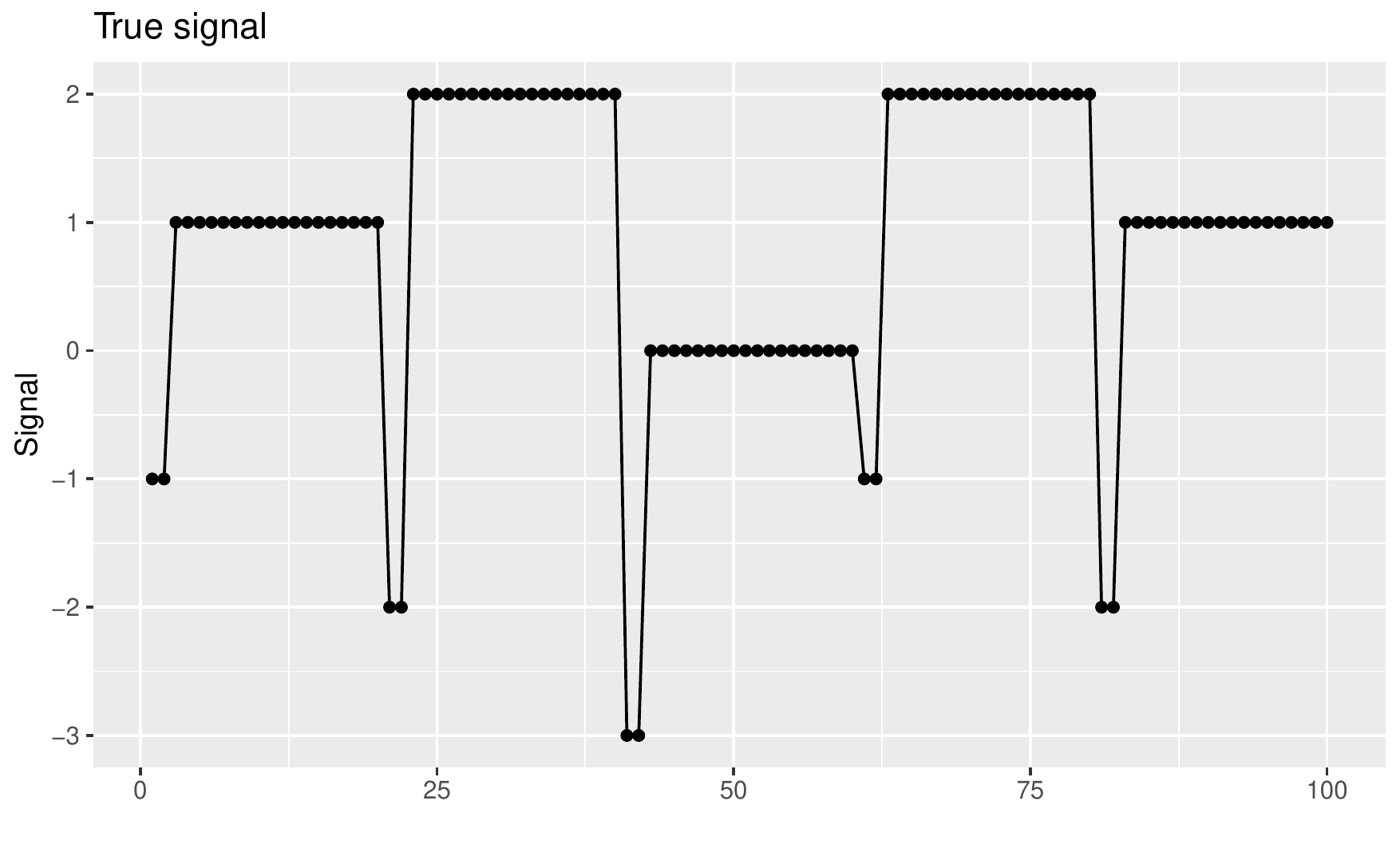} &   \includegraphics[width=75mm]{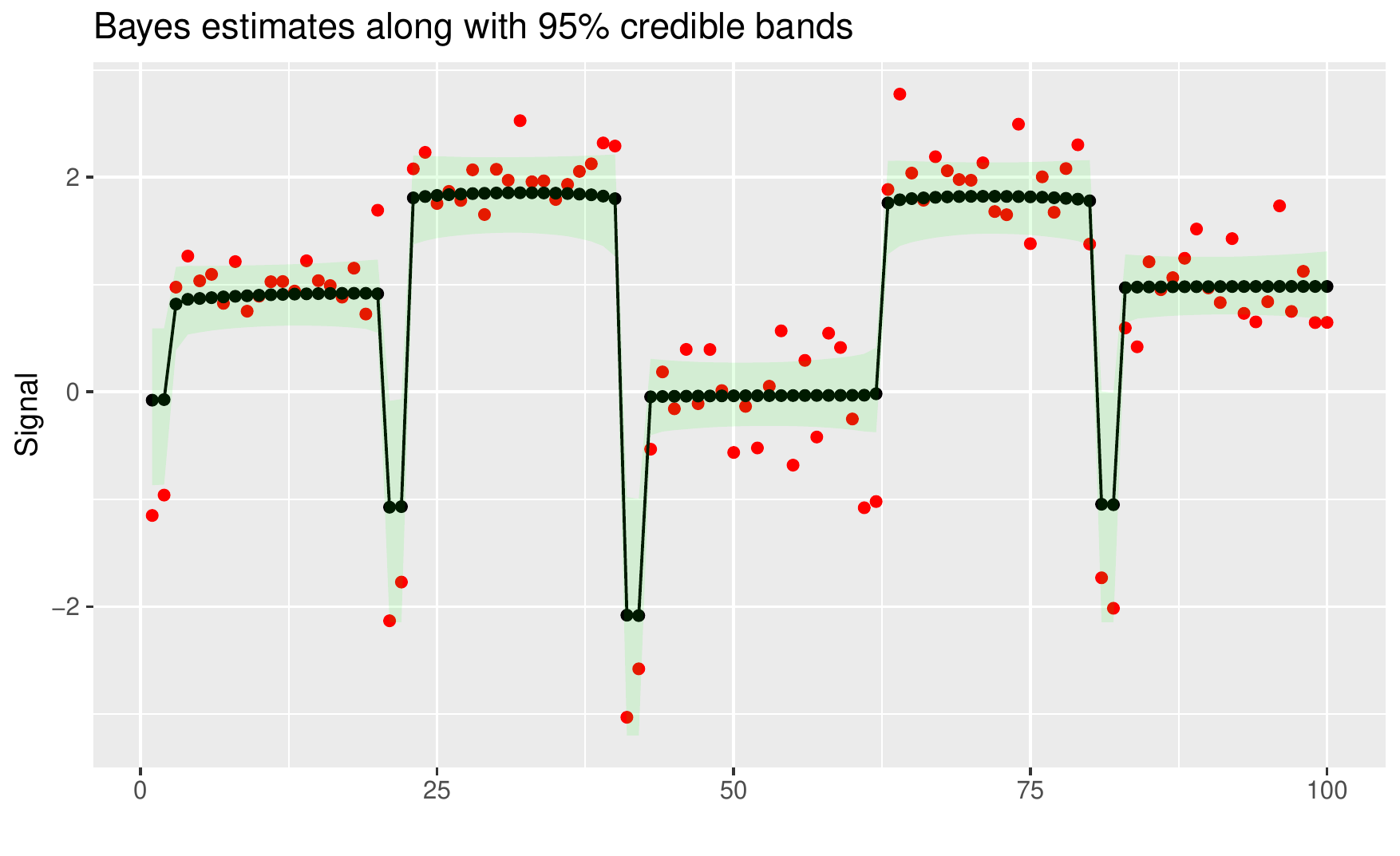} \\
		(a) True signal & (b) $t$-fusion estimates \\[6pt]
		\includegraphics[width=75mm]{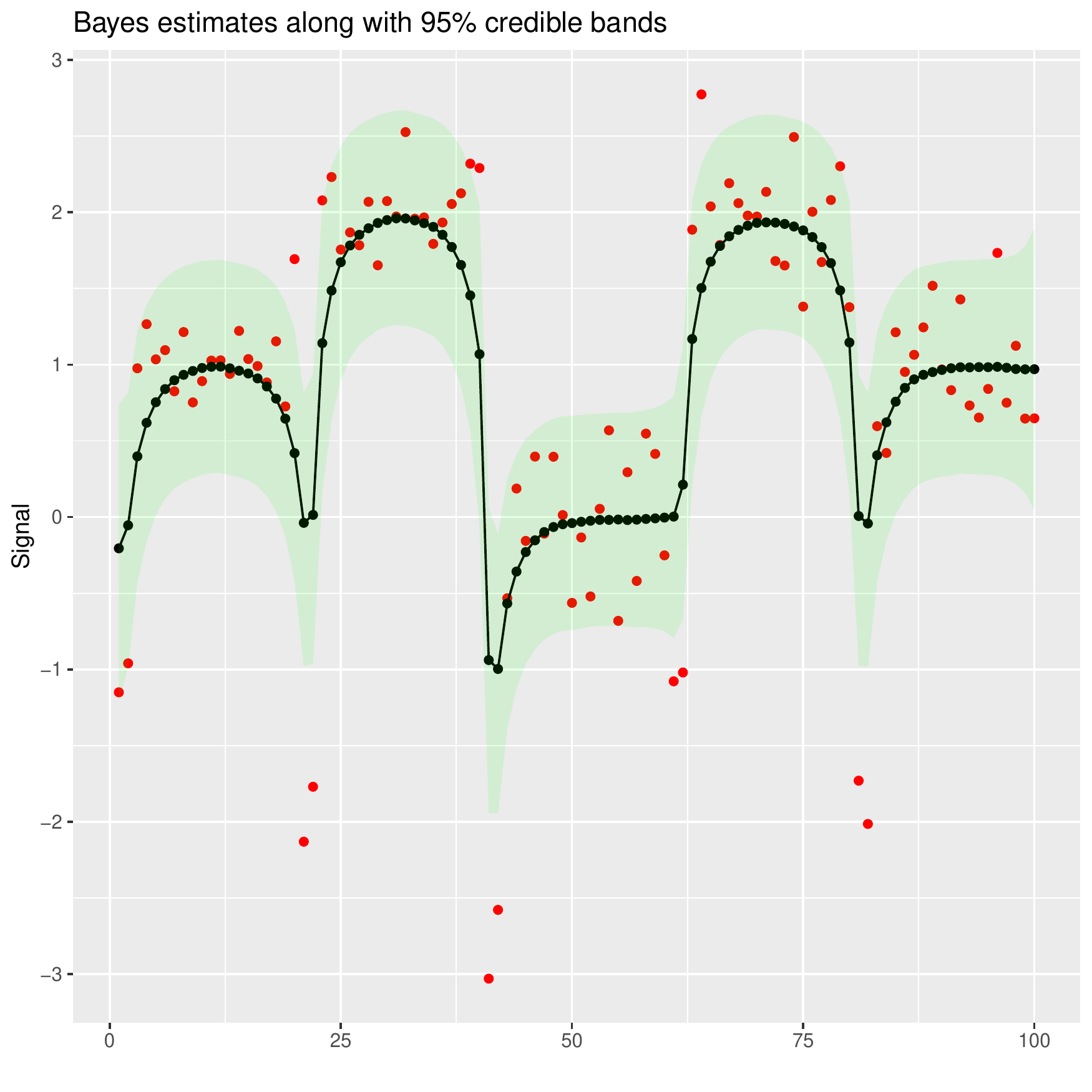} &   \includegraphics[width=75mm]{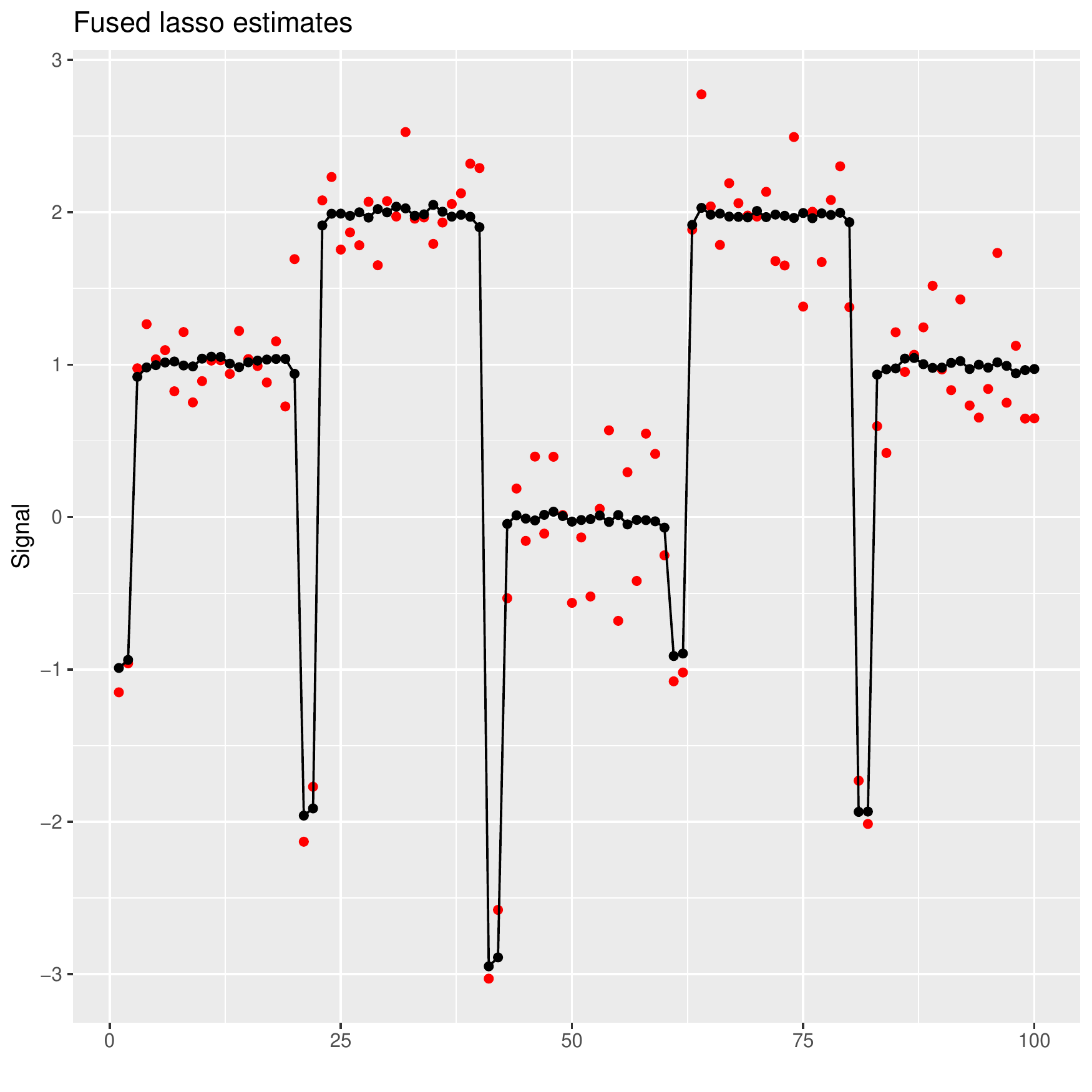} \\
		(c) Laplace fusion estimates & (d) $L_1$ fusion estimates\\[6pt]
	\end{tabular}
	\caption{Figure showing signal denoising performance for linear chain graphs in case of very unevenly spaced signals and error sd $\sigma = 0.3.$ 95\% credible bands (in green) are also provided for the Bayesian procedures. The red dots correspond to observations.}\label{sim:linear-veryuneven2}
\end{figure}

\begin{figure}
	\centering
	\begin{tabular}{cc}
		\includegraphics[width=65mm]{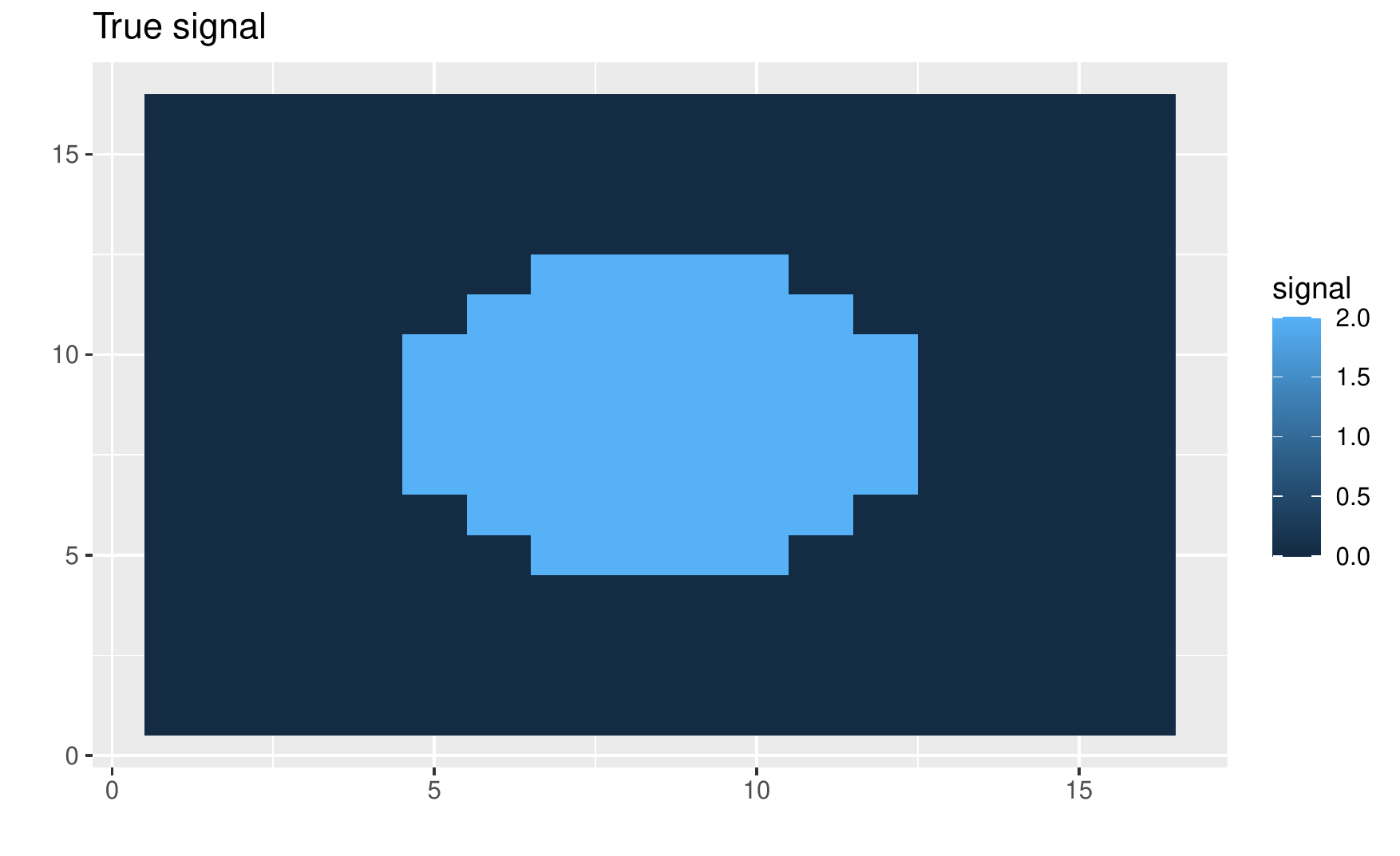} &   \includegraphics[width=65mm]{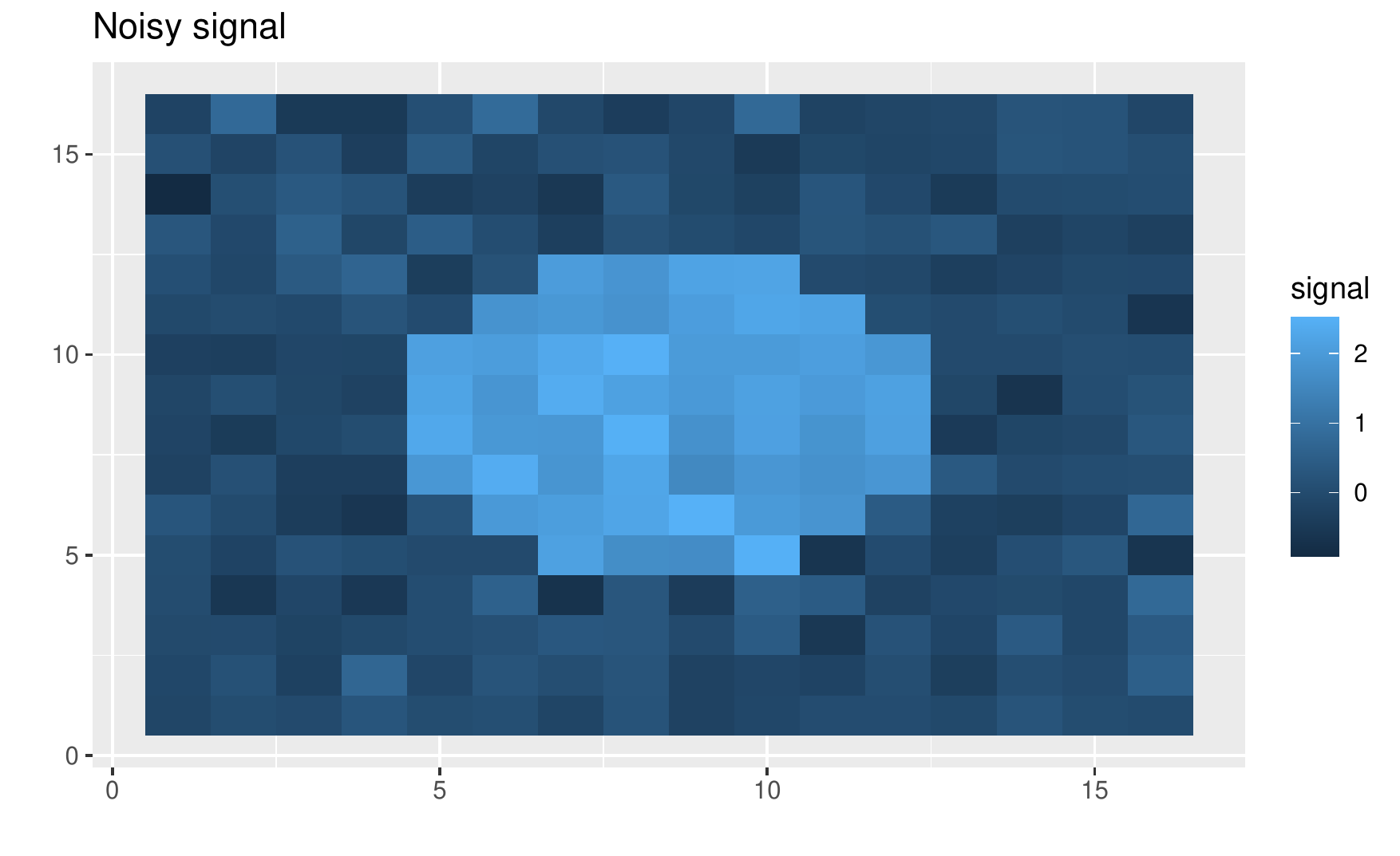} \\
		(a) True signal & (b) Noisy signal \\[1pt]
		\includegraphics[width=65mm]{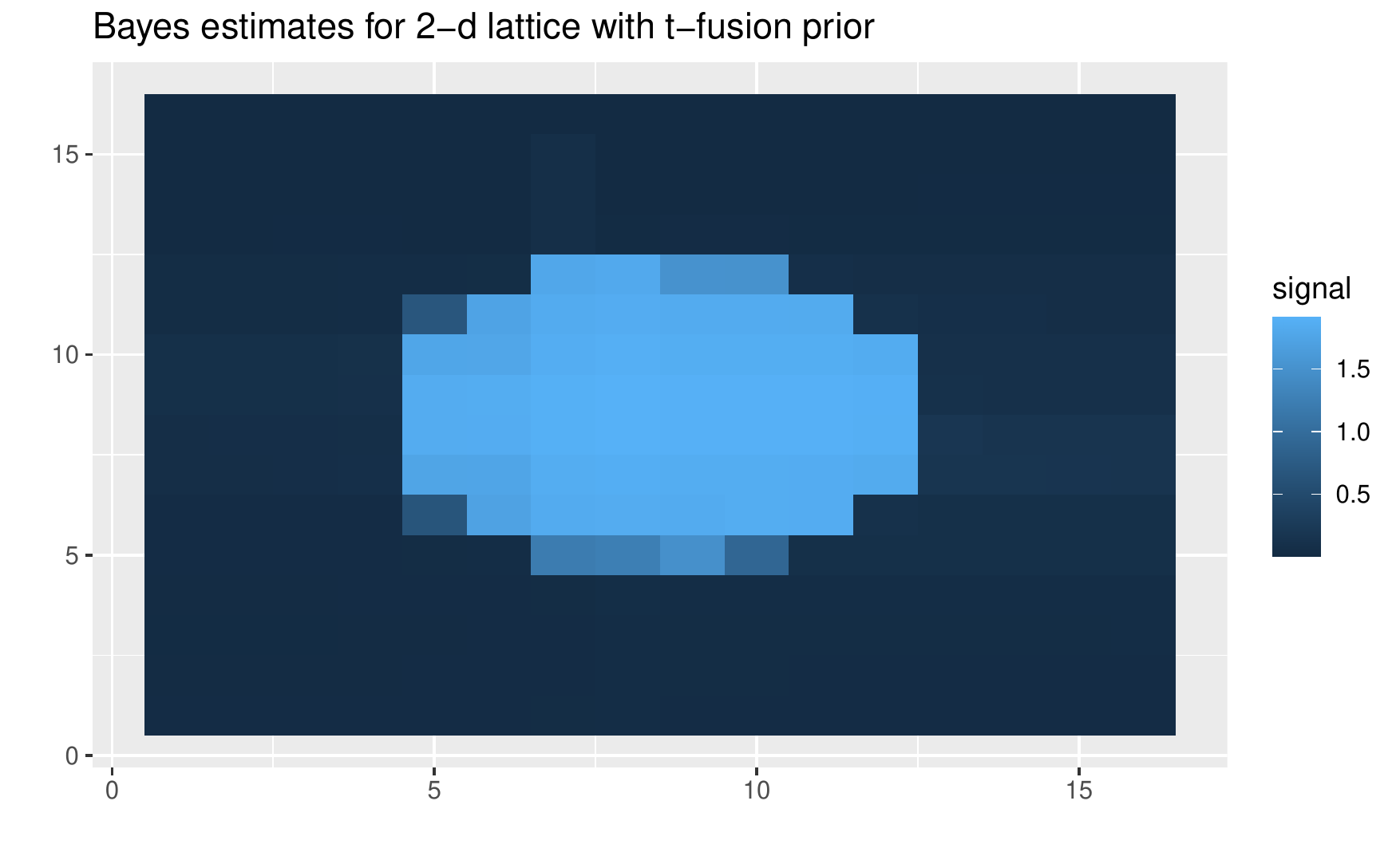} &   \includegraphics[width=65mm]{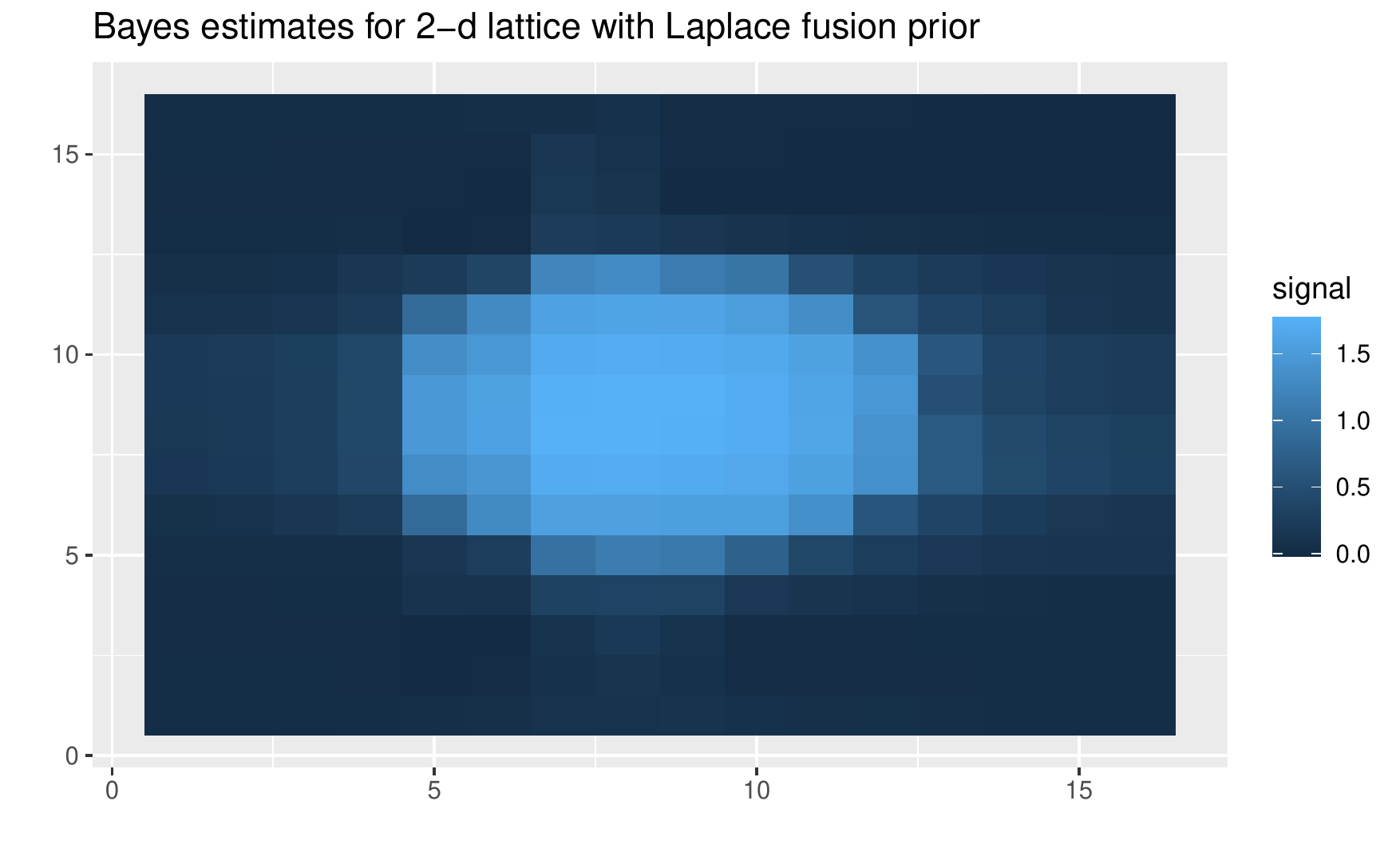} \\
		(c) $t$-fusion estimates & (d) Laplace fusion estimates\\[1pt]
		\multicolumn{2}{c}{\includegraphics[width=65mm]{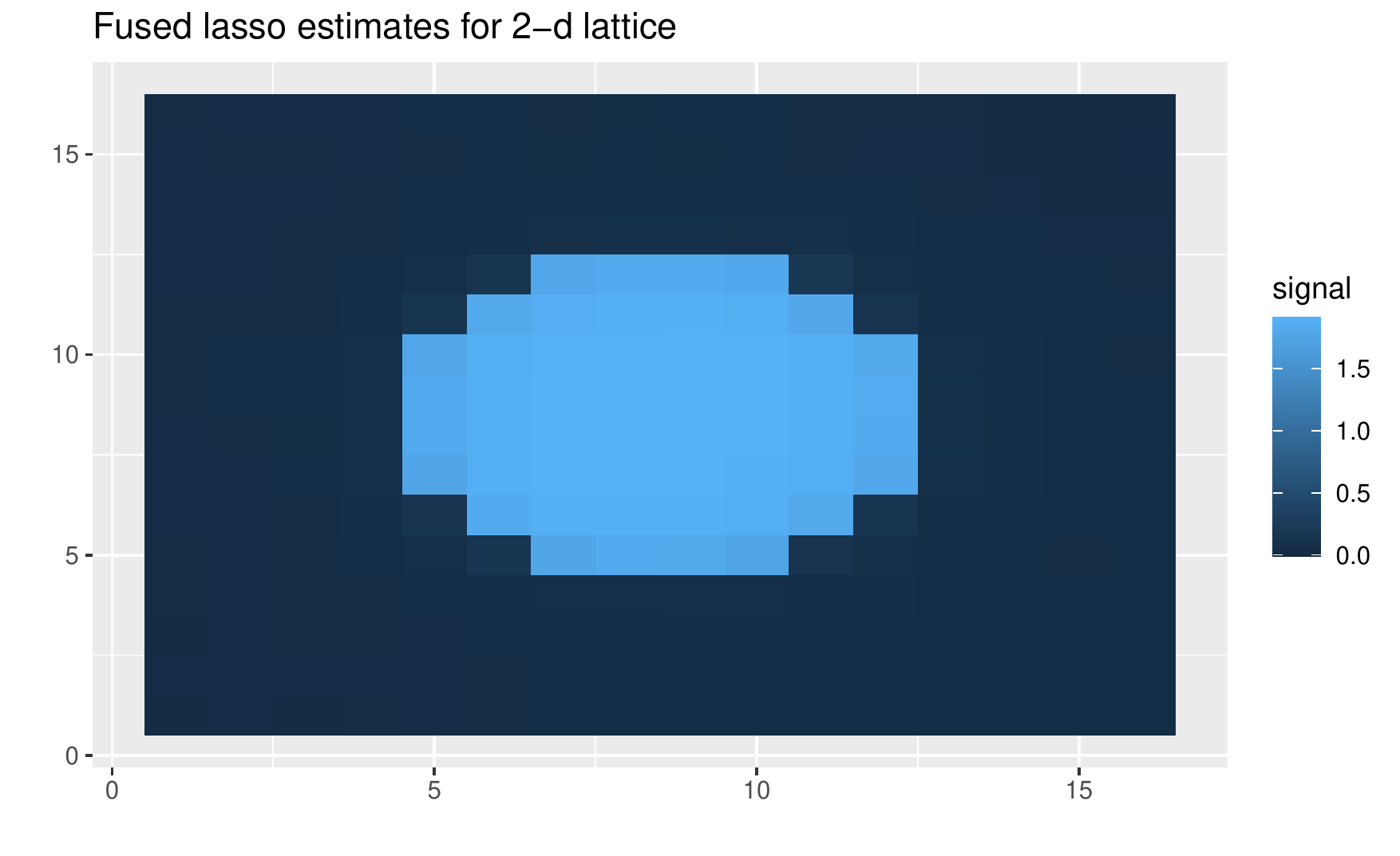} }\\
		\multicolumn{2}{c}{(e) $L_1$ fusion estimates}
	\end{tabular}
	\caption{Figure showing signal denoising performance for 2-d lattice graphs with weak signal strength.}\label{sim:2d-weak}
\end{figure}

\begin{figure}
	\centering
	\begin{tabular}{cc}
		\includegraphics[width=65mm]{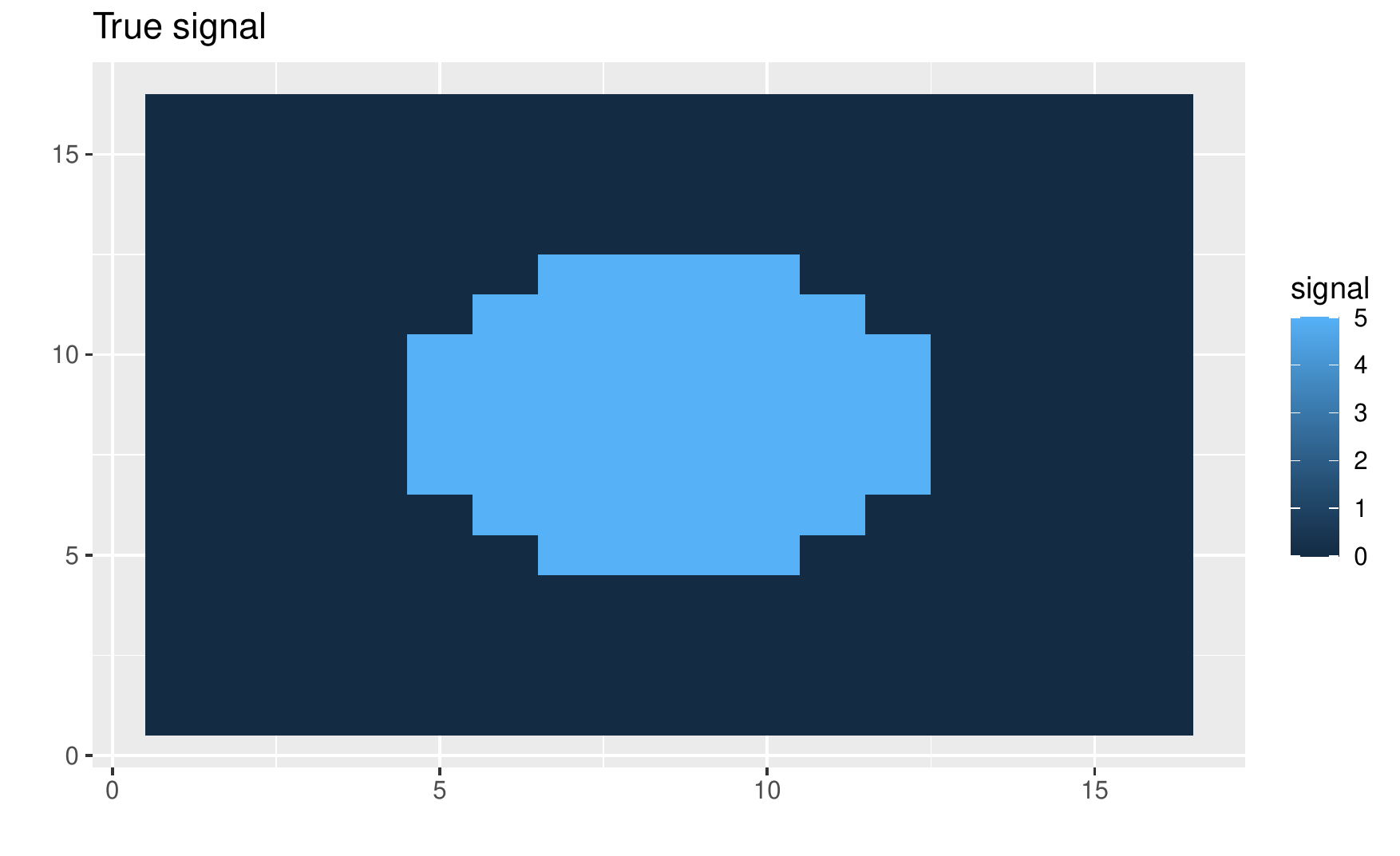} &   \includegraphics[width=65mm]{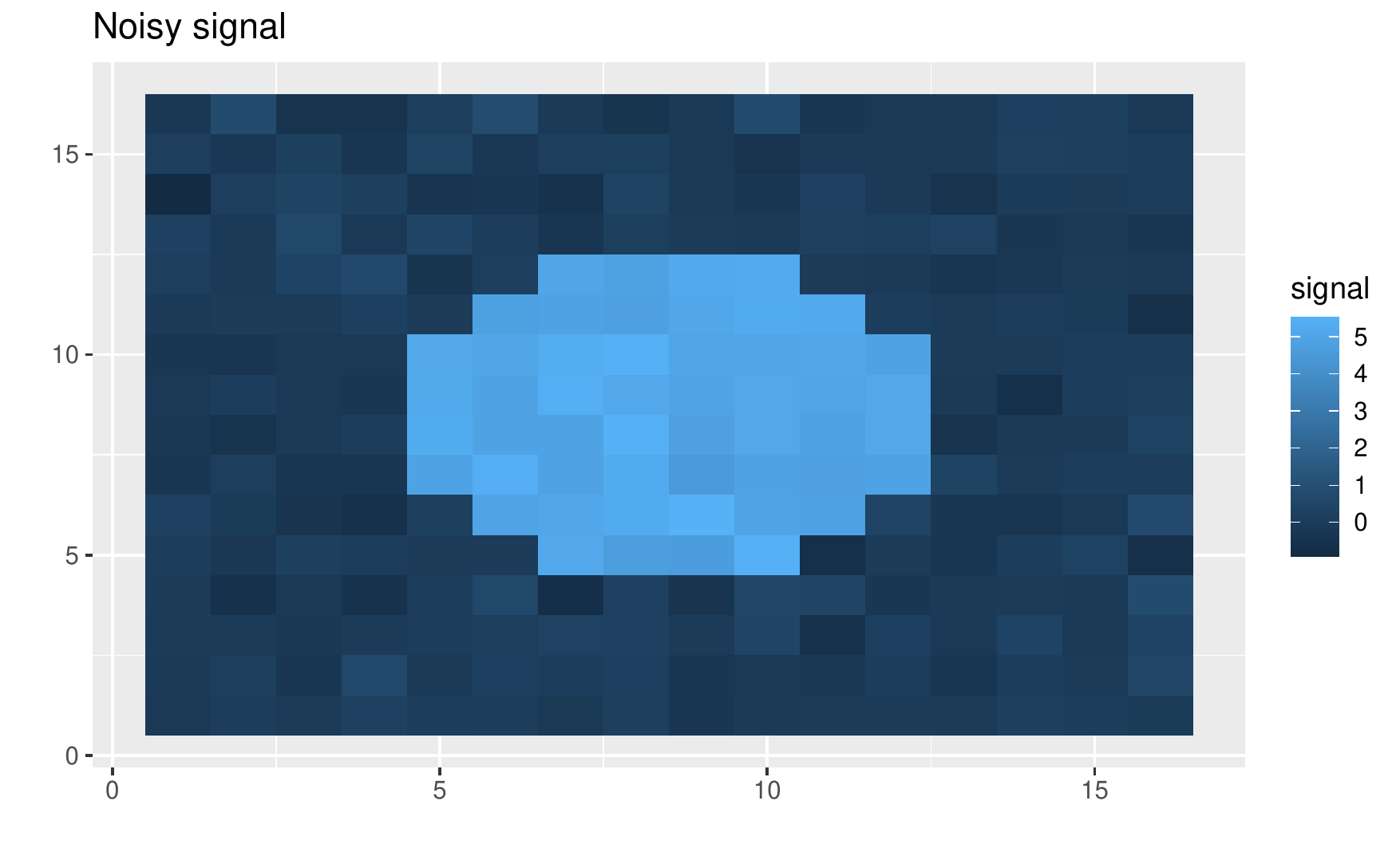} \\
		(a) True signal & (b) Noisy signal \\[1pt]
		\includegraphics[width=65mm]{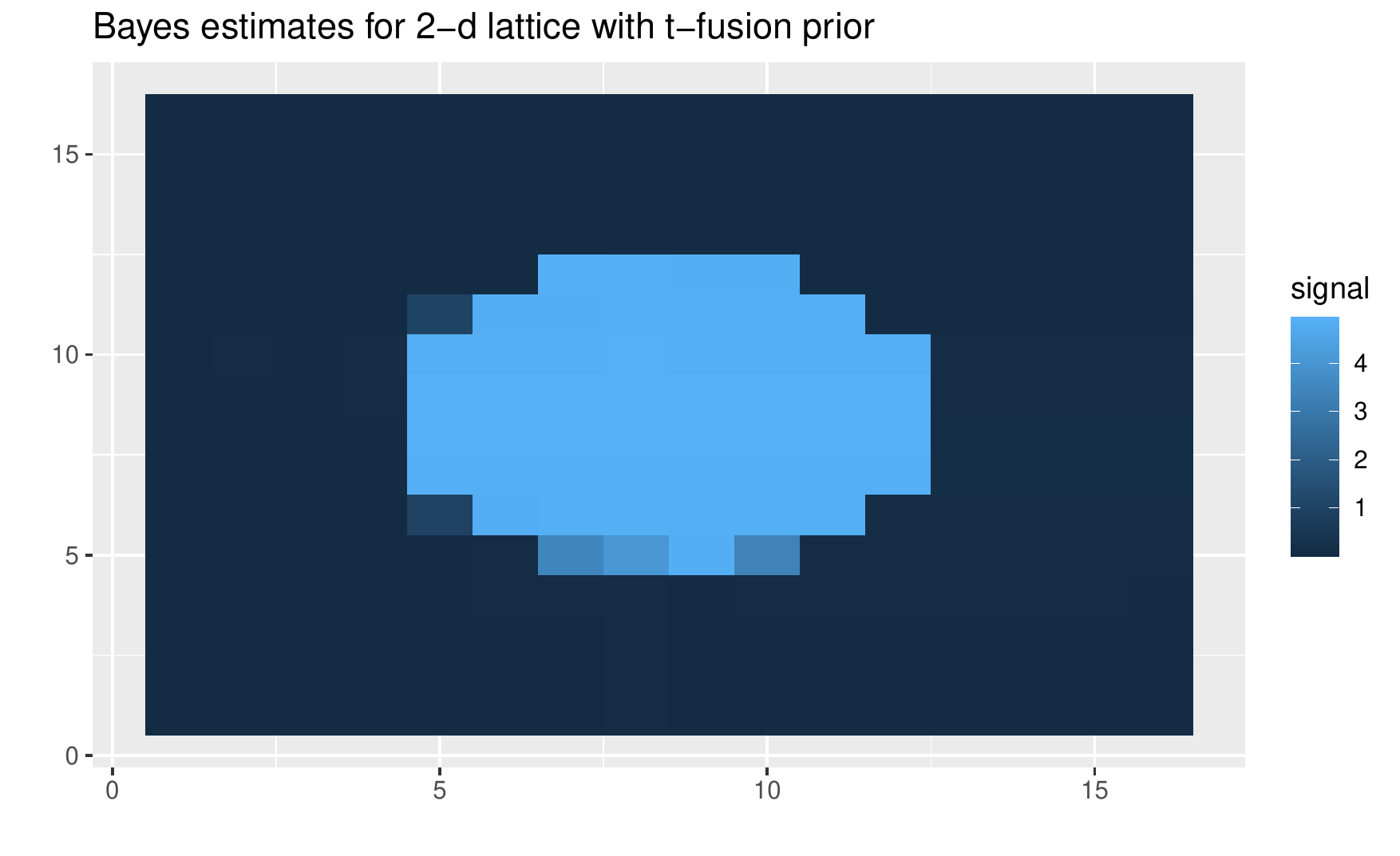} &   \includegraphics[width=65mm]{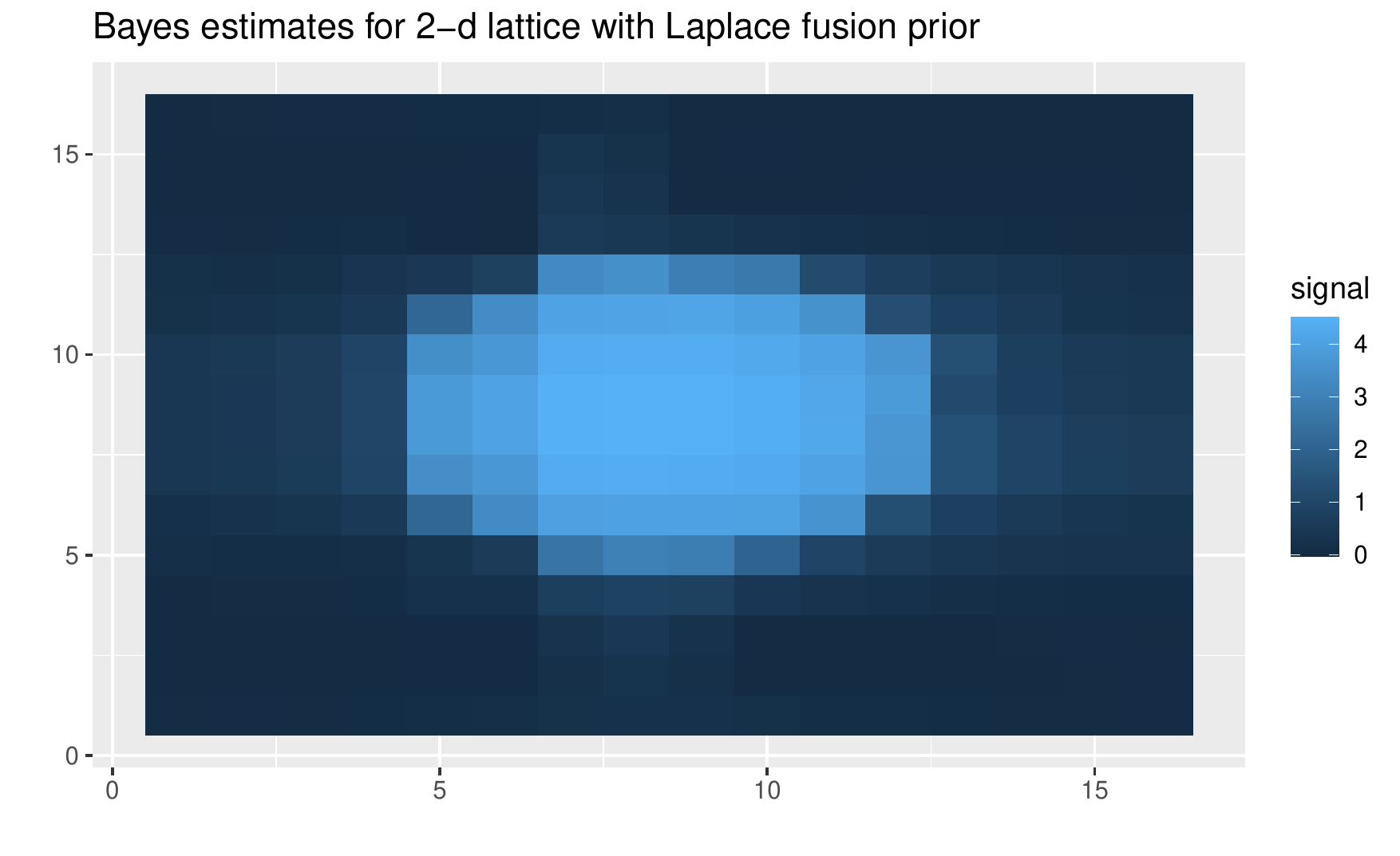} \\
		(c) $t$-fusion estimates & (d) Laplace fusion estimates\\[1pt]
		\multicolumn{2}{c}{\includegraphics[width=65mm]{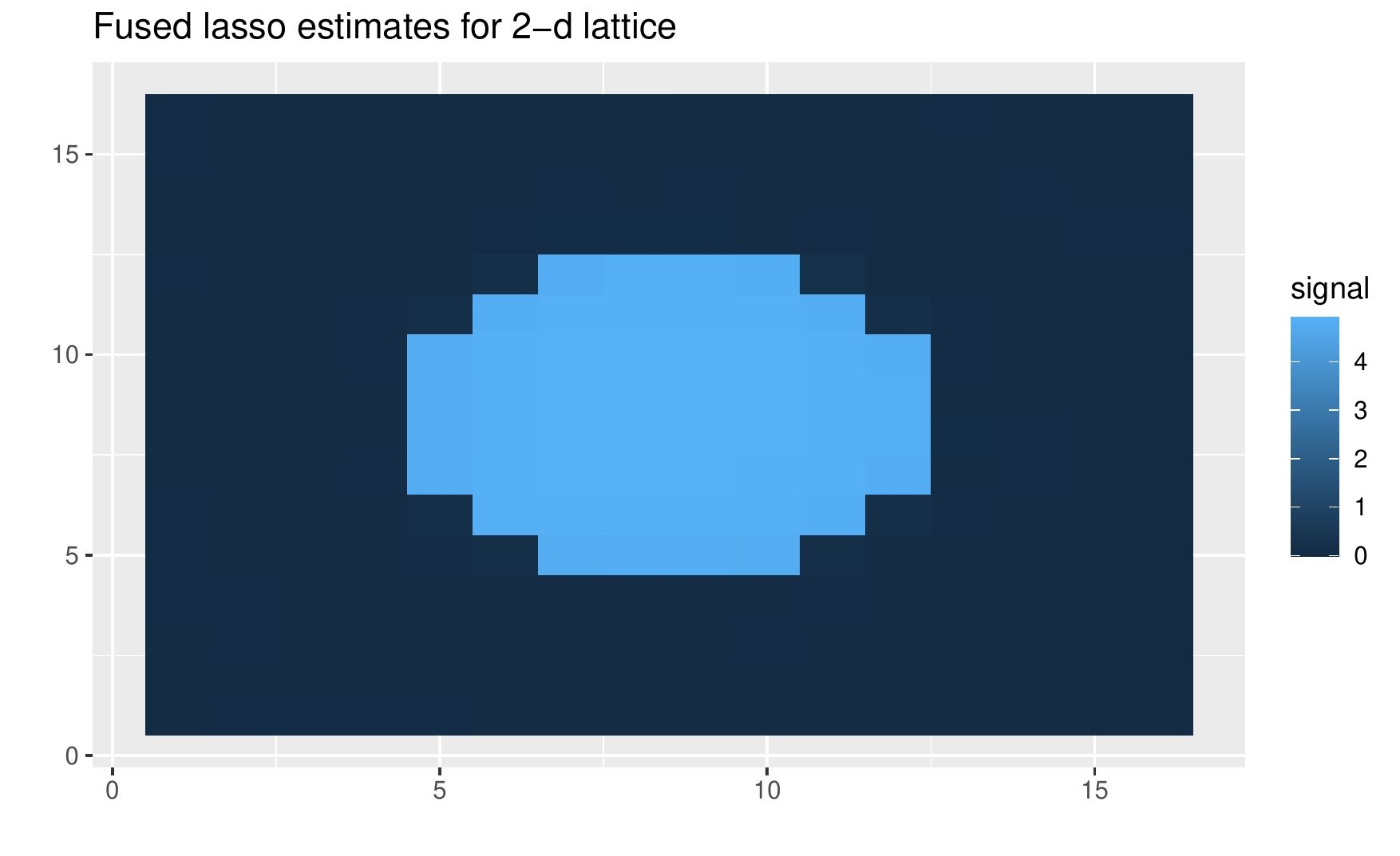} }\\
		\multicolumn{2}{c}{(e) $L_1$ fusion estimates}
	\end{tabular}
	\caption{Figure showing signal denoising performance for 2-d lattice graphs with medium signal strength.}\label{sim:2d-med}
\end{figure}

\end{document}